\documentclass[preprint,showpacs,preprintnumbers,amsmath,amssymb,superscriptaddress, nofootinbib]{revtex4}  
\usepackage{graphicx,color}
\usepackage{amsmath,amssymb}
\usepackage{url}
\usepackage{epstopdf}
\usepackage{slashed}

\newcommand{\be}{\begin{equation}}
\newcommand{\ee}{\end{equation}}
\newcommand{\bea}{\begin{eqnarray}}
\newcommand{\eea}{\end{eqnarray}}
\newcommand{\nn}{\nonumber}
\newcommand{\crn}{\nonumber \\}

\newcommand{\fr}{\frac}

\newcommand{\bc}{\begin{center}}
	\newcommand{\ec}{\end{center}}

\newcommand {\ba}{\begin{array}}
	\newcommand {\ea}{\end{array}}
\newcommand{\ben}{\begin{enumerate}}
	\newcommand{\een}{\end{enumerate}}
	
\usepackage{epsfig,graphicx} 
\usepackage{bm}
\usepackage{dcolumn}
\begin{document}
	\title{$(g-2)_{\ell}$ and LFV decays in a $U(1)_{L_{\mu} -L_{\tau}}$ left-right model with inverse seesaw neutrinos} 
\author{L. T. Hue} 
\email{lethohue@vlu.edu.vn}
\affiliation{Subatomic Physics Research Group, Science and Technology Advanced Institute, Van Lang University, Ho Chi Minh City, Vietnam}
\affiliation{Faculty of Applied Technology, Van Lang School of Technology,  Van Lang University, Ho Chi Minh City, Vietnam}

\author{Vo Quoc Phong} 
\affiliation{Department of Theoretical Physics, Faculty of Physics and Engineering Physics, University of Science, Ho Chi Minh City, Vietnam}
\affiliation{Vietnam National University, Ho Chi Minh City, Vietnam}
\email{vqphong@hcmus.edu.vn}
\author{T.D. Tham} 
\affiliation{Faculty of Natural Science Education, Pham Van Dong University, 509 Phan Dinh Phung, Quang Ngai Province, Vietnam}
\email{tdtham@pdu.edu.vn}
\author{N.H.T. Nha \footnote{corresponding author}}\email{nguyenhuathanhnha@vlu.edu.vn}
\affiliation{Department of Theoretical Physics, Faculty of Physics and Engineering Physics, University of Science, Ho Chi Minh City, Vietnam}
\affiliation{Vietnam National University, Ho Chi Minh City, Vietnam}
\affiliation{Subatomic Physics Research Group, Science and Technology Advanced Institute, Van Lang University, Ho Chi Minh City, Vietnam}
\affiliation{Faculty of Applied Technology, Van Lang School of Technology, Van Lang University, Ho Chi Minh City, Vietnam}

\begin{abstract}
In the framework of a $U(1)_{L_{\mu} -L_{\tau}}$ left-right model with inverse seesaw neutrinos  recently proposed, we show that one-loop contributions from heavy singly charged  Higgs bosons  to the anomalous magnetic moments of charged leptons  $(g-2)_{e,\mu}$ can accommodate the observed discrepancies, of $10^{-9}$ and $\mathcal{O}(10^{-13})$   for the  muon and electron,  respectively.  Meanwhile, all lepton-flavor violating decay rates for  $h\to e_b^\pm e_a^\mp $, $Z\to e_b^\pm e_a^\mp$, and   $e_b\to e_a \gamma$ are found to be highly suppressed as a consequence of  the gauged  $U(1)_{L_{\mu}-L_\tau}$  symmetry. 
%
\end{abstract}

\maketitle
\allowdisplaybreaks 
\section{Introduction}

We study in detail the one-loop contributions to the anomalous magnetic moments  $(g-2)_{e_a}$ of charged leptons $e_a$ and their lepton-flavor-violating (LFV) decays in the framework of the $U(1)_{L_{\mu} -L_{\tau}}$  left-right model (LRiss) with inverse-seesaw (ISS) neutrinos introduced in Refs. \cite{Majumdar:2020xws, Majumdar:2022jur}. The gauged $U(1)_{L_{\mu}-L_{\tau}}$  symmetry possesses several interesting features, including automatic anomaly cancellation \cite{He:1991qd, Foot:1994vd}, a  $\mu-\tau$ symmetry in the neutrino mass matrix \cite{Bell:2000vh, Joshipura:2003jh}, and a new neutral gauge boson that can give a sizable positive one-loop contribution to $(g-2)_{\mu}$ \cite{Baek:2001kca, Ma:2001md, Heeck:2011wj, Altmannshofer:2016brv, Zhou:2021vnf, Singh:2022tvz, Li:2025myw}.  The LRiss model is a left-right (LR) extension of the original LR models \cite{Pati:1974yy, Mohapatra:1974gc, Senjanovic:1978ev, Senjanovic:1975rk, FileviezPerez:2016erl} and retains the characteristic neutral gauge boson  $Z'_{\mu \tau}$ associated with the $U(1)_{L_{\mu}-L_{\tau}}$ symmetry \cite{Majumdar:2020xws, Majumdar:2022jur}. In this framework, LFV decay rates, such as $\tau \to \mu \gamma$, can be sizable. On the other hand, the mass $m_{Z'}$ is constrained by experimental data. For example, requiring equal left- and right-handed $U(1)_{L_{\mu}-L_{\tau}}$ couplings of $Z'$, $g^L_{\mu\tau}=g^R_{\mu\tau}$, leads to a stringent lower bound on $m_{Z'}$ \cite{Altmannshofer:2016brv}. Nevertheless, interesting regions of parameter space in simple $U(1)_{L_{\mu}-L_{\tau}}$ models remain compatible with a relatively heavy neutral gauge boson $Z'$ associated with the new $U(1)$ gauge symmetry \cite{Wang:2025kit}. It is therefore interesting to investigate whether a viable region accommodating both LFV and $(g-2)$ data can arise in a more general BSM framework such as the LRiss model.

In addition, the LRiss model contains singly charged Higgs bosons and ISS neutrinos, which can also give sizable one-loop contributions to both $ a_{e_a} \equiv (g-2)_{e_a}/2$ and LFV decay amplitudes, including charged-lepton-flavor-violating (cLFV) decays $(e_b\to e_a\gamma)$, LFV decays of the SM-like Higgs boson (LFV$h$), $h\to e_b^\pm e_a^\mp$, and LFV$Z$-boson decays, $Z\to e_b^\pm e_a^\mp$ \cite{Hue:2024rij, Zeleny-Mora:2025tiw}. More importantly, defining $\Delta a^{\mathrm{BSM}}_{e_a}\equiv  a^{\mathrm{BSM}}_{e_a} -a^{\mathrm{SM}}_{e_a}$,  as the discrepancy between the BSM and SM predictions, various BSM models incorporating specific structures of the ISS neutrino mass matrix and/or simple LFV couplings can predict stringent correlations between $\Delta a_{e_a}$ and cLFV decay rates \cite{Crivellin:2018qmi, Hong:2023rhg, Hue:2021xap, Huang:2024iip, Alvarado:2025zud, Nha:2026pvq, Nha:2026yat}.

  The current  experimental data  on $(g-2)_{\mu}$  \cite{Muong-2:2023cdq, Muong-2:2025xyk} are consistent with the  SM prediction \cite{Aliberti:2025beg}, with   $\Delta a_{\mu}\equiv \Delta a^{\mathrm{exp}}_{\mu}=  a^{\mathrm{exp}}_{\mu} -a^{\mathrm{SM}}_{\mu} =\left(3.8\pm 6.3 \right) \times 10^{-10}$, which still allows for sizable values of $\Delta a_{\mu} \simeq  10^{-9}$.  The experimental values of $(g-2)_e$, reported by different groups \cite{Hanneke:2008tm, Parker:2018vye, Morel:2020dww, Fan:2022eto}, exhibit discrepancies of the same order of magnitude, with $\left|\Delta a^{\mathrm{exp}}_e\right|=\mathcal{O}(10^{-13})$ \cite{Aoyama:2012wj, Laporta:2017okg, Aoyama:2017uqe, Terazawa:2018pdc, Volkov:2019phy, Gerardin:2020gpp}. Therefore, strong correlations between cLFV decay rates and $(g-2){e_a}$ in BSM scenarios may impose stringent constraints on both of them. In this work, we investigate all one-loop contributions predicted by the LRiss model to $\Delta a_{e_a}$ and LFV decay rates, including LFV$h$ and LFV$Z$ decays, which have not been discussed previously. The effects of $U(1)_{L_{\mu}-L_{\tau}}$ on the Yukawa couplings are investigated in detail to determine whether singly charged Higgs boson exchange can generate sizable one-loop contributions to  $\Delta a_{e_a}$ and LFV decay rates.

Our work is organized as follows. In Sec. \ref{sec:LISS}, we review the LRiss model, including the particle content, Higgs potential, and Yukawa Lagrangian introduced in Ref. \cite{Majumdar:2020xws}. We also determine the physical states and relevant mixing parameters, and analyze the ISS mechanism for generating active neutrino masses and the total neutrino mixing matrix. In Sec. \ref{sec:Feynrule}, we derive the relevant couplings and Feynman rules needed to calculate the one-loop contributions to $\Delta a_{e_a}$ and the LFV decay rates in the LRiss framework. We also provide qualitative estimates of the one-loop contributions from singly charged Higgs bosons to these quantities. In Sec. \ref{sec:MRij}, we consider all possible $U(1)_{L_{\mu}-L_{\tau}}$  charge assignments of the new neutral lepton singlets that realize the ISS mechanism and their implications for LFV processes. The numerical results and discussion are presented in Sec. \ref{sec:num}. Finally, we summarize our main results in Sec. \ref{sec:conclusion}. Two appendices provide detailed calculations of the Higgs sector and LFV decay amplitudes.

\section{ \label{sec:LISS} The $U(1)_{L_{\mu} -L_{\tau} }$ left-right model with ISS neutrinos }

The model was constructed based on the total gauge group 
\begin{equation}\label{eq:Stotal}
SU(2)_L \otimes SU(2)_R \otimes U(1)_{B-L} \otimes SU(3)_C\otimes U(1)_{L_{\mu}-L_{\tau}}.
\end{equation}
The particle content for the lepton and Higgs sector with quantum arrangement is listed in Table \ref{t:particle} \cite{Majumdar:2020xws}. 
\begin{table}[ht]
\centering 
\begin{tabular}{cccccc} 
\hline 
Fields& $SU(2)_L$& $SU(2)_R$& $U(1)_{B-L}$& $SU(3)_C$& $U(1)_{L_{\mu}- L_{\tau}}$\\
$\{L_{e}, L_{\mu }, L_{\tau }\}_{L(R)}$& 2(1) &1(2) &-1& 1& $\{0,1,-1\}$ \\
%
%
$\{ S_{1L},S_{2L}, S_{3L}\}$&  1& 1&0& 1&$\{ 0,1,-1\}$ \\
\hline 
$\Phi$& 2 &2 & 0& 1& 0\\
$H_{L(R)}$& 2(1) &1(2) & 1& 1& 0\\
%
%
$\chi$& 1 &1 & 0& 1& 1\\
$\chi'$& 1 &1 & 0& 1& 2\\
\hline 
\end{tabular}
\caption{Particle content in the LRiss model }\label{t:particle}
\end{table}

As we will see below, two Higgs singlets, $\chi$ and $\chi'$, are required to generate neutrino masses and mixing consistent with experimental data. The quark sector and $H_L$ were discussed in Refs. \cite{Majumdar:2020xws, Majumdar:2022jur}; however, they are not relevant to the present work and are therefore omitted here. Consequently, $H_L$ does not contribute to the lepton sector and is not considered further. The electric charge operator associated with the $U(1)_Q$ gauge symmetry in this model is given by $Q=T^3_L+T^3_R+\frac{1}{2}(B-L),$ where $B$ and $L$ denote the baryon and lepton numbers, respectively. The electric charges of all leptons and Higgs multiplets are given by
\begin{align}
L_{aL(R)}= & \begin{pmatrix}
\nu_a\\
e_a 
\end{pmatrix}_{L(R)},\; a=e,\mu,\tau;\; 
%
 \Phi= \begin{pmatrix}
\phi^0_1& \phi^+_1 \\
\phi^-_2& \phi^0_2
\end{pmatrix}
;
\;  H_{L(R)}= \begin{pmatrix}
	H^+_{L(R)}\\
	H^0_{L(R)}
\end{pmatrix}, \label{eq:Higgs}
\end{align}
and  $S_{aL}$, $\chi$ and $\chi'$ are neutral singlets. The vacuum expectation values (vevs) of neutral components are as follows:
\begin{align}
\label{eq:vevs}
\langle  \Phi \rangle = & \begin{pmatrix}
\frac{v_1}{\sqrt{2}}& 0 \\
0& \frac{v_2}{\sqrt{2}}
\end{pmatrix}
,
\;  \langle H_{L(R)} \rangle = \begin{pmatrix}
0\\
\frac{v_{L(R)}}{\sqrt{2}}
\end{pmatrix},\;  \langle \chi \rangle =  \frac{u}{\sqrt{2}} ,\;  \langle \chi' \rangle =  \frac{u'}{\sqrt{2}}. 
\end{align}

 The allowed Yukawa interactions for leptons are expressed in the following Lagrangian:
\begin{align}
\label{eq:LYlep}
-\mathcal{L}_{Y}^{\ell}=& \overline{L_{L}} \left[ Y^{\ell}\Phi +\tilde{Y}^{\ell} \tilde{\Phi} \right] L_{R} +  \overline{L_R} Y^{R}  \tilde{H}_RS_L +\frac{1}{2}\overline{(S_L)^C}\bar{\mu} S_L
\crn&+ \frac{1}{2}y^s_{12}\left( \overline{(S_{1L})^C} S_{2L} + \overline{(S_{2L})^C} S_{1L} \right)\chi^* + \frac{1}{2} y^s_{13}\left( \overline{(S_{1L})^C} S_{3L} + \overline{(S_{3L})^C} S_{1L} \right) \chi
\crn & +  \frac{1}{2}\left(y'^s_{22} \overline{(S_{2 L})^C} S_{2L}\chi'^* + y'^s_{33}\overline{(S_{3 L})^C} S_{3L}\chi' \right)+\mathrm{h.c.}
\end{align}
with $\tilde{\Phi}=\sigma_2\Phi^* \sigma_2$, $\tilde{H}_R=i\sigma_2H^*_R$, $L_{L(R)} \equiv (L_{e}, L_{\mu}, L_{\tau})^T_{L(R)}$, and $S_L=(S_1,S_2,S_3)^T_L$. 

 The Yukawa couplings for the symmetry invariant Lagrangian \eqref{eq:LYlep}  are all diagonal:
\begin{align}
\label{eq:YLR}
Y^{\ell}=& \mathrm{diag}\left(Y^1, Y^{2},Y^{3}\right)\;; \tilde{Y}^{\ell}= \mathrm{diag}\left(\tilde{Y}^1, \tilde{Y}^{2}, \tilde{Y}^{3}\right)\;; Y^{R}= \mathrm{diag}\left(Y^{R}_{1}, Y^{R}_{2},Y^{R}_{3}\right); 
\end{align}
and only three entries  $\bar{\mu}_{11},\bar{\mu}_{23}=\bar{\mu}_{32} $ of the Majorana mass matrix $\bar{\mu}$ are non-zero.  

The lepton masses are in the following form:
\begin{align}
\label{eq:Llep}
\mathcal{L}^{\ell}_{\mathrm{lep}}=& - \overline{e_L} \hat{\mathcal{M}}_{\ell} e_R-  \frac{1}{2} \left( \overline{(\nu_L)^c},\;\overline{\nu_R},\; \overline{(S_L)^c}\right) \mathcal{M}_\nu \begin{pmatrix}
\nu_L&
(\nu_R)^c
&
S_L
\end{pmatrix}^T
 +\mathrm{h.c.},
\end{align}
where the two charged and neutral lepton mass matrices are $\hat{\mathcal{M}}_{\ell}=\frac{v_2 Y^{\ell}  + v_1 \tilde{Y}^{\ell}}{\sqrt{2}}$ and 
\begin{align}
 \mathcal{M}_\nu= & \begin{pmatrix}
\mathcal{O}_{3\times 3}& m_D^T &  \mathcal{O}_{3\times 3}\\
m_D&\mathcal{O}_{3\times 3}  & M_R^T \\
\mathcal{O}_{3\times 3}&M_R  & \mu_L
\end{pmatrix}, \;  \mu_L=\begin{pmatrix}
\bar{\mu}_{11}&\frac{y^s_{12}u}{\sqrt{2}}  &\frac{y^s_{13}u}{\sqrt{2}}  \\
\frac{y^s_{12}u}{\sqrt{2}}& \frac{y'^s_{22}u'}{\sqrt{2}} & \bar{\mu}_{23} \\
\frac{y^s_{13}u}{\sqrt{2}}&\bar{\mu}_{23}  & \frac{y'^s_{33}u'}{\sqrt{2}}
\end{pmatrix}, \label{eq:Mnu9}
\\  m_D= &\frac{Y^{\ell T} v_1+\tilde{Y}^{\ell T}v_2}{\sqrt{2}}, \; M_R= Y^{R T}\frac{v_R}{\sqrt{2}} \nn .
\end{align}
The charged-lepton mass matrix is diagonal, yielding the following charged-lepton masses:  $m_{a}= (v_2 Y^{\ell}_{aa}  + v_1 \tilde{Y}^{\ell}_{aa})/\sqrt{2}$ with $a=1,2,3$.  For the neutral-lepton sector, the mass matrix in Eq.  \eqref{eq:Mnu9} is diagonalized by a $9\times9$ unitary matrix $U^{\nu}$, $U^{\nu T} \mathcal{M}_\nu U^{\nu} = \hat{\mathcal{M}}_\nu= \mathrm{diag}\left(\hat{m}_{\nu}, \; \hat{M}_S \right)$. Here $\hat{m}_{\nu}=\mathrm{diag}\left( m_{n_1},  m_{n_2},  m_{n_3}\right)$
and $\hat{M}_{S}=\mathrm{diag}\left( m_{n_4},\dots,  m_{n_9}\right)$ denote the active and heavy physical neutrino masses, respectively. The corresponding physical states are collected into a vector $n_{L(R)}=(n_1,n_2,\dots, n_9)_{L(R)}^T$.  These Majorana neutrinos satisfy $(n_{iL})^c=n_{iR}$ and a four-component Majorana state is $n_i=(n_{iL},(n_{iL})^c)^T$. The transformations between the flavor and physical bases are given by  $(\nu_L,\; (\nu_R)^c,\; S_L)^T=U^{\nu} n_{L}$ and $((\nu_L)^c,\; \nu_R,\; (S_L)^c)^T=U^{\nu *} n_{R}$.  Consequently, the heavy neutrino masses are approximately given by  $\hat{M}_S\simeq \mathrm{diag}(\hat{M}_R,\hat{M}_R)$, where  $M_R \simeq \hat{M}_R= \mathrm{diag}(M_1,M_2,M_3)$, implying that 
 \begin{align}
 \label{eq:M123}
 M_{1}=m_{n_4}=m_{n_7},\; M_{2}=m_{n_5}=m_{n_8},\; M_{3}=m_{n_6}=m_{n_9}.
 \end{align}
 
 In our numerical analysis, we take the entries of $\tilde{Y}^{\ell}$ as the free parameters, namely,
 \begin{align}
 \label{eq:YlmD}
 Y^{\ell}=\frac{\sqrt{2}\hat{\mathcal{M}}_{\ell} }{c_{\beta}v} -\tilde{Y}^{\ell} t_{\beta},\; m_D^T= \mathrm{diag}\left(m^1_D,\; m^2_D,m^3_D\right)= \hat{\mathcal{M}}_{\ell}t_{\beta} +\frac{v\tilde{Y}^{\ell} c_{2\beta}}{\sqrt{2}c_{\beta}},
 \end{align}
 where 
 \begin{equation}
 \label{eq:tb}
 t_{\beta} \equiv \tan \beta = \frac{v_1}{v_2},\; c_{\beta}=\cos\beta,\; s_{\beta}=\sin\beta,
 \end{equation}
 that satisfies $s_{\beta}^2 +c_{\beta}^2=1$. Throughout this work, we use the notation $
 s_x=\sin x,\; c_x=\cos x,\;t_x=\tan x,$  as well as $
 s_{2x}=\sin(2x),\;  s_{x\pm y}=\sin(x\pm y), \ldots $ for the corresponding trigonometric functions. The mixing parameters and the active-neutrino mass matrix associated with the ISS mechanism are given by \cite{Arganda:2004bz, Arganda:2014dta, Arganda:2015naa}; see also Refs. \cite{Thao:2017qtn, Hong:2022xjg} for detailed discussions. The important parameters are  
\begin{align}
\label{eq:nusector}
R^*=& \begin{pmatrix}
\mathcal{O}_{3\times 3},& R^{0*}
\end{pmatrix},\; R^{0*}=m_D^T \left(M_R\right)^{-1} =\mathrm{diag}\left( R^{0*}_1,\;  R^{0*}_2,\; R^{0*}_3\right), 
\crn m_{\nu}=&m_D^{\dagger} \left(M^*_R\right)^{-1} \mu_L(M_R^T)^{-1}m_D=U^{\nu*}_3 \hat{m}_{\nu}U^{\nu\dagger}_3, 
\end{align}
where $R^{0*}_{i} =m^i_D/M_i$ with $i=1,2,3$. 

All entries of the matrix $\mu_L$are determined in terms of the other parameters using Eq. \eqref{eq:nusector}. The matrix  $U^{\nu}$ can be written in the following form \cite{Hong:2022xjg, Hong:2024yhk}: 
\begin{align}
\label{eq:Usimeq}
U^{\nu}&\simeq  \begin{pmatrix}
\left(I_3 -\frac{R_0R_0^{\dagger}}{2}\right)U^{\nu}_3	& \frac{ R_0 }{\sqrt{2}}&  \frac{-iR_0}{\sqrt{2}}\\
\mathcal{O}_{3\times3}	& \frac{I_3}{\sqrt{2}}  &   \frac{iI_3}{\sqrt{2}} \\
-R_0^{\dagger} U^{\nu}_3& 	\frac{1}{\sqrt{2}}\left(I_3 -\frac{R_0^{\dagger}R_0}{2}\right)  &\frac{-i}{\sqrt{2}}  \left(I_3 -\frac{R_0^{\dagger}R_0}{2}\right) 
\end{pmatrix},
\end{align}

The diagonal structure of the three Yukawa coupling matrices $Y^{\ell}$, $\tilde{Y}^{\ell}$, and $Y^R$ shown in Eq. \eqref{eq:YLR} leads to the simple form of the active-neutrino mass matrix $m_{\nu}$  given in Eq. \eqref{eq:nusector}. If only one of the Higgs singlets,$\chi$ or $\chi'$, is present, at least one element of $m_{\nu}$ vanishes, as shown explicitly in Refs. \cite{Majumdar:2020xws, Majumdar:2022jur}.

In contrast, the experimental data on active-neutrino masses, $\hat{m}{\nu}$, and the lepton mixing matrix, $U^{\nu}_3\equiv U_{\mathrm{PMNS}}$, require a general neutrino mass matrix with all elements being nonzero. Here, $U_{\mathrm{PMNS}}$ denotes the lepton mixing matrix used to confront the theoretical predictions with experimental data; see, e.g., Ref. \cite{ParticleDataGroup:2024cfk} for its definition. Therefore, both $\chi$ and $\chi'$ are required to generate the general form of $\mu_L$ given in Eq. \eqref{eq:Mnu9}, thereby allowing the active-neutrino mass matrix $m_{\nu}$ to reproduce the experimentally allowed values.

The Higgs potential considered here is given by:
\begin{align}
\label{eq:vHiggsR}
V_S=& \mu_{R}^2 H_R^{\dagger} H_R + \mu^2_{\Phi} \mathrm{Tr}\left[\Phi^{\dagger} \Phi\right] + \tilde{\mu}^2_{\Phi} \mathrm{Tr}\left[(\tilde{\Phi}^{\dagger} \Phi) + (\tilde{\Phi} \Phi^{\dagger})\right]  + \lambda_2 \left(H_R^{\dagger} H_R\right)^2
\crn &+\mu_{\chi}^2 \chi^*\chi + \mu_{\chi'}^2 \chi'^*\chi' + \lambda_1 \left( \chi^*\chi\right)^2 + \lambda'_1 \left( \chi'^*\chi'\right)^2 +\lambda_{\chi \chi'}\left( \chi^*\chi\right) \left( \chi'^*\chi'\right)
+ \left[ \lambda'_{\chi \chi'}\chi^2 \chi'^* +\mathrm{h.c.} \right] 
%
%
\crn & + \lambda_3\left[ \mathrm{Tr}\left(\Phi^{\dagger} \Phi\right)\right]^2 +  \lambda_4 \mathrm{Tr}\left[ \left(\Phi^{\dagger} \Phi\right)^2 \right] 
%
+  \lambda_5 \left[ \left(\mathrm{Tr}\left( \tilde{\Phi} \Phi^{\dagger} \right) \right)^2 + \left(\mathrm{Tr}\left( \tilde{\Phi}^{\dagger} \Phi \right) \right)^2\right]
\crn &+   \lambda_6 \mathrm{Tr}\left(\Phi^{\dagger} \Phi\right)  \left[ \mathrm{Tr}\left( \tilde{\Phi} \Phi^{\dagger} \right)  + \mathrm{Tr}\left( \tilde{\Phi}^{\dagger} \Phi \right) \right] 
\crn&+ (\chi^*\chi) \left[  \lambda_7  H_R^{\dagger} H_R + \lambda_8  \mathrm{Tr}\left(\Phi^{\dagger} \Phi \right) + \lambda_9 \mathrm{Tr}\left(\tilde{\Phi}^{\dagger} \Phi+ \tilde{\Phi} \Phi^{\dagger} \right)\right]
\crn&+ (\chi'^*\chi') \left[  \lambda'_7  H_R^{\dagger} H_R + \lambda'_8  \mathrm{Tr}\left(\Phi^{\dagger} \Phi \right) + \lambda'_9 \mathrm{Tr}\left(\tilde{\Phi}^{\dagger} \Phi+ \tilde{\Phi} \Phi^{\dagger} \right)\right]
\crn&+ (H_R^{\dagger}H_R) \left[   \lambda_{10} \mathrm{Tr}\left(\Phi^{\dagger} \Phi \right) + \lambda_{11} \mathrm{Tr}\left(\tilde{\Phi}^{\dagger} \Phi+ \tilde{\Phi} \Phi^{\dagger} \right)\right]  +  \lambda_{12}  \left(H_R^{\dagger}\Phi^{\dagger} \Phi H_R \right),
\end{align}
and is generally consistent with the form presented in Ref. \cite{Ezzat:2021bzs}. We omit terms that can be expressed as linear combinations of the remaining terms, namely,
\begin{align*}
\mathrm{Tr}\left(\tilde{\Phi}^{\dagger} \Phi\right) \mathrm{Tr}\left(\tilde{\Phi} \Phi^{\dagger}\right)= & 2 \left[ \mathrm{Tr}\left(\Phi^{\dagger} \Phi\right)\right]^2 -2 \mathrm{Tr}\left[ \left(\Phi^{\dagger} \Phi\right)^2 \right],
\crn  \frac{1}{2}\left(H^{\dagger}_RH_R\right) \times \mathrm{Tr}\left(\tilde{\Phi}^{\dagger} \Phi+ \tilde{\Phi} \Phi^{\dagger} \right)=& \mathrm{Tr}\left(H_R^{\dagger}\Phi^{\dagger} \tilde{\Phi} H_R \right) +\mathrm{Tr}\left(H_R^{T}\Phi^{T} \tilde{\Phi}^* H_R^* \right),
\crn  (H_R^{\dagger}H_R) \mathrm{Tr}\left(\Phi^{\dagger} \Phi \right) =& H_R^{\dagger}\Phi^{\dagger} \Phi H_R + \left(H_R^{\dagger} \tilde{\Phi}^{\dagger} \tilde{\Phi} H_R \right). 
\end{align*}
The mixing and masses of physical Higgs bosons were presented in appendix \ref{app:Vhtotal}, where  neutral Higgs components $\phi^0_{k}$ ($k=1,2$), $H^0_R$, $\chi$, and $\chi'$ given in Eq. \eqref{eq:Higgs} are expanded around their vev as follows 
\begin{align}
\label{eq:H0expand}
\phi^0_k=\frac{v_k+r_k +i a_k}{\sqrt{2}},\; H_R^0= \frac{v_R+r_3 +i a_3}{\sqrt{2}},\; \chi=  \frac{u +r_4 +i a_4}{\sqrt{2}}, \; \chi'=  \frac{u' +r_5 +i a_5}{\sqrt{2}}.
\end{align} 

 The gauge boson masses and their mixing are derived from the following kinetic terms of the Higgs multiplets:
\begin{align}
\label{eq:LkH}
\mathcal{L}_{k}^H=&  \left(D_{\rho}H_R\right)^{\dagger} \left(D^{\rho}H_R\right) + \mathrm{Tr}\left[ \left(D_{\rho}\Phi\right)^{\dagger} \left(D^{\rho}\Phi\right)\right] +\sum_{s=\chi,\chi'} \left(D_{\rho} s\right)^{*} \left(D^{\rho}s\right), 
\end{align}
where the covariant derivatives for the Higgs multiplets are:
\begin{align}
\label{eq:DmuS}
D_{\rho} \Phi=& \partial_{\rho} \Phi -i \frac{g_L}{2} \sum_{a=1}^3\sigma^aW^a_{L \rho} \Phi +i \frac{g_R}{2} \Phi\sum_{a=1}^3\sigma^aW^a_{R \rho} , 
\crn D_{\rho} H_R=& \partial_{\rho} H_R -i \frac{g_R}{2} \sum_{a=1}^3\sigma^aW^a_{R \rho} H_R -i\frac{g_{B-L}}{2}B'_{\rho} H_R, 
\crn D_{\rho} \chi =& \partial_{\rho} \chi -i\frac{g_{\mu \tau}Y^{\chi(\chi')}_{\mu \tau}}{2}Z_{\mu \tau, \rho} \chi(\chi'),\; Y^{\chi(\chi')}_{\mu \tau}=1(2).
\end{align}
The masses, physical states, and mixing parameters of all leptons, gauge bosons, and Higgs bosons are derived as follows (details of calculations are provided in appendix \ref{app:Vhtotal}). Defining 
\begin{align}
\label{eq:tx}
t_{R}\equiv \frac{g_R}{g_L},\; t_{B-L}\equiv \frac{g_{B-L}}{g_L},
\end{align}
the squared mass matrix of singly charged gauge boson in the basis $(W^{\pm}_{L \mu}, W^{\pm}_{R \mu})$ is given by
\begin{align}
M^2_{W_{LR}}=\frac{g^2_L}{4}\left(
\begin{array}{cc}
v^2 & -s_{2\beta } t_R v^2 \\
-s_{2\beta } t_R v^2 & t_R^2 \left(v_R^2+v^2\right) \\
\end{array}
\right),
\end{align}
where $ v^2=v_1^2+v_2^2,\; v_1=v s_{\beta},\; v_2=vc_{\beta}.$
Correspondingly, the two physical states of singly charged gauge bosons, $W^{\pm}$ and $W'^{\pm}$, and their masses are related to states in the original basis as follows:
\begin{align}
\label{eq:thtaWLRpm}
\begin{pmatrix}
W^\pm_{L\mu}	\\
W^\pm_{R\mu}	
\end{pmatrix} = &\left(
\begin{array}{cc}
c_{\theta } & -s_{\theta } \\
s_{\theta } & c_{\theta } \\
\end{array}
\right) \begin{pmatrix}
W^\pm_{\mu}	\\
W'^\pm_{\mu}	
\end{pmatrix},
\crn  t_{2\theta}  \equiv &\tan(2\theta)=\frac{2 s_{2\beta } t_R v^2}{t_R^2 v_R^2+\left(t_R^2-1\right) v^2}
 \simeq \frac{2 s_{2\beta }  v^2}{t_R v_R^2},
\crn m_{W}^2=& \frac{g_L^2v^2}{4} \left( 1- s_{2\beta}t_Rt_{\theta} \right) , \; 
  m_{W'}^2=    \frac{g_L^2 c^2_{\theta}\left[  t_R^2 v_R^2+  v^2(t_R^2 -t^2_{\theta}) \right]}{4 c_{2\theta}}.
\end{align}
 We find that $0<t_{2\theta} \ll1$, implying  $0<c_{2\theta},s_{2\theta}\ll1$.  The results given in Eq. \eqref{eq:thtaWLRpm}  agree  with those of Ref. \cite{Hue:2024rij} with $t_R=1$. 
In the neutral gauge boson sector,  $Z_{\mu \tau}$ is already a physical state, with mass $m_{Z_{\mu \tau}}=g_{\mu \tau}\sqrt{u^2+4u'^2}/2$. On the other hand, the squared mass matrix of the neutral gauge bosons in the basis $(W^3_{R\mu}, B'_{\rho},W^3_{L\mu})$ is:
\begin{align}
\label{eq:M2g0}
\mathcal{M}^2_{RB'L}=&\frac{g_L^2}{4} \left(
\begin{array}{ccc}
t_R^2 \left(v_R^2+v^2\right) & -t_{B-L} t_R v_R^2 & -t_R v^2 \\
-t_{B-L} t_R v_R^2 & t_{B-L}^2 v_R^2 & 0 \\
-t_R v^2 & 0 & v^2 \\
\end{array}
\right),
\end{align}
where $t_{B-L}=g_{B-L}/g_L$ and $t_R=g_R/g_L$, as defined in Eq. \eqref{eq:tx}.  
The first stage of electroweak symmetry breaking, $SU(2)_R\otimes SU(2)_L\otimes U(1)_{B-L}\to SU(2)_L\times U(1)_Y$ of the SM, leads, in the limit  $v_R\neq 0$ and $v=0$, to the identification of the  massless  gauge bosons with the corresponding SM ones. In particular, $(W^3_{R \mu}, B'_{\rho},W^3_{L \mu})\to (Z'_{\mu},B_{\mu},W_{3\mu})$ where $W_{3\mu}$ and $B_{\mu}$ are gauge bosons associated with the diagonal generators of  $SU(2)_L$ and $U(1)_Y$ in the SM electroweak gauge group. In this limit, the corresponding transformation  is $C_{LR}\left. M^2_{RB'L}\right|_{v=0} C_{LR}^T= \mathrm{diag}\left( m_{Z'}^2,0,0\right)$, which gives:
\begin{align}
\label{eq:LRtoL}
 (W^3_{R \mu},B'_{\rho},\; W^3_{L \mu})^T=C_{LR}^T(Z'_{\mu},B_{\mu},W_{3\mu}),\; C_{LR}= \left(
\begin{array}{ccc}
c_{\zeta } & -s_{\zeta } & 0 \\
s_{\zeta } & c_{\zeta } & 0 \\
0 & 0 & 1 \\
\end{array}
\right),\; t_{\zeta}\equiv \tan \zeta= \frac{g_{B-L}}{g_{R}}. 
\end{align}
It is then matched to the SM gauge fields and couplings through the following covariant derivative:
\begin{align}
\label{eq:DLR}
D^{\mathrm{LR}}_{\rho} & =\partial_{\rho} -i g_L\sum_{a=1}^3T^a_LW^a_{L\rho}-i g_R\sum_{a=1}^3T^a_RW^a_{R \rho} -\frac{ig_{B-L}}{2}(B-L)B'_{\rho}-\frac{ig_{\mu \tau}}{2}X_{\mu{\tau}}Z_{\mu\tau, \rho}
\crn \Rightarrow& \partial_{\rho} -i g_L T^3_L W^3_{L \rho} -i\left(  g_Rs_{\zeta} T^3_{R}+ \frac{g_{B-L}}{2}(B-L) c_{\zeta} \right) B_{\rho} 
\crn &\equiv D^{\mathrm{SM}}_{\rho}= \partial_{\rho} -i g_2 T^3 W^3_{\rho} -i \frac{g_1}{2}Y B_{\rho},  
\end{align} 
where $g_{1}$ and $g_2$ are the gauge couplings of the $U(1)_Y$ and $SU(2)_L$ in the SM. 
Therefore, we identify $T^3_L =T^3,\; W^3_{L}=W^3,\;g\equiv g_2=g_L,\;$ and
\begin{align}
\label{eq:SMmatching}
 g_Rs_{\zeta} T^3_{R}+ \frac{g_{B-L}}{2}(B-L) c_{\zeta} =  \frac{g_1}{2}Y= \frac{g_2t_W}{2}Y,
\end{align}
where $g_1=g_2t_W= gt_W$. 
The matching of the electric charge operator between the LR and SM theories  gives $T^3_R+ \frac{B-L}{2}= \frac{Y}{2}.$ Combining this relation with Eq. \eqref{eq:SMmatching}, we obtain:
\begin{equation}\label{eq:gi}
\frac{1}{g_1^2}= \frac{1}{g_R^2} +\frac{1}{g_{B-L}^2} \to g_{B-L}=\frac{g_1}{c_{\zeta}}=\frac{gt_W}{c_{\zeta}},\; g_R=\frac{gt_W}{s_{\zeta}}.
\end{equation}
The mass matrix \eqref{eq:M2g0} can be transformed as follows:
\begin{align}
\label{eq:Cg}
&(C_{21}C_{LR})\mathcal{M}^2_{RB'L} (C_{21}C_{LR})^T= \frac{g_L^2}{4}\left(
\begin{array}{ccc}
\frac{t_W^2 \left(v^2 c_{\zeta }^4+v_R^2\right)}{c_{\zeta }^2 s_{\zeta }^2} & -\frac{ t_W v^2}{c_W t_{\zeta }} & 0 \\
-\frac{t_W v^2}{c_W t_{\zeta }} & \frac{v^2}{c_W^2} & 0 \\
0 & 0 & 0 \\
\end{array}
\right), \;
 C_{21}= \left(
\begin{array}{ccc}
1 & 0 & 0 \\
0 & -s_W & c_W \\
0 & c_W & s_W \\
\end{array}
\right), 
\end{align}
in which $C_{21}$ is exactly the SM transformation to diagonalize the neutral gauge boson mass matrix, and $s_{W}$ relates to the Weinberg angle $\theta_W$: $s_{W}=\sin \theta_W$. Finally, the $ZZ'$ mixing parameter, denoted as $\xi$, is used to diagonalize the mass matrix given in Eq. \eqref{eq:Cg}, namely
\begin{align}
\label{eq:czz'}
& C_g\mathcal{M}^2_{RB'L} C_g^T= \mathrm{diag}\left(m_{Z'}^2,\; m_{Z}^2,\; 0\right),
%
\; C_g= C_{ZZ'}C_{21}C_{LR},
\; C_{ZZ'}= \left(
\begin{array}{ccc}
c_{\xi } & -s_{\xi } & 0 \\
s_{\xi } & c_{\xi } & 0 \\
0 & 0 & 1 \\
\end{array}
\right),
\crn & 0< t_{2\xi}= \frac{2 c_{\zeta }^3 s_{\zeta } s_W v^2}{c_{\zeta }^4 s_W^2 v^2-c_{\zeta }^2 s_{\zeta }^2 v^2+s_W^2 v_R^2} \varpropto \mathcal{O}\left( \frac{v^2}{v_R^2}\right) \ll1,
\crn  & m_Z^2 =  \frac{g_L^2v^2}{4c_W^2}\left( 1 -\frac{2s_W t_{\xi}v^2}{ t_{\zeta } }\right),
\crn  &m_{Z'}^2 =  \frac{g_L^2c^2_{\xi} t_W^2 v_R^2}{s^2_{2\zeta}c_{2\xi}} + \frac{g_L^2v^2c^2_{\xi} t_W^2}{4c_{2\xi} t_{\zeta }^2} \left(1 -\frac{t^2_{\xi}t_{\zeta }^2}{s^2_W} \right).
\end{align}
Now, the transformation between the original and mass base of neutral gauge bosons  is
\begin{align}
\label{eq:Z'ZA}
\begin{pmatrix}
W^3_{R \mu}\\
B'_{\rho}\\
W^3_{L \mu}
\end{pmatrix}= C_g^T \begin{pmatrix}
Z'_{\mu}\\
Z_{\mu}\\
A_{\mu}
\end{pmatrix}, \; C_g^T= \left(
\begin{array}{ccc}
c_{\zeta } c_{\xi }+s_W s_{\zeta } s_{\xi } & c_{\zeta } s_{\xi }-c_{\xi } s_W s_{\zeta } & c_W s_{\zeta } \\
c_{\zeta } s_W s_{\xi }-c_{\xi } s_{\zeta } & -c_{\zeta } c_{\xi } s_W-s_{\zeta } s_{\xi } & c_W c_{\zeta } \\
-c_W s_{\xi } & c_W c_{\xi } & s_W \\
\end{array}
\right). 
\end{align}

\section{\label{sec:Feynrule} Feynman rules for one-loop contributions to LFV decays}

The sources of LFV processes are the couplings between different flavor leptons and the Higgs and gauge bosons, which arise from the Yukawa Lagrangian and the covariant kinetic terms of the leptons. The relevant Yukawa interactions are obtained from Eq. \eqref{eq:LYlep}. We omit the irrelevant terms in Eq. \eqref{eq:LYlep}, including the mass terms and the couplings of heavy neutral leptons to the Higgs singlets. The interactions involving the physical states are given by
\begin{align}
\label{eq:SU2Lexpand}
-\mathcal{L}_{Y}^{\ell \ell S}=&  \overline{e_L} \left[  \hat{\mathcal{M}}_{\ell} + \frac{h}{\sqrt{2}} \left( s_{\alpha} \tilde{Y}^{\ell}+c_{\alpha} Y^{\ell} \right) + \frac{h_2}{\sqrt{2}} \left( c_{\alpha} \tilde{Y}^{\ell} -s_{\alpha} Y^{\ell} \right) +  \frac{ia^0}{\sqrt{2}} \left( -c_{\beta} \tilde{Y}^{\ell}+s_{\beta} Y^{\ell} \right) \right]e_R 
\crn &+\overline{\nu_L} c_{\kappa}(s_{\beta} Y^{\ell} -c_{\beta}\tilde{Y}^{\ell})e_R h^+ + \overline{e_L} c_{\kappa} (c_{\beta} Y^{\ell}- s_{\beta} \tilde{Y}^{\ell}) \nu_Rh^-
\crn & +\overline{\nu_L} \left[  m_D^T + \frac{h}{\sqrt{2}} \left( c_{\alpha} \tilde{Y}^{\ell} +s_{\alpha} Y^{\ell} \right)  + \frac{h_2}{\sqrt{2}} \left( -s_{\alpha} \tilde{Y}^{\ell}+c_{\alpha} Y^{\ell} \right) +  \frac{ia^0}{\sqrt{2}} \left(-s_{\beta} \tilde{Y}^{\ell}+c_{\beta} Y^{\ell} \right) \right]  \nu_R 
\crn & +  \left[\overline{\nu_R} \left(M_R^T+ \frac{r_3 +ia_3}{\sqrt{2}}Y_R \right) +s_{\kappa} \overline{e_R} h^- Y_R\right]   S_L + \mathrm{h.c.}+\dots.,  
\end{align}
where $s_{\kappa}$ and $c_{\kappa}$  are defined in Eq. \eqref{eq:hpmi}.

Since the matrices $Y^{\ell}$ and $\tilde{Y^{\ell}} $ are diagonal, there are no tree-level LFV couplings between the neutral Higgs bosons and charged leptons in the model under consideration. In addition, the $h\overline{e_a}e_a$  couplings reduce to the SM form in the limit $\delta=\alpha-\beta \to 0$. Using the relations in Eq. \eqref{eq:YlmD}, the $h\overline{e_a}e_a$ couplings can be written as follows:
\begin{align}
\label{eq:hee}
\mathcal{L}_{h\ell \ell}=& -\frac{g}{2 m_W} \overline{e_L} \left[    \hat{\mathcal{M}}_{\ell} \left( c_{\delta}-t_{\beta}s_{\delta}  \right)  + \frac{v\tilde{Y}^\ell s_{\delta}}{\sqrt{2}c_{\beta}}\right]e_R h +\mathrm{h.c.}. 
\end{align}
 The Yukawa coupling of the SM-like Higgs boson with two neutrinos $h \overline{n_{i}} n_j$ from Eq. \eqref{eq:SU2Lexpand} is derived in the symmetric form  \cite{Dreiner:2008tw} with $g^{L}_{hij}=g^{L}_{hji}$, namely  
\begin{align}
 \mathcal{L}_{hnn}=& -\frac{h}{2}\sum_{i,j=1}^9 \overline{n_i} \left[ g^{L}_{hij} P_L + g^{L*}_{hij} P_R\right]n_j, \label{eq:hnn} 
\\  
g^L_{hij} = & \frac{1}{\sqrt{2}}\sum_{c=1}^3 \left( c_{\alpha} \tilde{Y}^{\ell} +s_{\alpha} Y^{\ell} \right)_{cc} \left[ U^{\nu}_{ (c+3)i}   U^{\nu}_{cj} + U^{\nu}_{ (c+3)j} U^{\nu}_{ ci}\right] . \label{eq:lahij}
\end{align}

The covariant kinetic terms of leptons are:
\begin{align}
\label{eq:LeD}
\mathcal{L}_D^{\ell}=& \sum_{a=1}^3 \left[ i \overline{L_{aL}}\gamma^{\rho}D_{\rho}L_{aL} + i \overline{L_{aR}}\gamma^{\rho}D_{\rho}L_{aR} + i \overline{S_{aL}}\gamma^{\rho}D_{\rho}S_{aL}\right],
\end{align} 
where $(B-L)$ charges are given in Table \ref{t:particle}  and the covariant derivatives of leptons are 
\begin{align}
\label{eq:Dell}
D_{\rho}L_{aL(R)}=&  \left[\partial_{\rho} -ig_L\sum_{b=1}^{3}\frac{\sigma_b}{2}W^b_{L(R)\rho} + \frac{ig_{B-L}}{2} B'_{\rho} -\frac{ig_{\mu \tau} Y^{\ell}_{\mu\tau,a}}{2} Z_{\mu\tau, \rho} \right]L_{a L(R)}
,
\crn D_{\rho}S_{aL}=&\left[\partial_{\rho} -i\frac{ig_{\mu \tau} Y^S_{\mu\tau,a}}{2} Z_{\mu\tau, \rho} \right]S_{a L},
\end{align}
where $Y^{\ell}_{\mu \tau,a}$ and  $Y^{S}_{\mu \tau,a}$ are the $U(1)_{L_{\mu} -L_{\tau}}$ of $L_{aL(R)}$ and $S_{aL}$ given in Tabel \ref{t:particle}, respectively.

The final results of physical couplings relevant to one-loop contributions to LFV decay amplitudes and $\Delta a_{e_a}$ are: 
\begin{align}
\label{eq:LDll}
\mathcal{L}^{V\ell \ell}=& \sum_{a} \left( -eA_{\rho}\overline{e_{a}}\gamma^{\rho} e_{a}\right) +\frac{g_{\mu \tau} }{2} Z_{\mu\tau, \rho} \left(  \overline{\mu}\gamma^{\rho}\mu\right)
-\frac{g_{\mu \tau} }{2} Z_{\mu\tau, \rho}  \left(\overline{\tau}\gamma^{\rho}\tau\right) 
\crn&+ \sum_{a} \frac{g}{2 c_W} \overline{e_{a}}\gamma^{\rho}  \left[   \left[ c_{\xi}(2s_W^2 -1) -s_Wt_{\zeta}s_{\xi}\right] P_L +   s_W\left(2s_Wc_{\xi}  -\frac{s_{\xi}c_{2\zeta}}{s_{\zeta} c_{\zeta}} \right)  P_R \right]   e_{a}Z_{\rho}
  \crn &  + \sum_{i,j=1}^9 \sum_{a} \frac{g}{2 c_W}\overline{n_{i}}\gamma^{\rho}\left[  \left( c_{\xi}+s_Wt_{\zeta}s_{\xi}\right) U^{\nu*}_{ai} U^{\nu}_{aj}P_L +  \frac{s_Ws_{\xi}}{s_{\zeta} c_{\zeta}} U^{\nu}_{(a+3)i} U^{\nu*}_{(a+3)j}P_R\right] n_{j} Z_{\rho}
  \crn&+ \sum_{i=1}^9 \sum_{a=e,\mu,\tau} \frac{g}{\sqrt{2}} \left[ W^{+}_{ \rho} \overline{n_{i}}\gamma^{\rho} \left(  c_{\theta}U^{\nu*}_{ai} P_L +\frac{t_Ws_{\theta}}{s_{\zeta}} U^{\nu}_{(a+3)i}P_R \right) e_{a} +\mathrm{h.c.}\right]
  \crn&+ \sum_{i=1}^9 \sum_{a=e,\mu,\tau} \frac{g}{\sqrt{2}} \left[ W'^{+}_{ \rho} \overline{n_{i}}\gamma^{\rho} \left(  -s_{\theta}U^{\nu*}_{ai} P_L +\frac{t_Wc_{\theta}}{s_{\zeta}}  U^{\nu}_{(a+3)i} P_R \right) e_{a} +\mathrm{h.c.}\right]   +\dots. 
\end{align}
We emphasize an important feature of the model: the neutral gauge boson $Z_{\mu\tau}$ does not contribute to LFV decays at the one-loop level. Large one-loop contributions to $\Delta a_{\mu}$ are generally accompanied by sizable contributions to $\Delta a_{\tau}$, which may provide an additional phenomenological signature that can be tested experimentally. It has been shown in several studies that a light $Z_{\mu\tau}$ with a mass below a few hundred MeV can generate sizable one-loop contributions to $\Delta a_{\mu}$, reaching values of up to $10^{-9}$, for $g_{\mu\tau}=\mathcal{O}(10^{-4})$ \cite{Majumdar:2020xws}. To focus on the one-loop contributions from singly charged Higgs boson exchange to $\Delta a_{\mu}$, we neglect the contributions from $Z_{\mu\tau}$ by considering the limit of a sufficiently small $g_{\mu\tau}$ \cite{Altmannshofer:2016brv}.

Using the  Majrojana property that $\overline{n_i}\gamma^{\rho} P_{L(R)}n_j=- \overline{n_j}\gamma^{\rho} P_{R(L)}n_i$ \cite{Dreiner:2008tw}, the couplings $Z\overline{n}n$ can be derived in the symmetric form with $g^L_{Zij}=g^{L*}_{Zji}$ as follows:
\begin{align}
\label{eq:Znn}
\mathcal{L}_{Znn} =&  \frac{e}{2} Z_{\rho} \sum_{i,j=1}^9 \overline{n_{i}} \gamma^{\rho}\left[ g^L_{Zij} P_L - g^{L}_{Zji} P_R  \right]n_j,
\\ 
g^L_{Zij} = &\frac{1}{2 s_Wc_W}\sum_{c=1}^3 \left[ \left( c_{\xi}+s_Wt_{\zeta}s_{\xi}\right) U^{\nu*}_{ci} U^{\nu}_{cj}-  \frac{s_Ws_{\xi}}{s_{\zeta} c_{\zeta}} U^{\nu}_{(c+3)j} U^{\nu*}_{(c+3)i}\right].  \label{eq:gLZij} 
\end{align}

In conclusion, the following LFV sources give one-loop contributions to both LFV decay amplitudes and $(g-2)_{e_a}$:
\begin{align}
\label{eq:LFVcouplings}
\mathcal{L}^{\mathrm{LFV}}=& \sum_{a=1}^3\sum_{i=1}^9 \overline{n_i}\left[   \left( g_{aih^+}^LP_L +g_{aih^+}^RP_R\right) h^+  + \sum_{V=W,W'}   \gamma^{\rho} \left( g_{aiV}^LP_L +g_{aiV}^RP_R\right)  V^+_{\rho}\right] e_a
\crn&
+\mathrm{h.c.}, 
\end{align}
where  the scalar factors $g^{L(R)}_{aiB}$ playing roles of LFV sources, with $B=h^+,W^+,W'^+$, are 
\begin{align}
\label{eq:gX}
g_{aih^+}^L=&  -  c_{\kappa} \left( \frac{\sqrt{2} m_{e_a}}{v} -2s_{\beta} \tilde{Y}^{a*}\right) U^{\nu}_{(a+3)i} ,
\crn g_{aih^+}^R=& -\left[ \frac{ c_{\kappa}}{c_{\beta}} \left(\frac{\sqrt{2}m_{e_a}s_{\beta}}{v} -\tilde{Y}^{a} \right) U^{\nu*}_{ai}
+  s_{\kappa}  \sum_{c=1}^{3}U^{\nu*}_{(c+6)i} \left(Y_R^*\right)_{ac}\right],
\crn g_{aiW}^L=& \frac{g}{\sqrt{2}}c_{\theta}U^{\nu*}_{ai}, \; 
g_{aiW}^R= \frac{g}{\sqrt{2}}\frac{t_Ws_{\theta}}{s_{\zeta}} U^{\nu}_{(a+3)i},
\crn g_{aiW'}^L= &\frac{g}{\sqrt{2}} \left( -s_{\theta}U^{\nu*}_{ai}\right), \;
 g_{aiW'}^R=  \frac{g}{\sqrt{2}} \frac{t_Wc_{\theta}}{s_{\zeta}}  U^{\nu}_{(a+3)i}.
\end{align}

\begin{figure}[ht] 
 	\includegraphics[width=11cm]{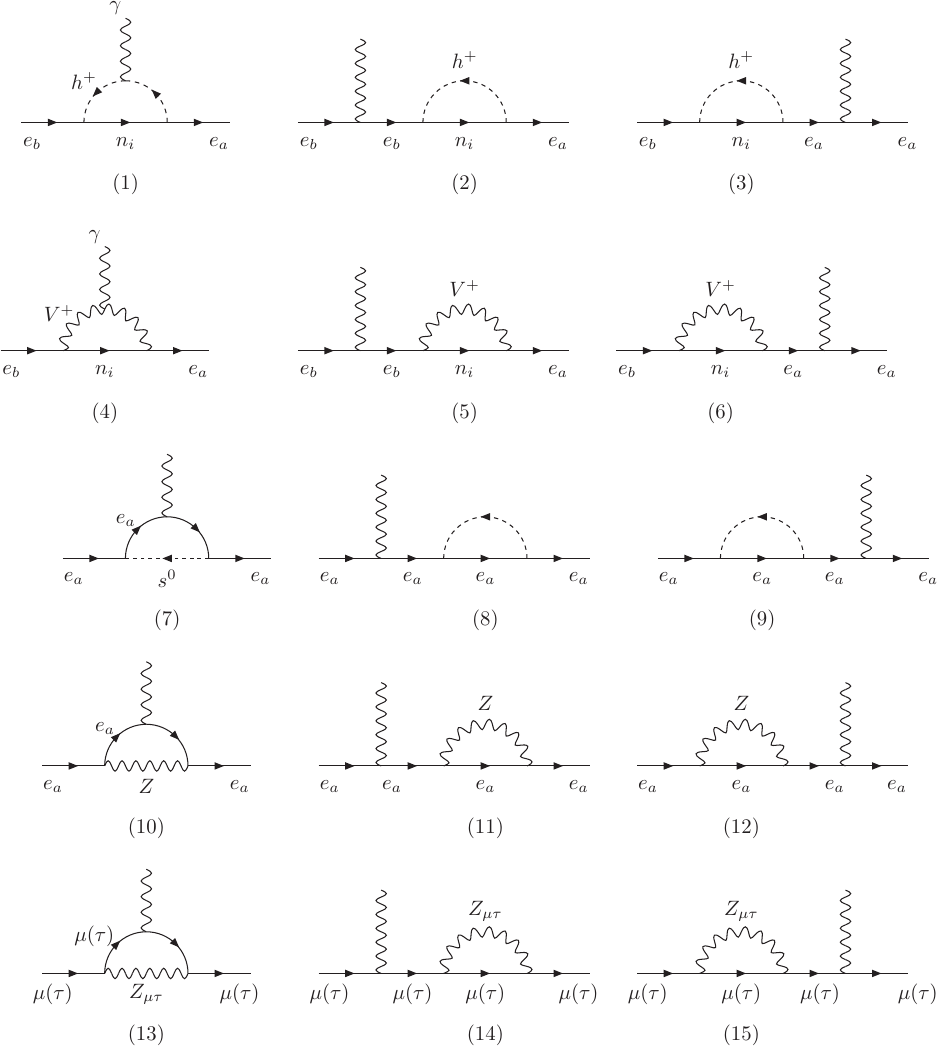}
 	\caption{ One-loop Feynman diagrams contributing to the decay amplitudes $e_b\to e_a\gamma$ and $(g-2)_{e_a}$ in the unitary gauge, with $V=W,W'$.}\label{fig:ebaU}
 \end{figure}

The  additional lepton flavor conserved (LFC) couplings, which give one-loop contributions to  $(g-2)_{e_a}$ but are not LFV amplitudes, are
 \begin{align}
 \label{eq:gax}
\mathcal{L}^{\mathrm{LFC}}=& \overline{e_L} \left[   \frac{h}{\sqrt{2}} \left( s_{\alpha} \tilde{Y}^{\ell}+c_{\alpha} Y^{\ell} \right) + \frac{h_2}{\sqrt{2}} \left( c_{\alpha} \tilde{Y}^{\ell} -s_{\alpha} Y^{\ell} \right) +  \frac{ia^0}{\sqrt{2}} \left( -c_{\beta} \tilde{Y}^{\ell}+s_{\beta} Y^{\ell} \right) \right]e_R 
\crn&+ \sum_{a} \frac{g}{2 c_W} \overline{e_{a}}\gamma^{\rho}  \left[   \left[ c_{\xi}(1-2s_W^2) -s_Wt_{\zeta}s_{\xi}\right] P_L +   s_W\left( \frac{s_{\xi}c_{2\zeta}}{s_{\zeta} c_{\zeta}} +c_{\xi}\right)  P_R \right]   e_{a}Z_{\rho}
\crn&+  \frac{g_{\mu \tau} }{2} Z_{\mu\tau, \rho} \left(  \overline{\mu}\gamma^{\rho}\mu\right)
-\frac{g_{\mu \tau} }{2} Z_{\mu\tau, \rho}  \left(\overline{\tau}\gamma^{\rho}\tau\right). 
 \end{align}
 Here we ignore the couplings of goldstone bosons,  irrelevant to the calculation in the unitary gauge. In this gauge, the one-loop Feynman diagrams give contributions to cLFV decays  $e_b\to e_a \gamma$ and $ (g-2)_{e_a}$ are shown in Fig.~\ref{fig:ebaU}.
Formulas for cLFV branching ratios  and $\Delta a_{e_a}=a^{\mathrm{LRiss}}_{e_a}-a^{\mathrm{SM}}_{e_a}$   are determined as follows \cite{Lavoura:2003xp, Lindner:2016bgg, Hue:2017lak, Crivellin:2018qmi, Tran:2022cwh,Hue:2023rks}:  
\begin{align}
\label{eq_brebaga}
&\mathrm{Br}(e_b\to e_a\gamma)= \frac{48\pi^2}{G_F^2 m_{e_b}^2}\left( \left| c^{\mathrm{LRiss}}_{(ab)R}\right|^2 + \left| c^{\mathrm{LRiss}}_{(ba)R}\right|^2\right) \mathrm{Br}(e_b\to e_a \overline{\nu_a}\nu_b),
\crn &\Delta a_{e_a} =-\frac{4m_{a}}{e} \mathrm{Re}\left[c_{(aa) R}^{\mathrm{LRiss}} -c_{(aa) R}^{\mathrm{SM}}\right], 
\end{align}
where $G_F=g^2/(4\sqrt{2}m_W^2)$, Br$(\mu\to e \overline{\nu_e}\nu_{\mu})\simeq 1$,  Br$(\tau\to e \overline{\nu_e} \nu_{\tau}) \simeq 0.1782$, Br$(\tau\to \mu \overline{\nu_\mu}\nu_{\tau})\simeq 0.1739$ \cite{ParticleDataGroup:2020ssz}, and
\begin{align}
\label{eq:cabR}
c^{\mathrm{LRiss}}_{(ab)R}=&c_{(ab)R}(h^\pm)+ c_{(ab)R}(W^{\pm}) +c_{(ab)R}(W'^{\pm}),
\crn c^{\mathrm{LRiss}}_{(ba)R}=&c_{(ba)R}(h^\pm)+ c_{(ba)R}(W^{\pm}) +c_{(ba)R}(W'^{\pm}),
\crn c_{(ab)R}(X)=& \frac{e}{16 \pi^2 m_X^2} \sum_{i=1}^9\left[ g^{LR}_{iXX,ab} m_{n_i}f_{X}(t_{i,X}) + \left( m_{e_b} g^{LL}_{iXX,ab} + m_{e_a} g^{RR}_{iXX,ab}\right) \tilde{f}_{X}(t_{i,X}) \right],
\crn  c_{(ba)R}(X)=&c_{(ab)R}(X) \left[ g^{LL} \leftrightarrow g^{RR},\; g^{LR} \leftrightarrow g^{RL}\right],
\end{align}
where $X=h^\pm,W,W'$; $t_{i,X}=m^2_{n_i}/m^2_X$ with $i=1,2,\dots,9$; and $g^{AB}_{iXX,ab}=g^{A*}_{aiX} g^{B}_{biX}$ with $A,B=L,R$ relating to the boson $X$ are derived from LFV coupling factors given in Eq. \eqref{eq:gX}. Precise analytic formulas are shown in appendix \ref{app:LFVB}. 
 Two form factors $f_X(x)= f_{S(V)}(x)$ and $\tilde{f}_X(x)= \tilde{f}_{S(V)}(x)$ for  $X=S(V)$ being scalar (gauge) boson exchanges in the loops of Feynman diagrams given in Fig. \ref{fig:ebaU}:
\begin{align}
\label{eq:fSV}
f_S(x)=& \frac{x^2 -1-2x\ln x}{4(x-1)^3}, 
\crn \tilde{f}_S(x)=& \frac{2x^2 +3x^2 -6x +1 -6x^2 \ln x}{24(x-1)^4},
\crn f_V(x)=& \frac{x^3 -12 x^2 +15 x -4 +6x^2 \ln x}{4(x-1)^3},
\crn \tilde{f}_V(x)=& \frac{-4x^4 +49x^3 -78 x^2 +43x -10 -18x^3 \ln x}{24(x-1)^4}.
\end{align}

We now focus on the "chiral-enhancement" terms proportional to $g^{LR}_{ih^+h^+,ab}$, which can give sizable contributions to $\Delta a_{e_a}$ and LFV decay amplitudes, as discussed in various works \cite{Crivellin:2018qmi, Dermisek:2020cod, Cherchiglia:2023utd}. For convenience,  the terms proportional to  $g^{XY}$ in analytic expressions of the one-loop contributions will be referred to as the $XY$ contributions through out this work.

 The expression of $U^{\nu}$ given in Eq. \eqref{eq:Usimeq} shows  that $g^{LR}_{ih^+h^+,ab}=0$ for $i\leq 3$, while  for $i>3$, it is proportional to  $\delta_{ac}$, where  $1\leq c=i-3,i-6\leq 3$. Explicitly:
 \begin{align}
\label{eq:gLRih}
g^{LR}_{ih^+h^+,ab}
=& c_{\kappa} \delta_{ac} \delta_{bc} \left( \frac{m_{e_a}}{v}- \sqrt{2}s_{\beta} \tilde{Y}^{a*}\right)
\\ &\times  \left[\frac{c_{\kappa}}{\sqrt{2} c_{\beta}M_c }\left( m_{e_b} t_{\beta} +\frac{\tilde{Y}^bc_{2\beta}v}{\sqrt{2} c_{\beta} } \right) \left(\frac{\sqrt{2} m_{e_b} s_{\beta}}{v} -\tilde{Y}^b\right) 
%
+\frac{s_{\kappa} M_c}{v_R } \left( 1- \frac{(m^b_D)^2}{2 M_c^2} \right) \right].\nn 
\end{align}
Consequently,  denoting  $c^{LR}_{(ab)R}(h^+) $ as the $LR$ contribution from $h^+$ exchange  to $c_{(ab)R}$ derived from Eq. \eqref{eq:cabR}, we obtain a simple formulas of $c^{LR}_{(ab)R}(h^+) \propto  g^{LR} \propto \sum_{c=1}^3\delta_{ac}\delta_{bc}=\delta_{ab}$. The most interesting feature  is that  $c^{LR}_{(ab)R} \propto \delta_{ab}$, implying that this contribution does not contribute to cLFV amplitudes. In contrast, it can give sizable contributions to $\Delta a_{e_a}$.  In particular, in the limit $m_{e_a}/v,s_{\kappa}\ll1$, the dominant contributions in Eq. \eqref{eq:gLRih} yields the following approximate expression, which is useful for cross-checking the numerical results:
\begin{align}
\label{eq:dah0}
a^{{LR}}_{e_a}({h^{\pm}}) 
 \simeq - \frac{c_{\kappa} m_{e_a} t_{\beta}}{2 \pi^2 m_{h^+}^2} \left[f_{S}(M_a^2/m^2_{h^+})  \right]  |\tilde{Y}^{a}|^2\left( m_{e_a} t_{\beta} +\frac{\tilde{Y}^ac_{2\beta}v}{\sqrt{2} c_{\beta} } \right).
\end{align}
It is evident that $a^{{LR}}_{e_a}({h^{\pm}}) $   can become sizable for increasing $t_{\beta}$ and $|\tilde{Y}^a|$, even for $m_{h^+}\propto \mathcal{O}(1)$. More importantly, this contribution is independent of the cLFV amplitudes and can therefore accommodate a contribution to $\Delta a_{\mu}$ of value $10^{-9}$.

 We compare our results with those of previous works \cite{Majumdar:2020xws, Majumdar:2022jur}. The general analytic formulas used in these works are given in Ref. \cite{Lindner:2016bgg} and coincide with our results; see also Ref. \cite{Hue:2023rks} for a detailed discussion. Except for the contributions from the singly charged Higgs bosons, our calculations of the remaining one-loop contributions involving the two singly charged gauge bosons, neutral Higgs bosons, and neutral gauge bosons are generally consistent with those presented in Refs. \cite{Majumdar:2020xws, Majumdar:2022jur}. In particular, in the limit $s_{\theta}=0$, the LR contributions from the two gauge bosons vanish, while the LL and RR contributions coincide with those given in Refs. \cite{Majumdar:2020xws, Majumdar:2022jur}. On the other hand, the contributions from singly charged Higgs-boson exchange obtained in those works omit an LR term proportional to $\epsilon_f=M_c/m_{e_a}$, which corresponds to $a^{{LR}}_{e_a}(h^\pm)$ in our result.

\begin{figure}[ht] 
  	\includegraphics[width=12cm]{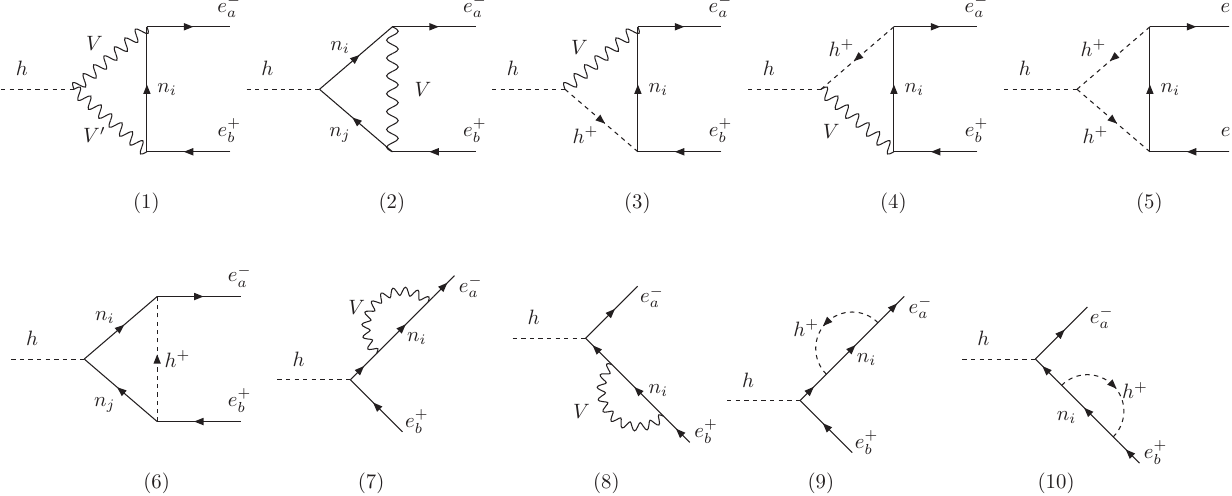}
  	\caption{ One-loop Feynman diagrams contributing to the LFV$h$ decay amplitudes in the unitary gauge, with $V,V'=W,W'$.}\label{fig:hebaU}
  \end{figure}

  In the unitary gauge, the one-loop Feynman diagrams give contributions to LFV$h$ decays  $h\to e_b e_a$   are shown in Fig. \ref{fig:hebaU}.
Correspondingly, Lagrangian for couplings of the SM-like Higgs boson $h$ with Higgs and gauge bosons giving one-loop contributions to LFV$h$ decays are included the kinetic terms of the Higgs multiplets
\begin{align}
\label{eq:LDS}	
\mathcal{L}_D^S=\mathrm{Tr}\left[ (D_{\mu}\Phi)^{\dagger}(D^{\mu}\Phi)\right] + (D_{\mu}H_R)^{\dagger}(D^{\mu}H_R)=\mathcal{L}_{hBB'}+ \mathcal{L}_{ZBB'}+\dots,
\end{align}
where $\mathcal{L}_{hBB'}$ and $\mathcal{L}_{ZBB'}$ consist of couplings that generate one-loop Feynman diagrams for LFV$h$ and LFV$Z$ decays, respectively.  Expanding $\mathcal{L}_D^S$ leads to the following needed couplings: 
\begin{align}
\label{eq:hZBB}
\mathcal{L}_{hBB'}=& g_{hWW} \times hW^+_{\mu}W^{-\mu} + g_{hWW}  \times hW'^+_{\mu}W'^{-\mu} +g_{hWW'} \times h (W^+_{\mu}W'^{-\mu}+ W^-_{\mu}W'^{+\mu})
\crn&+  g_{W^-h^+h}  \times  \left[  (p_+-p_0)_{\mu}  (hh^+W^{-\mu}) + (p_0 -p_-)_{\mu}  (h h^-W^{+\mu})\right] 
\crn&+g_{W'^-h^+h}  \times\left[ (p_+-p_0)_{\mu}  (hh^+W'^{-\mu}) +(p_0 -p_-)_{\mu} (hh^-W'^{+\mu})  \right] ,
\end{align}
where
\begin{align}
\label{eq:ghBB}
g_{hWW}= & \frac{g^2 v \left(c_{\delta } s_{\theta }^2 s_W^2+c_{\delta } c_{\theta }^2 c_W^2 s_{\zeta }^2- s_{\zeta } c_{\theta }  s_{\theta } s_{2W} s_{(\alpha+\beta)}\right)}{2 c_W^2 s_{\zeta }^2},
\crn g_{hW'W'}= &  \frac{g^2 v \left(c_{\delta } c_{\theta }^2 s_W^2+c_{\delta } c_W^2 s_{\zeta }^2 s_{\theta }^2+ c_{\theta } s_{\zeta } s_{\theta } s_{2W} s_{(\alpha+\beta)} \right)}{2 c_W^2 s_{\zeta }^2},
\crn g_{hWW'}= &g_{hW'W}=\frac{g^2 v \left[c_{\delta } c_{\theta } s_{\theta } \left(s_W^2-c_W^2 s_{\zeta}^2\right)-c_W s_{\zeta } s_W c_{2 \theta } s_{(\alpha+\beta)}\right]}{2 c_W^2 s_{\zeta }^2},
\crn g_{W^-h^+h} =& \frac{g c_{\kappa}}{2}\left[- c_{\theta } s_{\delta}  + \frac{ s_{\theta } s_W c_{(\alpha+\beta)} }{  c_W s_{\zeta }} \right], 
\crn g_{W'^-h^+h} =&  \frac{g c_{\kappa }}{2} \left[ s_{\theta } s_{\delta} + \frac{ c_{\theta } s_W c_{(\alpha+\beta)}}{  c_W s_{\zeta }} \right]. 
\end{align}
 We confirm that the coupling $hZZ$, $hhZZ$, and $hhWW$ agree with the SM results in the limits $\theta \varpropto \mathcal{O}(v^2/v_R^2)\to0$ and $\delta=\alpha -\beta  \varpropto \mathcal{O}(v^2/v_R^2) \to 0$. 
 
  The triple Higgs coupling  factor for one-loop contribution to the LFV$h$ decay amplitude:
  \begin{align}
  \label{eq:la3h}
  \lambda_{hh^+h^-}=& -\frac{2 s_{2\kappa }  c_{\delta} \lambda_{11}  v_R }{t_{2 \beta }}
  %
  + 2 c_{\kappa }^2 \lambda _3 v c_{\delta} + 2 c_{\kappa }^2 \lambda _4 v \left[c_{\alpha } c_{\beta }\left(1+2  s_{\beta }^2\right) +s_{\alpha } s_{\beta } \left(c_{\beta
  }^2+1\right)\right]
\crn& -8 c_{\beta } c_{\kappa }^2 \lambda _5 s_{\beta } v s_{\alpha +\beta} +2 c_{\kappa }^2 \lambda _6 v c_{2\beta } s_{\delta }
%
-\frac{4 \lambda _{11} s_{\kappa }^2 v  c_{(\alpha +\beta)}}{t_{2\beta }},
  \end{align}
which  is derived from the Higgs potential $V_h=-\mathcal{L}_{hh^+h^-} +\dots$ with $\mathcal{L}_{hh^+h^-}= - \lambda_{hh^+h^-}\times hh^+h^-$ following the notations given in Ref. \cite{Hue:2024rij}.

\begin{figure}[ht] 
  	\includegraphics[width=12cm]{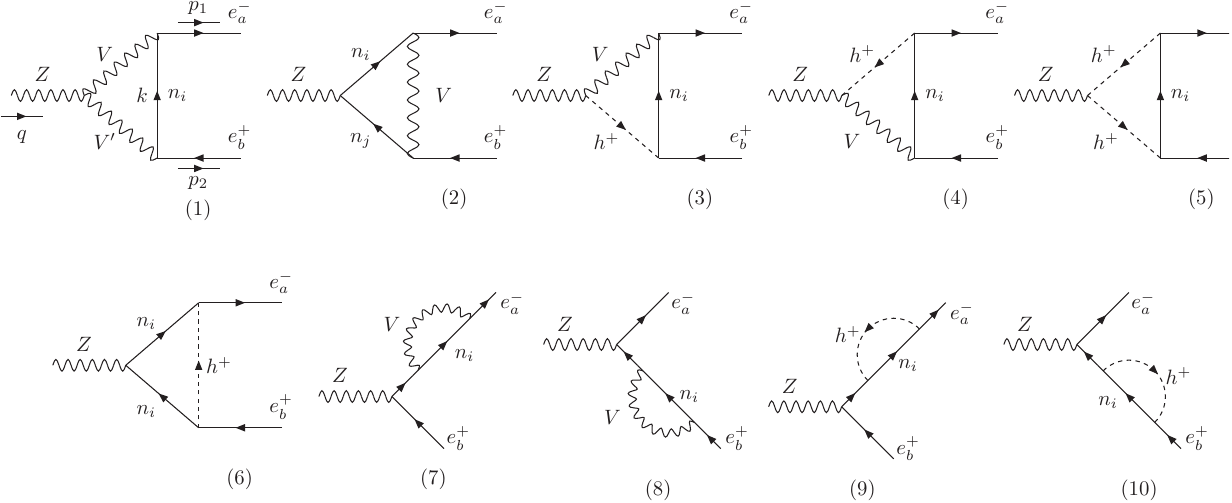}
  	\caption{ One-loop Feynman diagrams contributing to the LFV$Z$ decay amplitudes in the unitary gauge, with $V,V'=W,W'$.}\label{fig:ZebaU}
  \end{figure}

In the unitary gauge, the one-loop Feynman diagrams give contributions to LFV$Z$ decays  $Z\to e_b e_a$   are shown in Fig. \ref{fig:ZebaU}.  

The Lagrangian  $\mathcal{L}_{ZBB'}$ with $B,B' = W, W', h^\pm$ can be written as follows:
\begin{align}
\label{eq:LZBB}
\mathcal{L}_{ZBB'}=& e (p_{+\mu } -p_{-\mu }) g_{Zh^+h^-}  Z^{\mu} h^+h^{-} + e g_{h^-W^+Z}  Z^{\mu} \left(   h^+W^{-}_\mu  + h^-W^{+}_\mu \right) 
\crn& 
+e g_{h^-W'^+Z}   Z_{\mu} \left( h^+W'^{-\mu}  +  h^-W'^{+\mu}  \right) ,
\end{align}
where 
\begin{align}
\label{eq:gZBB}
g_{Zh^+h^-}=& \frac{(c_{\zeta }^2  -s_{\zeta }^2 s_{\kappa }^2  ) s_{\xi }}{2 c_W c_{\zeta }  s_{\zeta }} + \frac{  c_{\xi } \left(c_{\kappa }^2-2 s_W^2\right)}{2 c_W s_W}, 
\crn g_{h^-W^+Z}=& \frac{g s_{\theta } \left[ s_{\kappa } s_W v_R (c_{\zeta } c_{\xi } s_W+s_{\zeta } s_{\xi })+c_{\zeta } c_{\kappa } c_{\xi } c_W^2  v  c_{2\beta} \right]}{2 c_{\zeta } c_W^2 s_{\zeta }},
\crn g_{h^-W'^+Z}=& \frac{ g c_{\theta } \left[  s_{\kappa } s_W v_R (c_{\zeta } c_{\xi } s_W+s_{\zeta } s_{\xi })+c_{\zeta }  c_{\kappa } c_{\xi } c_W^2 v c_{2\beta} \right]}{2 c_{\zeta } s_{\zeta } c_W^2 }. 
\end{align}
Finally the triple couplings of $Z$ with charged gauge bosons are derived from the covariant kinetic terms of the gauge bosons, namely 
\begin{align}
\label{eq:LDV}
\mathcal{L}_{D}^V=& -\frac{1}{4}\sum_{a=1}^3F_{\mu\nu}F^{\mu\nu} -\frac{1}{4}\sum_{a=1}^3F'_{\mu\nu}F'^{\mu\nu} 
\crn =& \sum_{V,V'=W,W'} \left( -e g_{ZVV'}\right) \Gamma^{\mu \nu \sigma}(p_0,p_+,p_-) \times (Z_{\mu} V^+_{\nu} V'^-_{\sigma})+\dots, 
\end{align} 
where $\Gamma^{\mu \nu \sigma}(p_0,p_+,p_-)=g^{\mu \nu}(p_0 -p_+)^{\sigma} + g^{ \nu \sigma}(p_+ -p_-)^{\mu} + g^{ \sigma \mu}(p_- -p_0)^{\nu}$ after applying the transformation that  $\partial_{\mu}V^{0,\pm}(p)= -ip_{0,\pm\mu}$. The scalar factors $g_{ZVV'}$ are listed as follows
\begin{align}
\label{eq:gZVV}
g_{ZWW}=& t_W^{-1}\left[c_{\theta }^2 c_{\xi }  + \frac{ s_{\theta }^2 s_W \left( s_{\xi } t^{-1}_{\zeta }  -c_{\xi }s_W \right) }{c_W^2 } \right] ,
\crn g_{ZW'W'}  =&   t_W^{-1} \left[s_{\theta }^2 c_{\xi }  +\frac{  c_{\theta }^2  s_W \left( s_{\xi} t^{-1}_{\zeta }  -c_{\xi } s_W \right)}{c_W^2 }\right],
\crn g_{ZWW'}=&g_{ZW'W}= \frac{ c_{\theta } s_{\theta } (t^{-1}_{\zeta } s_{\xi } s_W-c_{\xi})}{s_W c_W}. 
\end{align}
  Formulas for the LFV decay rates of the two bosons SM-like Higgs $h$  and $Z$ are available  in Refs. \cite{Pilaftsis:1992st, Arganda:2004bz, Arganda:2014dta} and \cite{DeRomeri:2016gum, Jurciukonis:2021izn}, respectively. The notation used in this work follows Refs. \cite{Hue:2024rij, Hong:2024yhk, Hong:2023rhg}. Detailed formulas are provided in Appendix \ref{app:LFVB}.
 
 \section{\label{sec:MRij} Non-diagonal form of $M_R$ for LFV decay rates and $\Delta a_{e_a}$}
 
 We start with the leptonic Yukawa sector in Eq. \eqref{eq:LYlep}. Since the $U(1)_{L_{\mu}-L_{\tau}}$ charges of the SM leptons are fixed, the Yukawa coupling matrices $Y^{\ell}$ and $\tilde{Y}^{\ell}$ remain diagonal. On the other hand, the $U(1)_{L_{\mu}-L_{\tau}}$ charges of the three left-handed SM singlets $S_L=(S_1,S_2,S_3)_L$ are not fixed a priori, but are constrained by the requirement that Eq. \eqref{eq:LYlep} be invariant under $U(1)_{L_{\mu}-L_{\tau}}$. Consequently, the second term in Eq. \eqref{eq:LYlep} allows six possible assignments of the $U(1)_{L_{\mu}-L_{\tau}}$ charges of the $S_{iL}$ fields. The neutrino mass matrix $M_R$ therefore need not be diagonal in general. Introducing two unitary matrices $V^{L,R}$ to diagonalize $M_R$, such that $\hat{M}_R=V^{LT}M_RV^R$ \cite{Dreiner:2008tw}, the general form of $U^{\nu}$ is given by
\begin{align}
\label{eq:Usimeq1}
V=& \frac{1}{\sqrt{2}}\begin{pmatrix}
V^R& iV^R \\ 
V^L& -iV^L 
\end{pmatrix} ,\; 
%
U^{\nu} \simeq  \begin{pmatrix}
\left(I_3 -\frac{R^{0}R^{0\dagger}}{2}\right)U^{\nu}_3	& \frac{ R^{0} V^L}{\sqrt{2}}&  \frac{-iR^{0} V^L}{\sqrt{2}}\\
\mathcal{O}_{3\times3}	&  \frac{V^{R}}{\sqrt{2}}  &   \frac{iV^{R}}{\sqrt{2}} \\
-R^{0\dagger} U^{\nu}_3& 	\left(I_3 -\frac{R^{0\dagger}R^{0}}{2}\right) \frac{V^L}{\sqrt{2}}  &-i  \left(I_3 -\frac{R^{0\dagger}R^{0}}{2}\right) \frac{V^L}{\sqrt{2}}
\end{pmatrix}.
\end{align}

The $U(1)_{L_{\mu}-L_{\tau}}$ charges of the three $S_{iL}$ fields, with $i=1,2,3$, admit the following six possible assignments, including the one shown in Table \ref{t:particle}: $U(1)_{L_{\mu} -L_{\tau}} \{S_{1},S_{2},S_{3} \}=\{0,1,-1\}$. The other five assignments are obtained by the permutations $\{(1,2),(1,3),(2,3),(1,2,3),(3,2,1)\}$, which lead to $U(1)_{L_{\mu} -L_{\tau}} \{S_{1},S_{2},S_{3} \}=\{1,0,-1\}$, $\{-1,1,0\}$, $\{0,-1,1\}$, $\{-1,0,1\}$, and $\{1,-1,0 \}$, respectively.  The corresponding mass matrices $M_R$ take the following forms:
\begin{align}
\label{eq:MRs}
M_R=&(1):\; \mathrm{diag} \left(M_1,\;M_2,\;M_3\right);\; (2):\begin{pmatrix}
0&M_{12}  &0  \\
M_{21}& 0 & 0 \\
0&0  &M_{33} 
\end{pmatrix} ;\; (3):\begin{pmatrix}
0&0  &M_{13}  \\
0& M_{22} & 0 \\
M_{31}&0  &0
\end{pmatrix};
\crn &\; (4):\begin{pmatrix}
M_{11}&0  &0  \\
0& 0 & M_{23} \\
0&M_{32}  &0
\end{pmatrix};
\;(5):\begin{pmatrix}
0&M_{12}  &0  \\
0& 0 & M_{23} \\
M_{31}&0  &0
\end{pmatrix};\;(6):\begin{pmatrix}
0&0 &M_{13}  \\
M_{21}& 0 & 0 \\
0&M_{32}  &0
\end{pmatrix}. 
\end{align}

In all six cases, the matrices satisfy $M_RM_R^{\dagger}=M_R^{\dagger}M_R=\mathrm{diag}(|M_1|^2,\;|M_2|^2,\;|M_3|^2)$. Therefore, the two unitary matrices $V^L$  and $V^R$ that diagonalize $M_R$ can be determined straightforwardly, with one of them being the identity matrix. Specifically, if $V^L=I_3$  ($V^R=I_3$), the corresponding $V^R$ ($V^L$) can be obtained from $M_R$ ($M_R^\dagger$) using the relations $M_RM_R^{\dagger} (M_R^{\dagger}M_R)=\mathrm{diag}(|M_1|^2,\;|M_2|^2,\;|M_3|^2)$, namely: 
\begin{align}
\label{eq:VRs}
V^{R\dagger}=& (2):\begin{pmatrix}
0&1  &0  \\
1& 0 & 0 \\
0&0  &1 
\end{pmatrix} ;\; (3):\begin{pmatrix}
0&0  &1  \\
0& 1 & 0 \\
1&0  &0
\end{pmatrix};
\; (4):\begin{pmatrix}
1&0  &0  \\
0& 0 & 1 \\
0&1  &0
\end{pmatrix};
\;(5):\begin{pmatrix}
0&1  &0  \\
0& 0 & 1 \\
1&0  &0
\end{pmatrix};\;(6):\begin{pmatrix}
0&0 &1  \\
1& 0 & 0 \\
0&1  &0
\end{pmatrix},
\end{align}
where we assume all non-zero elements of $M_{R}$ are positive. 

In the general case, the entries of  $V^{RT}$  may have phases with signs opposite to those of the corresponding entries in $M_R$, but the final conclusions remain unchanged. Similarly, when $V^{R}=I_3$, $V^L$ takes the following possible forms, with the correspondence $V^{R\dagger}\to V^{L*}$. It should be noted that, for a non-diagonal $M_R=V^{L*}\hat{M}_RV^{R\dagger}$, the couplings $g^R_{aih^+}$ given in Eq. \eqref{eq:gX} are modified according to $ s_{\kappa} U^{\nu*}_{(a+6)i} \left(Y_R^*\right)_{aa} \to s_{\kappa}  \sum_{c=1}^{3}U^{\nu*}_{(c+6)i} \left(Y_R^*\right)_{ac}$. In addition, $R^{0}$ given in Eq. \eqref{eq:nusector} is modified as $R^0=m_D^{\dagger} V^{R*} \hat{M}_R^{-1}V^{L\dagger}$.   Using the approximate expression for $U^{\nu}$ given in Eq. \eqref{eq:Usimeq}, the couplings $g^{L(R)}_{aih^+}$ in Eq. \eqref{eq:gX} can be derived explicitly in terms of  $V^R$. For example, $g^{L}_{aih^+}$ contains $ \left(R^0V^L\right)_{ac}= \left( V^*_{R}\right)_{ac} m^a_D/M_c \forall a,c=1,2,3$  for $1\leq c=i-3,i-6\leq 3$, with $i\geq 4$.  Similarly, $g^R_{aih^+}$ contains the combinations $M_R^TV^L= V^{R*} \hat{M}_R$ and $M^T_RR^{0\dagger} R^0 V_L= m_Dm_D^{\dagger}V^{R*} \hat{M}_R^{-1}$, which yield $  \left[M_R^T 	\left(I_3 -(R^{0\dagger}R^{0})/2\right)V^L  \right]^*_{bc}= V^{R}_{bc} M_c \left( 1 - (m_D^b)^2M_c^{-2}/2 \right).$  Hence,  $g^{LR}_{ih^+h^+,ab}=0$ for $1\leq i\leq 3$, whereas  $ g^{LR}_{ih^+h^+,ab}= g^{LR}_{(i+3)h^+h^+,ab}$ for $4\leq i=c+3\leq6$,  leading to  the following dominant contribution from $ g^{LR}_{ih^+h^+,ab}$ in the limit $s_{\kappa}\ll1$:
\begin{align}
\label{eq:LRpart}
g^{LR}_{ih^+h^+,ab} =
 \frac{ c^2_{\kappa} V^{R}_{bc} V^{R*}_{ac}}{\sqrt{2}c_{\beta}M_c} \left( m_{e_b}t_{\beta} + \frac{\tilde{Y}^{b}c_{2\beta}v}{\sqrt{2} c_{\beta}}\right)\left(\frac{\sqrt{2}m_{e_b}s_{\beta}}{v} - \tilde{Y}^{b}\right)  \left( \frac{ m_{e_a}}{v} -\sqrt{2}s_{\beta} \tilde{Y}^{a*}\right).
\end{align}
This expression is independent of $V^L$ but depends on the nonzero entries of $V^R$, leading to
\begin{align}
\label{eq:ax}
\Delta a_{e_a}^{{LR}} \simeq  &-\frac{c^2_{\kappa} t_{\beta} m_{e_a}}{2 \pi^2 m^2_{h^+}} \left( m_{e_a}t_{\beta} + \frac{\tilde{Y}^{a}c_{2\beta}v}{\sqrt{2} c_{\beta}}\right)|\tilde{Y}^{a}|^2 \sum_{c=1}^3|V^{R}_{ac}|^2 f_{S}(M_c^2/m^2_{h^+}),
\\ \; \mathrm{Br}^{{LR}}(e_b \to e_a\gamma) \propto & \left(  \left|\left( t_{\beta} + \frac{\tilde{Y}^{b}c_{2\beta}v}{\sqrt{2} c_{\beta}m_{e_b}} \right)\left( \tilde{Y}^{b} \tilde{Y}^{a*}\right) \sum_{c=1}^3V^{R*}_{ac} V^{R}_{bc}f_{S}(M_c^2/m^2_{h^+})\right|^2
\right.
\crn& \left. \quad+\left|\left( \frac{m_{e_a}}{m_{e_b}}t_{\beta} + \frac{\tilde{Y}^{a}c_{2\beta}v}{\sqrt{2} c_{\beta}m_{e_b}}\right)\left( \tilde{Y}^{a} \tilde{Y}^{b*}\right) \sum_{c=1}^3V^{R*}_{bc} V^{R}_{ac}f_{S}(M_c^2/m^2_{h^+})\right|^2\right) \nn,
\end{align}
The nonzero entries $V^{R*}_{ab}=1$, with $a,b=1,2,3$, are listed in Table \ref{t:Vca} for all permutations of the $U(1)_{L_{\mu}-L_{\tau}}$ charge assignments of the $S_{aL}$ fields.
\begin{table}[ht]
	\centering 
	\begin{tabular}{cccc} 
		\hline 
	Non-zero $V^R_{ab}$	&$\Delta a_{e}$& $\Delta a_{\mu}$&$\Delta a_{\tau}$ \\
		\hline 
		(1): $V^R_{11}, V^R_{22},V^R_{33}=1$ &$|V^{R}_{11}|^2$& $|V^{R}_{22}|^2$& $|V^{R}_{32}|^2$\\	
		(2): $V^R_{12}, V^R_{21},V^R_{33}=1$ &$|V^{R}_{12}|^2$& $|V^{R}_{21}|^2$& $|V^{R}_{33}|^2$\\
		(3): $V^R_{13}, V^R_{31},V^R_{22}=1$&$|V^{R}_{13}|^2$& $|V^{R}_{22}|^2$& $|V^{R}_{31}|^2$\\
		(4): $V^R_{11}, V^R_{23},V^R_{32}=1$ &$|V^{R}_{11}|^2$& $|V^{R}_{23}|^2$& $|V^{R}_{32}|^2$\\	
		(5): $V^R_{32}, V^R_{21}, V^R_{13}=1$ &$|V^{R}_{13}|^2$& $|V^{R}_{21}|^2$& $|V^{R}_{32}|^2$\\
		(6): $V^R_{31}, V^R_{12},V^R_{23}=1$ &$|V^{R}_{13}|^2$& $|V^{R}_{31}|^2$& $|V^{R}_{33}|^2$\\	
	\end{tabular}
	\caption{Nonzero entries of $V^{R}$ for six possible permutations of  the $U(1)_{L_{\mu}- L_{\tau}}$ charge assignments, yielding sizable  "chirally-enhanced" one-loop contributions to $\Delta a_{e_a}$, while vanishing in LFV decay amplitudes.}\label{t:Vca}
\end{table}
 We also list precisely the nonzero contributions to $\Delta a_{e_a}$. In contrast,  we always have $\mathrm{Br}^{{LR}}(e_b \to e_a\gamma) \varpropto V^{R*}_{ac}V^R_{bc}=0$ for all $1\leq a<b\leq3$ because at least one of the two elements $ V^{R}_{ac}$ or $V^R_{bc}$must vanish. This property persists even for the subdominant terms in the $LR$ contributions. Therefore, the one-loop contributions to the LFV decay rates arise only from the $g^{LL}$ and $g^{RR}$ parts. The explicit expressions for these factors in the corresponding contributions are given in Appendix \ref{app:LFVB}. Qualitatively, all LFV couplings given in Eq. \eqref{eq:gX} exhibit the following property for $i>3$:
 \begin{align}
 \label{eq:gXXab}
 g^{XX}_{i,ab}\propto \left\{ U^{\nu*}_{(a+3)i}U^{\nu}_{(b+3)i},\; U^{\nu*}_{ai}U^{\nu}_{bi} \right\}\propto  \left\{ V^{R}_{ac}V^{R*}_{bc},\;  V^{R}_{ac}V^{R*}_{bc} \right\} \propto \delta_{ab},
 \end{align} 
 where $c=\left\{i-3,i-6\right\}$, for $U^{\nu}$ given in Eq. \eqref{eq:Usimeq1}. Consequently, these products vanish for $a\neq b$, and hence give no one-loop contributions to LFV amplitudes. The only nonzero products occur for $1\leq i\leq3$, for which $ g^{XX}_{i,ab}\propto \left(U^{\nu*}_3\right)_{ai} \left(U^{\nu}_3\right)_{bi}\neq0$, corresponding to the contributions of active neutrinos to the LFV decay amplitudes. These contributions  are highly suppressed by the tiny active neutrino masses. Our numerical calculations show that only the $RR$ contribution from singly charged Higgs-boson exchange is nonzero, but it gives a tiny branching ratio, Br$(e_b\to e_a\gamma)<10^{-50}$, consistent with the qualitative estimate based on Eq. \eqref{eq:cRR} in Appendix \ref{app:LFVB}. The contributions from singly charged gauge-boson exchange to the cLFV decay rates, given by Eq. \eqref{eq:cLL} in Appendix \ref{app:LFVB}, yield
  \begin{equation}
 \label{eq:cLFVrates}
 \mathrm{Br}(\mu\to e \gamma)\simeq 3.77\times 10^{-54},\; \mathrm{Br}(\tau \to e\gamma) \simeq 7.3\times 10^{-55},\; \mathrm{Br}(\tau\to \mu\gamma)\simeq  1.4\times 10^{-53}.
 \end{equation}
 These results are fully consistent with the pioneering works \cite{Petcov:1976ff, Bilenky:1977du, Cheng:1980tp}. The dominant contributions arise from the $LL$ part of $W^\pm$ exchange involving only active neutrinos, whereas the $RR$ contributions from singly charged Higgs-boson exchanges are highly suppressed relative to the current experimental sensitivities to$\Delta a_{e,\mu}$. These results are confirmed by our numerical calculations using the general formulas given in Eq. \eqref{eq_brebaga}.
 
 To conclude the qualitative discussion, we present an alternative way to demonstrate the suppression of the LFV decay rates. Starting from the Yukawa Lagrangian in Eq. \eqref{eq:LYlep}, we redefine the basis of $S_L$ according to  $S_L\to V^RS_L$, which eliminates $V^R$ from the diagonalization procedure: $V^{LT}Y^RV^R\to V^{LT}Y^R=\sqrt{2}\hat{M}_R/v_R$. Working in this basis for $S_L$, $V^R$ does not appear explicitly in any subsequent step of the calculation. Since $V^L$ also drops out of the final analytic expressions, the resulting physical predictions are exactly equivalent to those obtained for case (1), in which $M_R$ is diagonal. The numerical analysis presented below will identify the regions of parameter space that accommodate sizable values of $\Delta a_{e_a}$ consistent with the current experimental constraints.

\section{ \label{sec:num}Numerical discussions}
 The numerical values of experimental data are  taken from Ref. \cite{ParticleDataGroup:2024cfk}, including the neutrino oscillation parameters, the  charged lepton masses; the masses of two gauge bosons $W$,  $Z$, and the SM-like Higgs bosons, namely 
\begin{align}
\label{eq_ex}
g &=0.652,\; G_F=1.166\times 10^{-5} \;\mathrm{GeV}^{-2},\;\alpha_e=\frac{1}{137}= \frac{e^2}{4\pi} ,\; s^2_{W}=0.231, 
\crn  m_W& =80.377 \; \mathrm{GeV},\;  m_Z =91.1876 \; \mathrm{GeV},\; m_h=125.25\; \mathrm{GeV}, \; 	 \Gamma_Z=2.4955 \; \mathrm{GeV}, 
\crn 	m_e&=5\times 10^{-4} \;\mathrm{GeV},\; m_{\mu}=0.105 \;\mathrm{GeV} ,\; m_{\tau}=1.776 \;\mathrm{GeV}. 
\end{align}

We focus on the best-fit values  of the neutrino oscillation parameter corresponding to  the normal ordering (NO) scheme  with $m_{n_1}<m_{n_2}<m_{n_3}$, namely  $s^2_{12}=0.32,\;   s^2_{23}= 0.547,\; s^2_{13}= 0.0216 $, $\Delta m^2_{21}\equiv m^2_{n_2} - m^2_{n_1}=7.55\times 10^{-5} [\mathrm{eV}^2], \;$, and $	\Delta m^2_{32} \equiv m^2_{n_3} - m^2_{n_2}=2.424\times 10^{-3} [\mathrm{eV}^2]$.  For simplicity, we fix the Dirac CP phase to $ \delta_D= 180 \;[\mathrm{Deg}] $  in the numerical analysis. Consequently, the neutrino masses and mixing matrix are fixed as follows  
\begin{align}
\hat{m}_{\nu}&=  \left( \hat{m}^2_{\nu}\right)^{1/2}= \mathrm{diag} \left( m_{n_1}, \; \sqrt{ m_{n_1}^2 + \Delta m^2_{21}},\; \sqrt{m_{n_1}^2  +\Delta m^2_{21} +\Delta m^2_{32}} \right), 	\label{eq:NOmnu}
\\  U_{\mathrm{PMNS}} &=f(\theta_{12},\theta_{13},\theta_{23},\delta_D) =\left(
\begin{array}{ccc}
c_{12} c_{13} & c_{13} s_{12} & s_{13} e^{-i \delta_D} \\
-c_{23} s_{12}-c_{12} s_{13} s_{23} e^{i \delta_D } & c_{13} c_{23}-s_{12} s_{13} s_{23} e^{i \delta_D } & c_{13} s_{23} \\
s_{12} s_{23}-c_{12} c_{23} s_{13} e^{i \delta_D } & -c_{23} s_{12} e^{i \delta_D } s_{13}-c_{13} s_{23} & c_{13} c_{23} \\
\end{array}
\right),\nn 
\end{align}
where $\sum m_{n_i} \leq0.186$ eV based on PDG 2025 data combing both the data of Planck 2018 \cite{Planck:2018vyg} and ACT DR6 CMB ($<$0.077 eV) \cite{DESI:2025gwf}, implying that $m_{n_1}\leq 0.011$ eV.

The set of free parameters in the LRiss  framework is given by  the $SU(2)_R$ breaking scale $v_R$, $t_{\beta}$, the mass $m_{Z_{\mu\tau}}$ of the $Z_{\mu\tau}$ boson, the mixing parameter $\zeta$ given in Eq. \eqref{eq:LRtoL}, three entries of $\tilde{Y}^{\ell}$, and three heavy neutrino masses  $M_{1,2,3}$.  These parameters determine $m_D$, $R^0$, and $\mu_L$ as functions of the entries of $\tilde{Y}^{\ell}$ and $M_{1,2,3}$, which are then required to satisfy perturbativity constraints. There are additional free parameters in the Higgs sector, namely the singly charged Higgs mass $m_{h^+}$, the small mixing $\delta$ in Eq. \eqref{eq:C01}, and Higgs-self couplings. Finally, the non-unitary of the active neutrino mixing matrix given Eq. \eqref{eq:Usimeq},  denoted as $\eta\equiv \frac{1}{2}R^0R^{0\dagger}$,  is also subject to experimental constraints \cite{Blennow:2023mqx, Fernandez-Martinez:2016lgt, Yu:2024nkc}. Eq. \eqref{eq:nusector} shows that $\eta$ is diagonal with constraints on nonzero entries are $\eta_{11,33}\leq 10^{-3},\eta_{22}\leq 10^{-4}$, where 
\begin{align}
\label{eq:etaii} \eta_{ii}=(R^0_i)^2= \frac{1}{2}\left[\frac{m_D^i}{M_i}\right]^2 ,\; i=1,2,3.
\end{align}

The cLFV decay rates are constrained by recent experimental results as follows \cite{BaBar:2009hkt, MEG:2016leq, Belle:2021ysv, MEGII:2023ltw, MEGII:2025gzr}:  	$\mathrm{Br}(\mu\rightarrow e\gamma) < 1.5\times 10^{-13}$, $\mathrm{Br}(\tau\rightarrow e\gamma) <3.3\times 10^{-8}$, and  $\mathrm{Br}(\tau\rightarrow \mu\gamma) <4.2\times 10^{-8}$.   The latest experimental constraints on LFV$h$ decay rates are $	\mathrm{Br}(h\rightarrow \tau \mu) <1.5\times 10^{-3}$, $	\mathrm{Br}(h\rightarrow \tau e) <2\times 10^{-3}$, and  $\mathrm{Br}(h\rightarrow \mu e) <4.4\times 10^{-5}$ \cite{CMS:2021rsq, ATLAS:2019xlq, CMS:2023pte, ATLAS:2023mvd}.  
The latest experimental constraints on LFV$Z$ decay rates  are $	\mathrm{Br}(Z\rightarrow \tau^\pm \mu^\mp) <6.5\times 10^{-6} $,  $\mathrm{Br}(Z\rightarrow \tau^\pm e^\mp) <5.0\times 10^{-6} $, and $	\mathrm{Br}(Z\rightarrow \mu^\pm e^\mp) <1.9 \times 10^{-7}$ \cite{ATLAS:2021bdj, ATLAS:2022uhq, CMS:2025wqy}.

All allowed parameter points retained in our numerical analysis satisfy the above constraints on the LFV decay rates, as well as the current experimental constraints on $\Delta a_{e_a}$. In particular, the allowed ranges considered for the electron, muon, and tau are  $ -2.5 \times 10^{-10} \leq 	\Delta a_{\mu}\leq 10^{-9}$ and $|\Delta a_{\mu}|>10^{-11}$; $5\times 10^{-15} \leq |\Delta a_{e}| \leq 5\times 10^{-13}$;  while $|\Delta a_{\tau}|$ is left unconstrained because the experimental sensitivity to $\Delta a_\tau$ is much weaker than the SM prediction, with $\Delta a_{\tau}\leq \mathcal{O}(10^{-3})$  \cite{DELPHI:2003nah, ATLAS:2026wrz, ATLAS:2022ryk, CMS:2024qjo}.  We fix $s_{\delta}=0$, such that all SM-like couplings reproduce their SM values. We also set  $g_L=g_R$, corresponding to $t_{\zeta}=1$ for simplicity. The remaining scan ranges of the free parameters are:
\begin{align}
\label{eq:freeRange}
&5 \leq v_R [\mathrm{TeV}]\leq 50 ; \; 0.02 \leq t_{\beta} \leq 50;\; 10^{-3}\leq m_{n_1} [\mathrm{eV}] \leq 0.011;\; 
\crn & 0.5 \leq m_{h^+}, M_{1,2,3}[\mathrm{TeV}]\leq 5;\;  \left|\tilde{Y}^a\right| \leq \sqrt{4\pi},\; a=1,2,3.
\end{align} 
In addition, throughout the numerical analysis, we verify that all Yukawa couplings, $Y^{\ell}$ and $Y_{R}$, as well as the Higgs self-couplings remain within the perturbative regime. The lower bound on $v_R$ is taken from Ref. \cite{BhupalDev:2014hro, Karmakar:2023ixo}.

\begin{figure}[ht]
	\centering
	\begin{tabular}{ccc}
		\includegraphics[width=5.5cm]{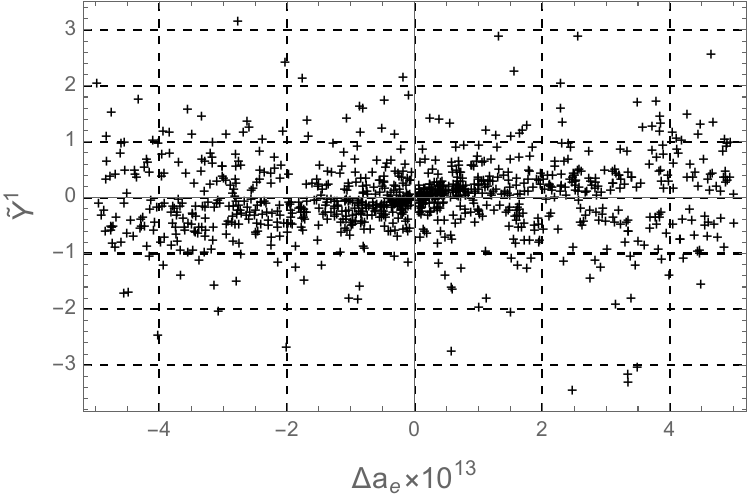} &\includegraphics[width=5.5cm]{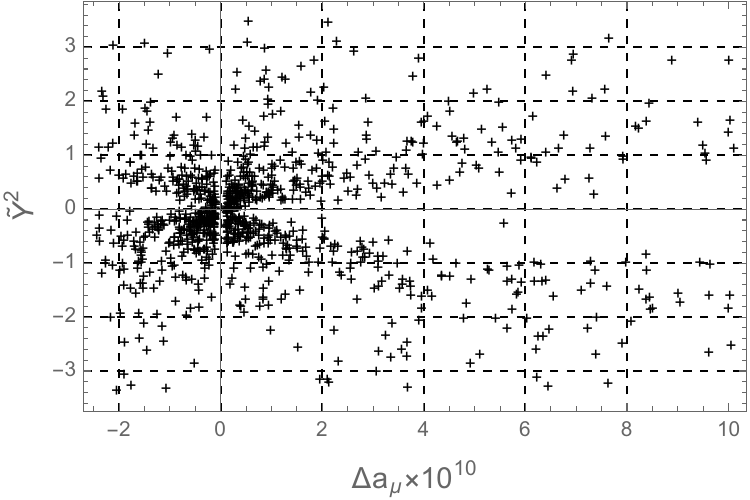}& 	\includegraphics[width=5.5cm]{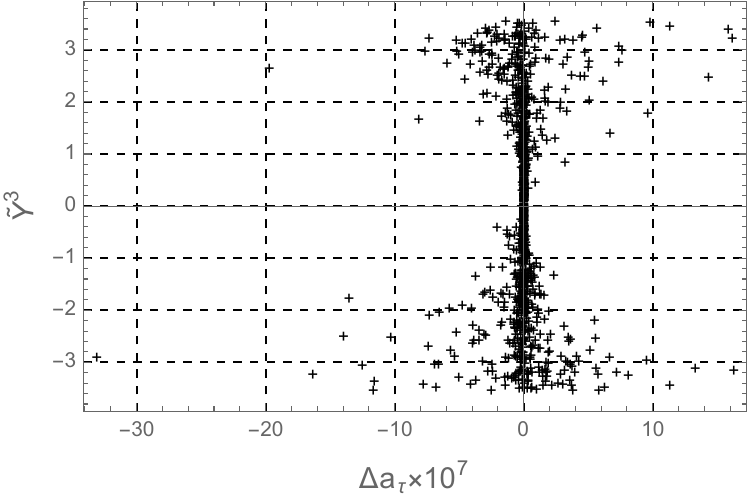} \\
	\includegraphics[width=5.5cm]{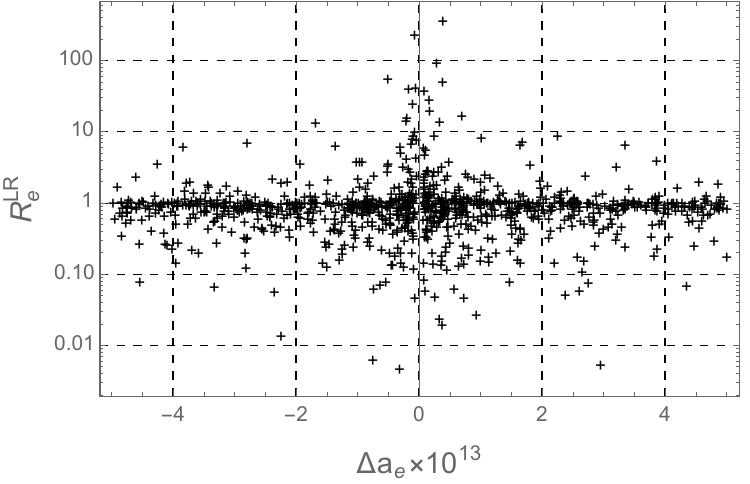} &\includegraphics[width=5.5cm]{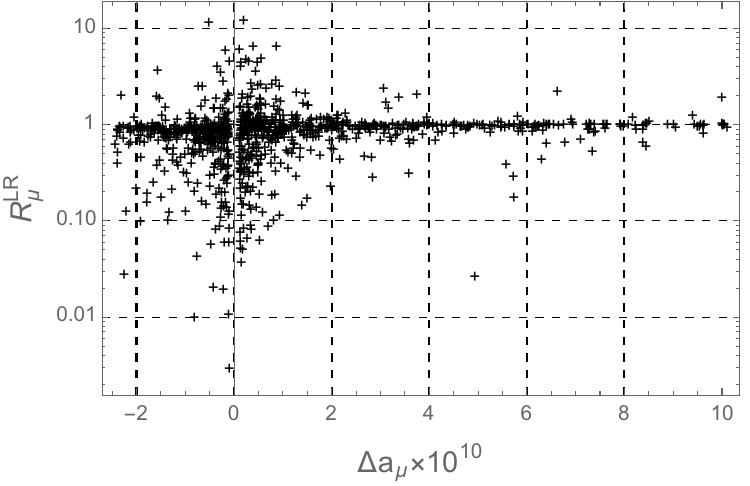}& 	\includegraphics[width=5.5cm]{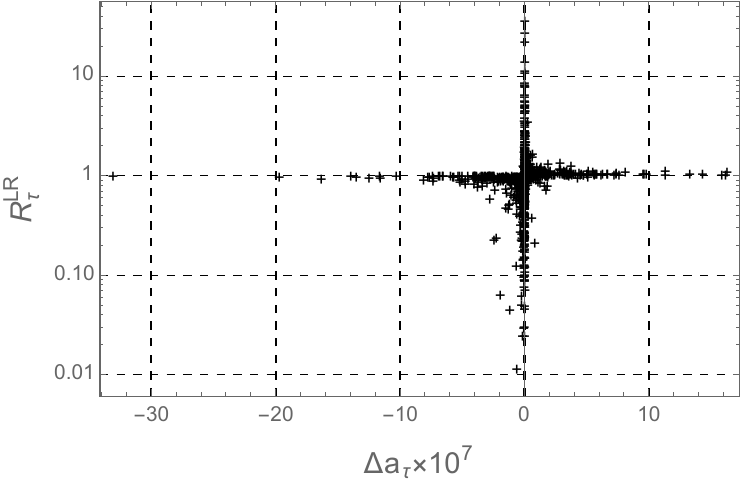} \\
	\end{tabular}
	\caption{$\tilde{Y}^a$ and $R_{e_a}^{{LR}}= \Delta a_{e_a}^{{LR}}/\Delta a_{e_a}$  as functions of  $\Delta a_{e_a}$.}\label{fig:Yell_aea}
\end{figure}

First, Fig. \ref{fig:Yell_aea} shows the strong correlation between the entries $\tilde{Y}^{a}$ of $\tilde{Y}^{\ell}$ and $\Delta a_{e_a}$, 
allowing all large values of $\tilde{Y}^a$ in the ranges given in Eq. \eqref{eq:freeRange}.  We also investigate the relative contribution of the LR term in Eq. \eqref{eq:ax} to the total $\Delta a_{e_a}$, defined as $R_{e_a}^{{LR}}= \Delta a_{e_a}^{{LR}}/\Delta a_{e_a}$ vs. $\Delta a_{e_a}$, as a function of $\Delta a_{e_a}$. The effects of, for example,  $\tilde{Y}^{1}$ on $\Delta a_{\mu,\tau}$ are found to be insignificant. Interestingly, sizable values of $|\Delta a_e|$, of order $\mathcal{O}(10^{-13})$, are allowed even for nonzero $\tilde{Y}^{1}\to0$. In contrast, large values of $|\Delta a_{\mu}|>6\times10^{-10}$ and $|\Delta a_{\tau}|>5\times10^{-7}$ require sizable values of $|\tilde{Y}^{2}|$ and $|\tilde{Y}^{3}|$, respectively. This clearly demonstrates that large values of $|\Delta a_{\mu,\tau}|$  are dominated by the LR contributions from $h^+$ exchange, with $|R_{\mu,\tau}^{\mathrm{LR}}|\to1$. This behavior requires sizable values of $|\tilde{Y}^{2,3}|$, as indicated by Eqs.\eqref{eq:dah0} and  \eqref{eq:LRpart}. In contrast, large values of $|\Delta a_e|\propto \mathcal{O}(10^{-13})$  can be obtained for $\tilde{Y}^{1}\to 0$, provided that $t_{\beta}$ is sufficiently large.

\begin{figure}[ht]
	\centering
	\begin{tabular}{ccc}
		\includegraphics[width=5.5cm]{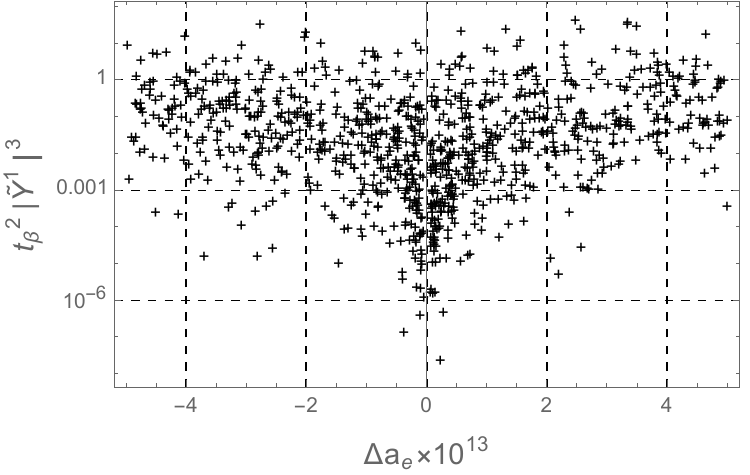} &\includegraphics[width=5.5cm]{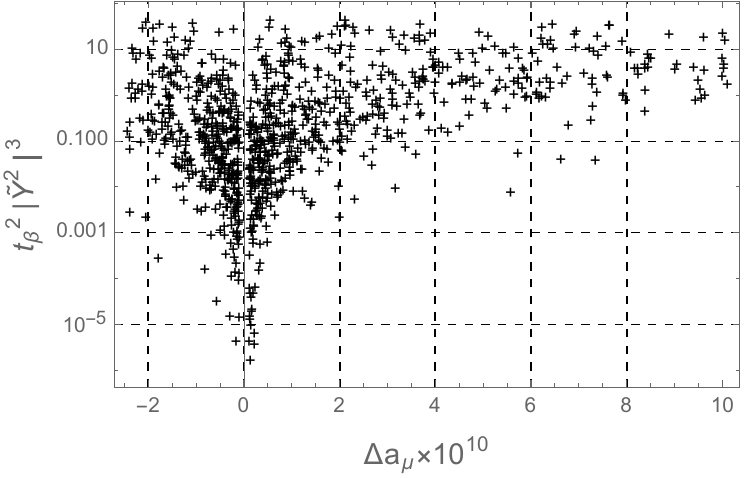}& 	\includegraphics[width=5.5cm]{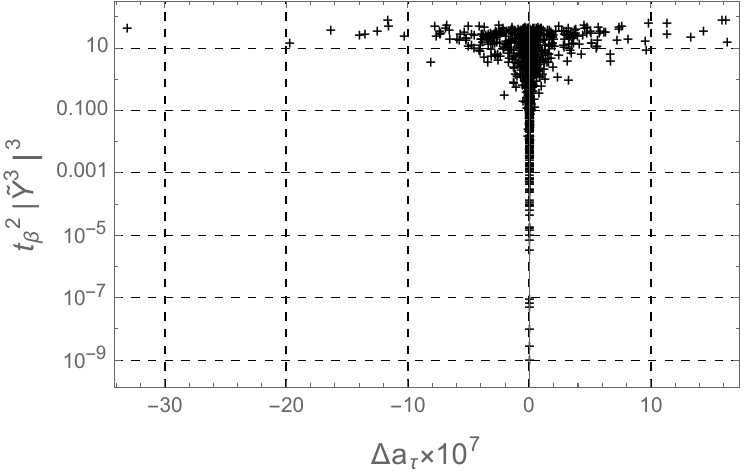} \\
			\includegraphics[width=5.5cm]{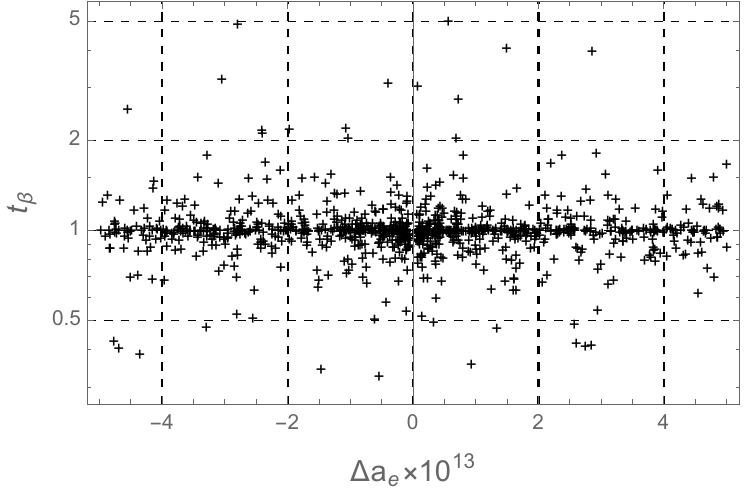} &\includegraphics[width=5.5cm]{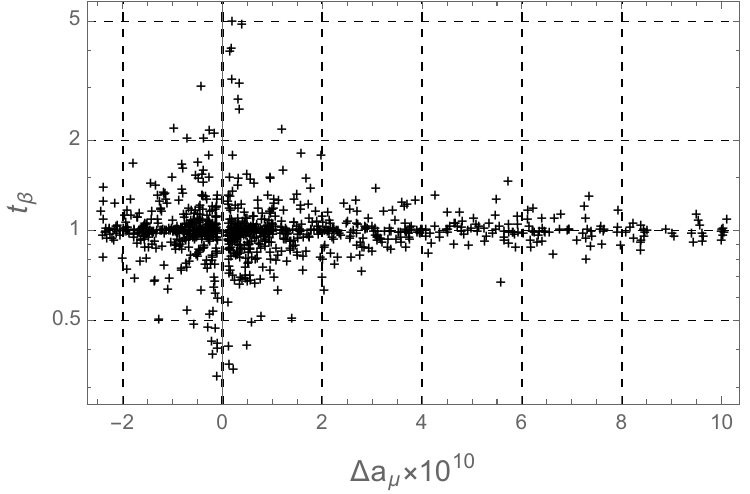}& 	\includegraphics[width=5.5cm]{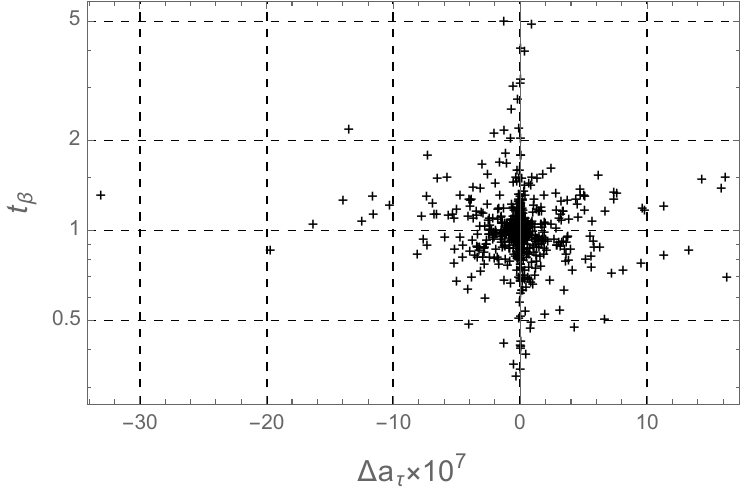}
	\end{tabular}
	\caption{ $|\tilde{Y}^a|^3 t^2_{\beta}$ and $t_{\beta}$ as functions of $\Delta a_{e_a}$.}\label{fig:tb_aea}
\end{figure}

Fig. \ref{fig:tb_aea} show the correlations  of $|\tilde{Y}^a|^3 t^2_{\beta}$, and $t_{\beta}$ with $\Delta a_{e_a}$.

 For large $|\tilde{Y}^a|$, one has $|\tilde{Y}^a|^3 t^2_{\beta}\propto a^{{LR}}_{e_a}({h^{\pm}})$, as indicated by Eq. \eqref{eq:dah0}. This further confirms the behavior observed in the second row of Fig. \ref{fig:Yell_aea}, namely, that $a^{\mathrm{LR}}{e_a}(h^\pm)$ becomes the dominant contribution for sufficiently large $|\Delta a_{e_a}|$. 
 
 The second row of Fig. \ref{fig:tb_aea} shows that large values of $|\Delta a_{e_a}|$ favor small $t_{\beta}$. This also implies that large values of $|\tilde{Y}^{a}|$ require small $t_{\beta}$, in agreement with the constraints from the non-unitarity parameter $\eta$. In particular, $\tilde{Y}^{a}t_{\beta} \propto \eta_{aa}= (m^a_D/(2M_a))^2$, which is constrained by limits on the non-unitarity of the active-neutrino mixing matrix \cite{Blennow:2023mqx, Fernandez-Martinez:2016lgt, Yu:2024nkc}. 

\begin{figure}[ht]
	\centering
	\begin{tabular}{ccc}
		\includegraphics[width=5.5cm]{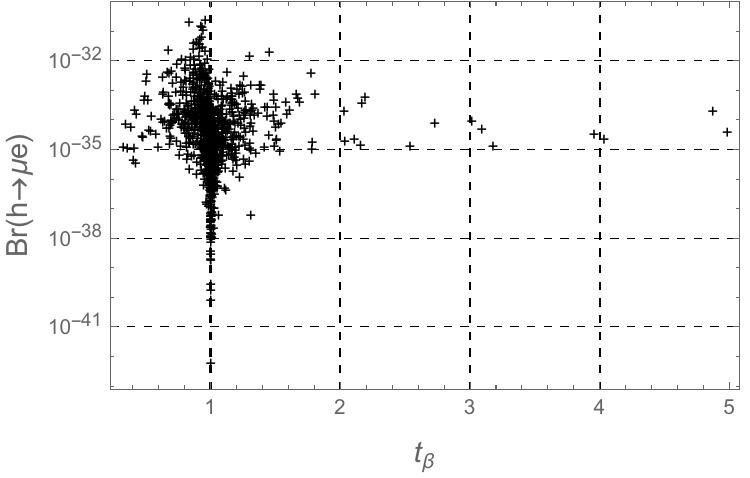} &\includegraphics[width=5.5cm]{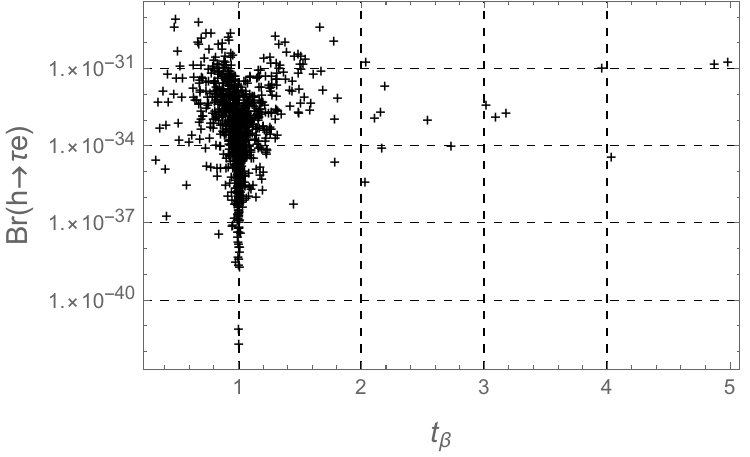}& 	\includegraphics[width=5.5cm]{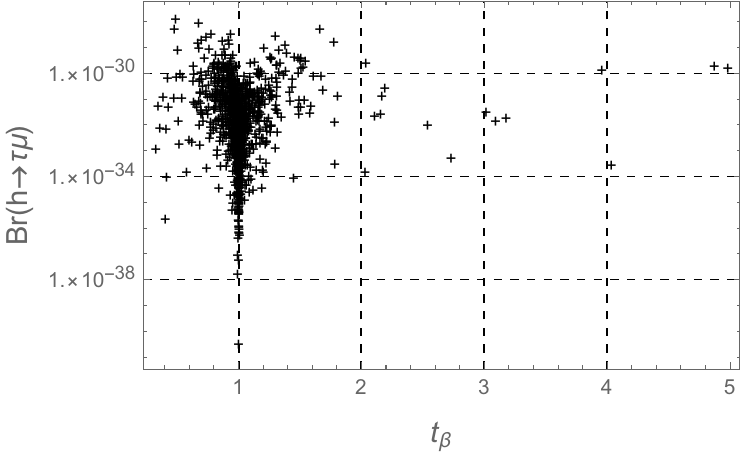} \\
	\includegraphics[width=5.5cm]{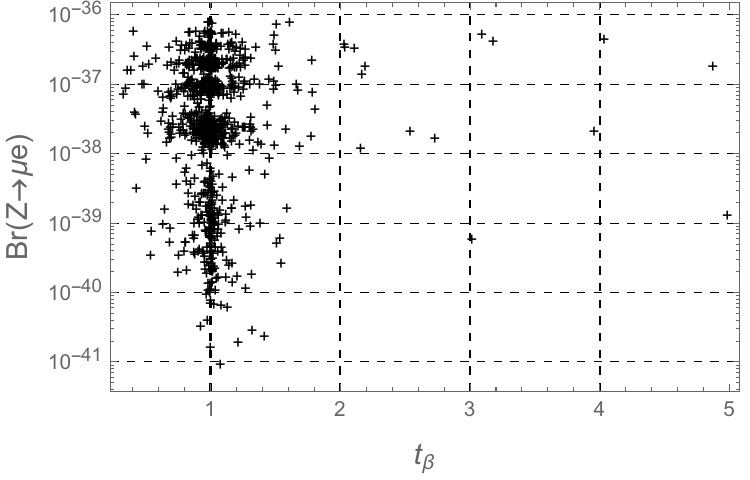} &\includegraphics[width=5.5cm]{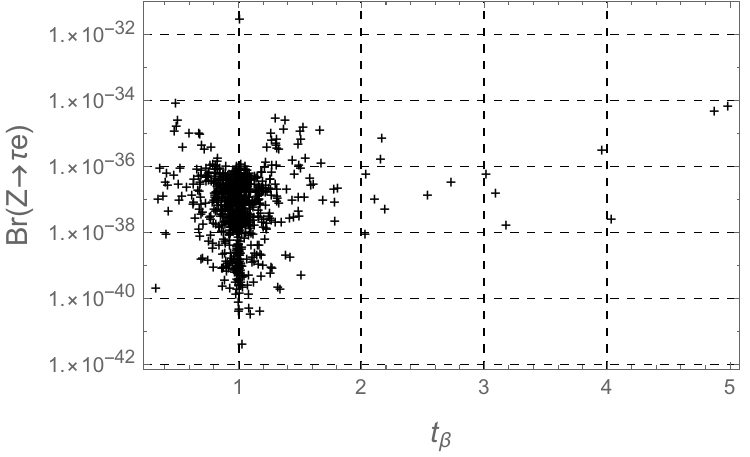}& 	\includegraphics[width=5.5cm]{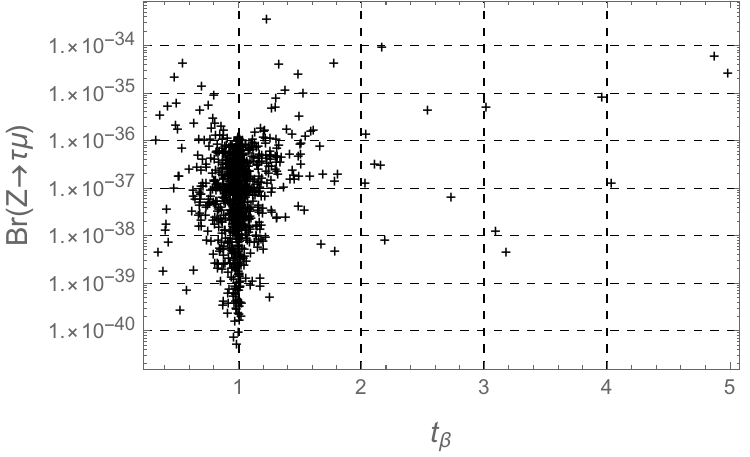} 
	\end{tabular}
	\caption{ LFV$h$ and LFV$Z$ decay rates as functions of $t_{\beta}$. }\label{fig:am_LFVhZ}
\end{figure}

To complete the discussion of all LFV decay rates, we consider the LFV$h$ and LFV$Z$ decays. The LRiss model predicts highly suppressed rates, Br$(h\to e_be_a)< \mathcal{O}(10^{-28})$ and  Br$(Z\to e_b^\pm e_a^{\mp})<\mathcal{O}(10^{-34})$, as shown in Fig. \ref{fig:am_LFVhZ}, 
where LFV$Z$ and LFV$h$ decay rates are shown as functions of $t_{\beta}$.

 We recall that the cLFV decay rates receive no one-loop contributions from $h^=$ exchange or heavy neutral-neutrino exchange; see Eq. \eqref{eq:cLL} and the detailed discussion in Appendix \ref{app:LFVB}. The corresponding decay rates are therefore nearly constant, as shown in Eq. \eqref{eq:cLFVrates}.

\begin{figure}[ht]
	\centering
	\begin{tabular}{ccc}
		\includegraphics[width=5.5cm]{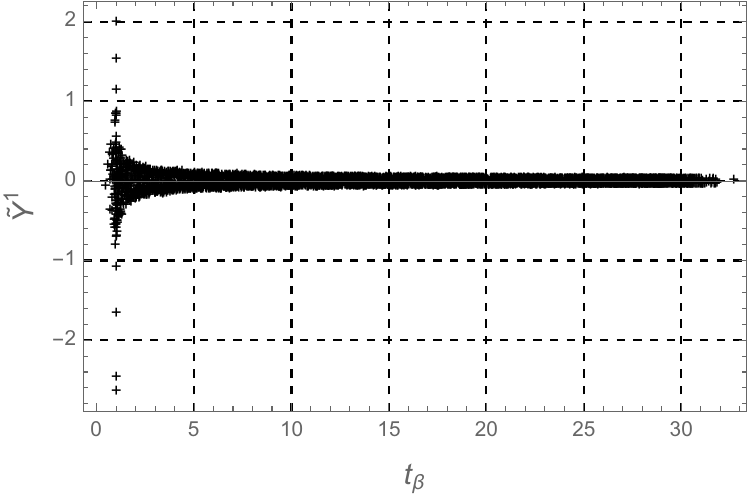} &\includegraphics[width=5.5cm]{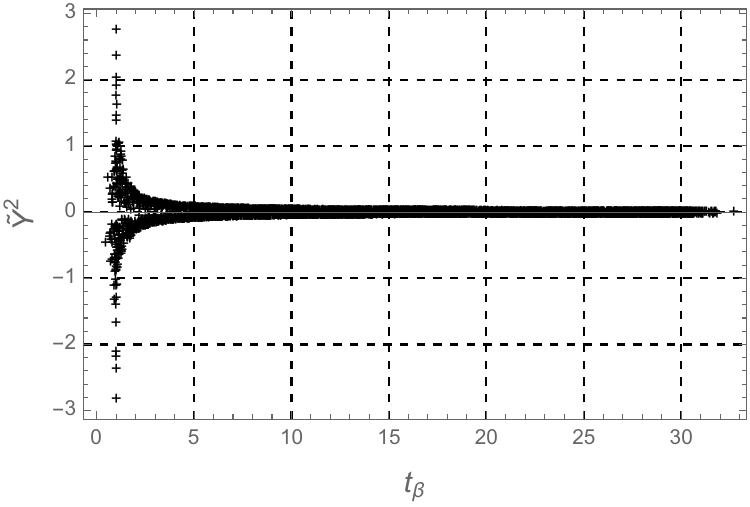}& 	\includegraphics[width=5.5cm]{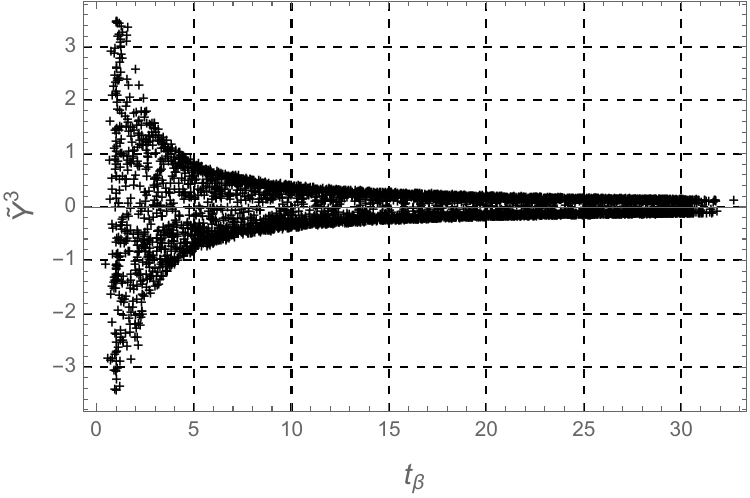} \\
	\includegraphics[width=5.5cm]{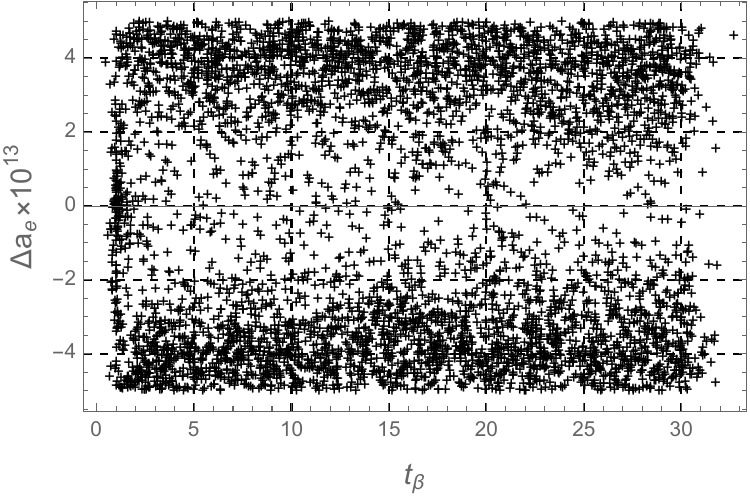} &\includegraphics[width=5.5cm]{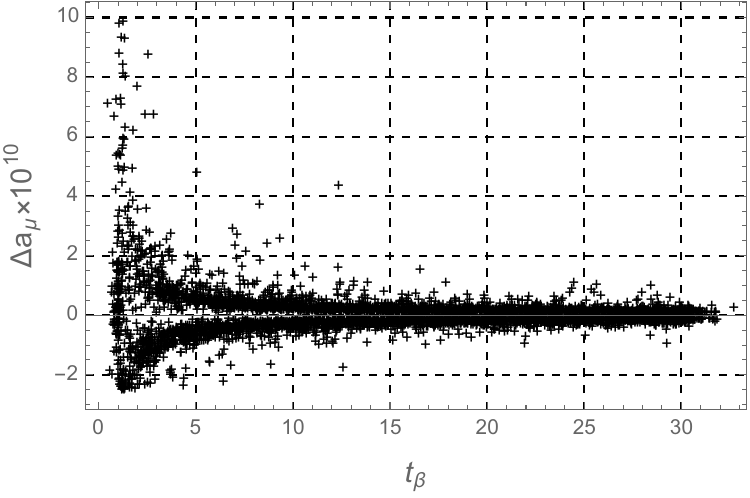}& 	\includegraphics[width=5.5cm]{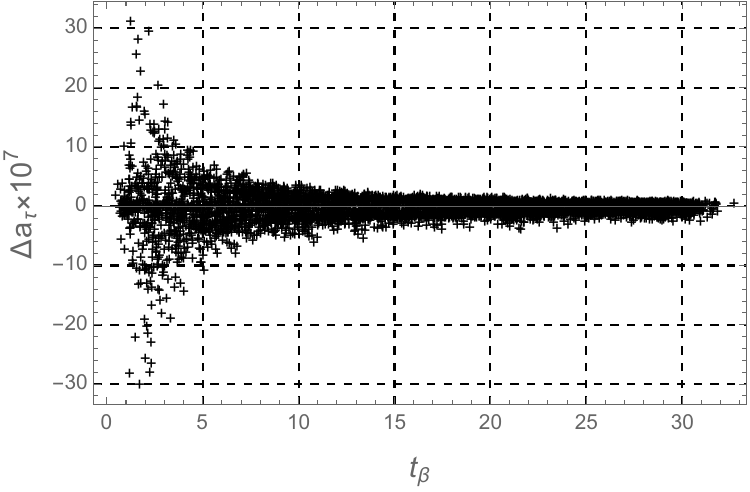} \\
	\end{tabular}
	\caption{$\Delta a_{e_a}$ and $\tilde{Y}^a$ as functions of $\eta_{22}$  in the parameter regions including   large values of  $t_{\beta}$}\label{fig:tb1_aeaYa}
\end{figure}

In the numerical analysis above, the allowed parameter points favor only small values of $t_{\beta}<5$. We now focus on the region of large $t_{\beta}$ to examine more clearly the behavior of $\Delta a_{e_a}$.  Fig. \ref{fig:tb1_aeaYa} show correlations between  $\Delta a_{e_a}$ and $\tilde{Y}^a$ with $t_{\beta}$.
The panels in the first row exhibit the same behavior: large $t_{\beta}$ requires all three entries of the matrix $\tilde{Y}^{\ell}$ to be small. On the other hand, the left panel in the second row shows that $\Delta a_e$ can reach a maximum of $\mathcal{O}(10^{-13})$ over the entire range of $t_{\beta}$, indicating that $\Delta a_e$ is not significantly affected by the current upper bound on $\eta_{11}$. The two remaining panels show a strong decrease in $\Delta a_{\mu,\tau}$ with increasing $t_{\beta}$.

\begin{figure}[ht]
	\centering
	\begin{tabular}{cc}
		\includegraphics[width=7.5cm]{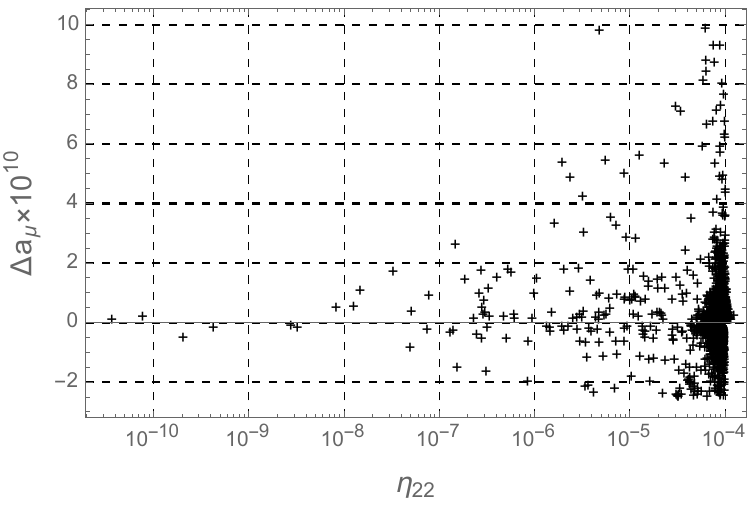} & 	\includegraphics[width=7.5cm]{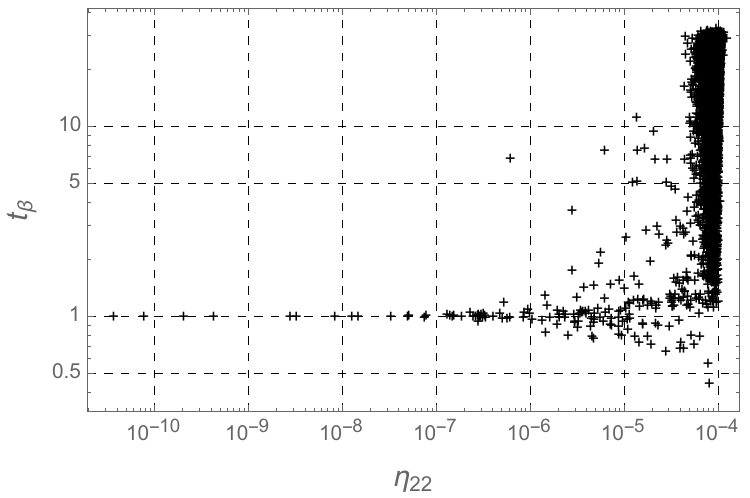} \\
		\end{tabular}
	\caption{ $\Delta a_{\mu}$ and $t_{\beta}$ as functions of $\eta_{22}$  in the parameter regions including   large values of  $t_{\beta}$.}\label{fig:eta22}
\end{figure}

In particular, large and positive values of $\Delta a_{\mu}$ require simultaneously small $t_{\beta}$ and sizable values of $\tilde{Y}^2$. Therefore, the non-unitary parameter $\eta_{22}\propto(t_{\beta}\tilde{Y}^2/M_2)^2$ has a strong impact on $\Delta a_{\mu}$; see Fig. \ref{fig:eta22} for illustration. 
The current upper bound on $\eta_{22}$ still allows $\Delta a_{\mu}$ to reach values of order $10^{-9}$, but the allowed values decrease significantly when the projected future constraint on $\eta_{22}$ is imposed.  The small allowed values of  $\eta_{22}$ also favor $t_{\beta}$ values around unity, consistent with the behavior illustrated in the second row of Fig. \ref{fig:tb_aea}.  The same conclusion applies to $\Delta a_{\tau}$, which should therefore also be investigated once experimental measurements become sufficiently sensitive to this quantity.

\section{\label{sec:conclusion} Conclusions}   

In this work, we study the one-loop contributions to $\Delta a_{e_a}$  and LFV decay amplitudes in the framework of the LRiss model, uncovering several interesting results that have not been reported previously. First, the gauged $U(1)_{L{_\mu}-L_{_\tau}}$ symmetry leads to a strict ISS  structure of the total neutrino mass matrix, with a naturally diagonal form of $m_D$, while $M_R$ can take six possible forms arising from different permutations of the $U(1)_{L_{\mu}-L_{\tau}}$ charge assignments of the three new neutral lepton singlets $S_{aL}$. Consequently, neither the LR one-loop contributions nor the heavy-neutrino exchange diagrams contribute to the LFV amplitudes at the one-loop level, resulting in LFV rates that are well below the current experimental sensitivities. In contrast, the LR contributions from one-loop diagrams involving singly charged Higgs bosons can give sizable contributions to $\Delta a_{e,\mu}$, reaching values close to the current experimental sensitivities, namely $\Delta a_{\mu}\simeq 10^{-9}$ and $\Delta a_e\sim\mathcal{O}(10^{-13})$. Therefore, in addition to the one-loop contributions from the light gauge boson $ Z_{\mu\tau}$, which can accommodate the experimental $(g-2)_\mu $ data as discussed previously, we find that the LR one-loop contributions from singly charged Higgs bosons predicted by the LRiss model can reach the same order of magnitude and should therefore not be neglected. Furthermore, the $U(1)_{L_{\mu}-L_{\tau}}$ models without singly charged Higgs bosons cannot accommodate a sizable contribution of $\Delta a_e\sim\mathcal{O}(10^{-13})$. Hence, future experimental measurements of $(g-2)_e$ could provide an important means of distinguishing between these $U(1)_{L_{\mu}-L_{\tau}}$ models.
\section*{Acknowledgments}
 This research is funded by Vietnam National Foundation
for Science and Technology Development (NAFOSTED) under the grant number 103.01-2025.04.
  
 \appendix 
 \section{\label{app:Vhtotal} Higgs potential, mass spectrum, and scalar mixing parameters}
 
The linear Eqs.  corresponding to the Higgs potential given in Eq. \eqref{eq:vHiggsR} are:
 \begin{align}
 \label{eq:Higgslinear}
0=& v_1^3 (\lambda_3+\lambda_4)+3 \lambda_6 v_1^2 v_2+\frac{1}{2} v_1 \left[  \lambda_8 u^2+ \lambda'_8 u'^2 +2 v_2^2 (\lambda_3+4 \lambda_5)+\lambda_{10} v_R^2+2 \mu_{\Phi }^2\right] 
\crn &+v_2 \left(\lambda_9 u^2+ \lambda'_9 u'^2+\lambda_6 v_2^2+\lambda_{11} v_R^2+2 \tilde{\mu }_{\Phi}^2\right),
 \crn 0=& \lambda_6 v_1^3+v_1^2 v_2 (\lambda_3+4 \lambda_5)+v_1 \left( \lambda_9 u^2+ \lambda'_9 u'^2+3 \lambda_6 v_2^2+\lambda_{11} v_R^2+2 \tilde{\mu }_{\Phi }^2\right) 
 \crn &+\frac{1}{2} v_2 \left[  \lambda_8 u^2 + \lambda'_8 u'^2 +2 v_2^2 (\lambda_3+\lambda_4)+\lambda_{10} v_R^2+\lambda_{12} v_R^2+2 \mu_{\Phi }^2\right],
 \crn 0=& 2 \mu_R^2   +2 \lambda_2 v_R^2+\lambda_7 u^2+ \lambda'_7 u'^2+\lambda_{10} \left(  v_1^2 + v_2^2\right)+ 4 \lambda_{11} v_1 v_2 +\lambda_{12} v_2^2,
  \crn 0=& 2 \mu_{\chi }^2 -4 f_{\chi \chi '} u' + 2 \lambda_1u^2 +\lambda_7 v_R^2 +\lambda_8 (v_1^2 +v_2^2) +4 \lambda_9 v_1 v_2 +\lambda _{\chi \chi '} u'^2,
  \crn 0=& 2 \mu_{\chi '}^2 - 2 f_{\chi \chi '} \frac{ u^2}{u'}+2 \lambda '_1 u'^2 +\lambda '_7 v_R^2 +\lambda '_8 (v_1^2 +v_1^2)+4 \lambda '_9 v_1 v_2 +\lambda _{\chi \chi '} u^2. 
 \end{align}
Substituting these equations from Eq. \eqref{eq:Higgslinear} into the Higgs potential given in Eq. \eqref{eq:vHiggsR}, we obtain the Higgs potential in terms of only the independent Higgs-sector parameters, while the dependent parameters are chosen to be $\mu^2_{\chi}$, $\mu^2_{\chi'}$, $\mu^2_{R}$, $\mu^2_{\Phi}$, and $\lambda_{12}$.

The squared mass matrix of singly charged  Higgs bosons in the basis $(\phi^{\pm}_1, \phi^{\pm}_2, H^{\pm}_R)$ is:
\begin{align}
\label{eq:M2c}
\mathcal{M}^2_{\pm}:
\; (11)&=-t_{\beta } \left[ v^2 (\lambda_6-c_{\beta } s_{\beta } (\lambda_4-4 \lambda_5))+\lambda_9 u^2+\lambda_{11} v_R^2+2 \tilde{\mu
}_{\Phi }^2\right],
\crn (22)&= \frac{(11)}{t^2_{\beta }}, \; (33)= (11) \times \frac{v^2 \left(c_{\beta }^2-s_{\beta }^2\right)^2}{s_{\beta }^2 v_R^2}, \; (12) = (21)= \frac{(11)}{t_{\beta }},
 \crn (13)&= (31)= (11) \times \frac{v \left(s_{\beta }^2-c_{\beta }^2\right)}{s_{\beta } v_R}, \; 
 (23)= (32)= (11) \times \frac{c_{\beta } v \left(s_{\beta }^2-c_{\beta }^2\right)}{s_{\beta }^2 v_R}.
\end{align}

Correspondingly, the  squared mass matrices of neutral  Higgs bosons in the base  $(a_1,a_2, a_3,a_4,a_5)$ (CP-odd) and $(r_1,r_2,r_3,r_4,r_5)$ (CP-even) are:
\begin{align}
\mathcal{M}^2_{a}: \; (11)&= -\frac{\lambda_9 u^2+v^2 (\lambda_6+4 \lambda_5 s_{2\beta })+\lambda_{11} v_R^2+2 \tilde{\mu }_{\Phi }^2}{t_{\beta }}, 
\crn (22)&= -t_{\beta } \left(\lambda_9 u^2+v^2 (\lambda_6+4 \lambda_5 s_{2\beta })+\lambda_{11} v_R^2+2 \tilde{\mu }_{\Phi }^2\right) =(11) \times t_{\beta}^2, 
\crn (12)&=(21)= -\lambda_9 u^2-v^2 (\lambda_6+4 \lambda_5 s_{2\beta })-\lambda_{11} v_R^2-2 \tilde{\mu }_{\Phi }^2=(11) \times t_{\beta}, 
\crn (44)&= 4 f_{\chi \chi '} u';\; (55)= \frac{f_{\chi \chi '} u^2}{u'};\; (45) =(54)=-2 f_{\chi \chi '} u,
\crn (ij)&=0\; (i>2,j>2,(ij)\neq (44),(55),(45),(54)),
\label{eq:M2a}
\\ \mathcal{M}^2_{r}: \; (11)  &= -\frac{\lambda_{11} v_R^2}{t_{\beta }} -\frac{\lambda_9 u^2}{t_{\beta }} -\frac{2 \tilde{\mu }_{\Phi }^2}{t_{\beta }} + \frac{c_{\beta }^2 v^2 \left(2 t_{\beta }^3 (\lambda_3+\lambda_4)+\lambda_6 \left(3 t_{\beta }^2-1\right)\right)}{t_{\beta }}, 
\crn    (22)&=- \lambda_{11} t_{\beta } v_R^2 - \lambda_9 t_{\beta } u^2 -2 t_{\beta } \tilde{\mu }_{\Phi }^2 + c_{\beta }^2 v^2 \left(2 \lambda_3+2 \lambda_4-\lambda_6 t_{\beta } \left(t_{\beta }^2-3\right)\right), 
\crn (33)&= 2 \lambda_2 v_R^2, \; 
(44)=2 \lambda_1 u^2,
\crn (12)&=(21)= \lambda_9 u^2+v^2 (3 \lambda_6+s_{2\beta } (\lambda_3+4 \lambda_5))+\lambda_{11} v_R^2+2 \tilde{\mu }_{\Phi }^2, 
\crn (13)&=(31)= v_R v (2 c_{\beta } \lambda_{11}+\lambda_{10} s_{\beta }),
\;  (14) =(41)= u v (2 c_{\beta } \lambda_9+\lambda_8 s_{\beta }),
\crn (23)&=(32)= \frac{ v}{s_{\beta } v_R} \left\{ \frac{1}{2} v_R^2 \left(4 c_{\beta }^2 \lambda_{11}+\lambda_{10} s_{2\beta }\right) -2 c_{\beta }^2 \lambda_9 \left(t_{\beta }^2-1\right) u^2 -4 c_{\beta }^2 \left(t_{\beta }^2-1\right) \tilde{\mu }_{\Phi }^2  
\right. \crn & \left. \frac{ }{}\hspace{2.7cm} -2 c_{\beta }^4 \left(t_{\beta }^2-1\right) v^2 \left(\lambda_6+\lambda_6 t_{\beta }^2-\lambda_4 t_{\beta }+4 \lambda_5 t_{\beta
}\right) \right\} ,
\crn (24)&=(42)= u v (c_{\beta } \lambda_8+2 \lambda_9 s_{\beta }),
 \; (34)=(43)= \lambda_7 v_R u.\label{eq:M2r}  
\end{align}
The neutral CP-odd sector contains two neutral CP-odd Higgs $A^0_1$ and $A^0_2$ and three massless states $G^0_{1}$, $G^0_{2} $, and $G^0_3 \equiv a_3$. Their linear combinations give rise to three Goldstone bosons absorbed by $Z,Z'$, and $Z_{\mu \tau}$, respectively. The transformation between the flavor and physical bases is given by
\begin{align}
\label{eq:AG03}
\begin{pmatrix}
a_1\\
a_2
\end{pmatrix}&= \left(
\begin{array}{cc}
c_{\beta } & -s_{\beta } \\
s_{\beta } & c_{\beta } \\
\end{array}
\right) \begin{pmatrix}
A^0_1\\
G^0_1
\end{pmatrix}=\begin{pmatrix}
c_{\beta}A^0_1 -s_{\beta} G^0_1\\
s_{\beta}A^0_1 +c_{\beta} G^0_1
\end{pmatrix}, 
\crn \begin{pmatrix}
a_4\\
a_5
\end{pmatrix}&= \left(
\begin{array}{cc}
c_{u } & s_{u } \\
-s_{u } & c_{u } \\
\end{array}
\right) \begin{pmatrix}
A^0_2\\
G^0_2
\end{pmatrix}=\begin{pmatrix}
c_{u}A^0_2 +s_{u} G^0_2\\
-s_{u}A^0_2 +c_{u} G^0_2
\end{pmatrix}, 
\crn m_{A_1}^2&= -\frac{v^2 (8 c_{\beta } \lambda_5 s_{\beta }+\lambda_6) +\lambda_9 u^2 +\lambda'_9 u'^2+\lambda_{11} v_R^2+2 \tilde{\mu }_{\Phi }^2}{c_{\beta } s_{\beta }},
\crn m_{A_2}^2&= \frac{f_{\chi \chi '} \left(4 u'^2+u^2\right)}{u'},
\end{align}
where $u/(2u')=t_u\equiv s_u/c_u$ satisfying $s_u^2+c_u^2=1$. 

The singly charged Higgs sector contains one physical state $h^\pm$ and two massless states corresponding to two Goldstone bosons $G^\pm_W$ and  $G^\pm_{W'}$ associated with the $W^\pm$ and $W'^{\pm}$ bosons, respectively. Defining $t_{\kappa}\equiv \frac{vc_{2\beta}}{v_R} \ll1$ , the corresponding transformation is given by:
\begin{align}
\label{eq:hpmi}
\begin{pmatrix}
\phi^\pm_1\\
\phi^\pm_2\\
H^\pm_{R}
\end{pmatrix}= & \left(
\begin{array}{ccc}
s_{\beta } s_{\kappa } & c_{\beta } & c_{\kappa } s_{\beta } \\
c_{\beta } s_{\kappa } & -s_{\beta } & c_{\beta } c_{\kappa } \\
c_{\kappa } & 0 & -s_{\kappa } \\
\end{array}
\right) \begin{pmatrix}
G^\pm_1\\
G^\pm_2\\
h^\pm
\end{pmatrix},
\\ m^2_{h^+}=&-\frac{\left(c_{2\beta }^2 v^2+v_R^2\right) \left[v^2 (2 \lambda_6 +(4 \lambda_5 -\lambda_4) s_{2\beta })+2 \left(\lambda_9 u^2+\lambda_{11} v_R^2+2 \tilde{\mu }_{\Phi
	}^2\right)\right]}{2 c_{\beta } s_{\beta } v_R^2}.\nn  
\end{align}
Finally, the neutral CP-even Higgs sector consists of four physical states, including at least one light Higgs boson that is identified with the SM-like Higgs boson consistent with the  LHC measurements. In particular, in the limit $v=0$, the squared mass matrixtakes a block-diagonal form, consisting of two $2\times2$ submatrices $M^2_r=\mathrm{diag}\left( M^2_{r,1}, \;M^2_{r,2}\right)$, where 
\begin{align}
\label{eq:Mr212}
M^2_{r,1} = -\frac{\lambda_9 u^2+\lambda_{11} v_R^2+2 \tilde{\mu }_{\Phi }^2}{t_{\beta }} \times \left(
\begin{array}{cc}
1 & -t_{\beta } \\
-t_{\beta } & t_{\beta }^2 \\
\end{array}
\right),\; M^2_{r,2}= \left(
\begin{array}{cc}
2 v_R^2 \lambda_2 & v_R \lambda_7 u \\
v_R \lambda_7 u & 2 \lambda_1 u^2 \\
\end{array}
\right).
\end{align}
The first matrix yields a zero value $m^2_{h_1}=0$, which can be identified with the SM-like Higgs boson in the limit $v=0$, while the other eigenvalue is  $m^2_{h_2}=-\frac{\lambda_9 u^2+\lambda_{11} v_R^2+2 \tilde{\mu }_{\Phi }^2}{s_{\beta }c_{\beta}}$. For simplicity, and to derive the mass and physical state of the SM-like Higgs boson explicitly, we assume a block-diagonal form of $M^2_r$, with the two submatrices given in Eq. \eqref{eq:Mr212}. Requiring $(M^2_{r})_{13}$$=(M^2_{r})_{31}=(M^2_{r})_{14}=(M^2_{r})_{41}=$  $(M^2_{r})_{23}=(M^2_{r})_{32}=(M^2_{r})_{24}=(M^2_{r})_{42}=0$,  leads to the following conditions on the couplings::
\begin{align}
\label{eq:Higgscondition}
\lambda_8=\lambda_9 =\lambda'_8=\lambda'_9 =0,\; \lambda_{10}=-\frac{2 c_{\beta } \lambda_{11}}{s_{\beta }},\;  \tilde{\mu}_{\Phi}= \frac{v^2}{2} \left[ c_{\beta } s_{\beta } (\lambda_4-4 \lambda_5)-\lambda_6 \right]. 
\end{align}
The SM-like Higgs boson is a linear combination of the two states $(r_1,r_2)$ corresponding to the following $2\times 2$ squared mass matrix:
\begin{align}
\label{eq:mr12}
M^2_{r,1}: &(11)= -\frac{c_{\beta } \lambda_{11} v_R^2}{s_{\beta }} + v^2 \left(c_{\beta }^2 (-(\lambda_4-4 \lambda_5))+4 c_{\beta } \lambda_6 s_{\beta }+2 s_{\beta }^2 (\lambda_3+\lambda_4)\right) 
\crn & (22) = -\frac{\lambda_{11} s_{\beta } v_R^2}{c_{\beta }}+  v^2 \left(2 c_{\beta }^2 (\lambda_3+\lambda_4)+4 c_{\beta } \lambda_6 s_{\beta }-s_{\beta }^2 (\lambda_4-4 \lambda_5)\right), 
\crn & (12) = (21)=  \lambda_{11} v_R^2 +v^2 \left(2 \lambda_6 +c_{\beta } s_{\beta } (2 \lambda_3+\lambda_4+4 \lambda_5)\right).
\end{align}
This matrix satisfies the following transformation:
\begin{align}
\label{eq:C01}
&C^0_{1} M^2_{r,1} C^{0T}_{1} =M^2, \; C^0_{1}= \left(
\begin{array}{cc}
s_{\beta } & c_{\beta } \\
c_{\beta } & -s_{\beta } \\
\end{array}
\right),
\\ & M^2_{11}=v^2 \left(\left(c_{2\beta }^2+1\right) \lambda_4 +4 \left(1 - c_{2\beta }^2\right) \lambda_5+8 c_{\beta }  s_{\beta }\lambda_6 +2 \lambda_3\right) \varpropto \mathcal{O}(v^2),
\crn &M^2_{22}= -c_{2\beta }^2 v^2 (\lambda_4-4 \lambda_5)-\frac{\lambda_{11} v_R^2}{c_{\beta } s_{\beta }} \varpropto \mathcal{O}(v^2_R),
\crn& M^2_{12}= M^2_{21}= -2 c_{2\beta } v^2 (c_{\beta } s_{\beta } (\lambda_4-4 \lambda_5) -\lambda_6) \varpropto \mathcal{O}(v^2),
\crn & C_{\delta} =\left(
\begin{array}{cc}
c_{\delta } & s_{\delta } \\
-s_{\delta } & c_{\delta } \\
\end{array}
\right),\; C_{\delta} M^2C^T_{\delta}=\mathrm{diag}\left( m^2_{h},\; m^2_{h_2}\right),
\crn t_{2\delta}  &= -\frac{2 c_{2\beta } s_{2\beta } v^2 (s_{2\beta } (\lambda_4-4 \lambda_5)-2 \lambda_6)}{s_{2\beta } v^2 \left(2 c_{2\beta }^2 \lambda_4-8 c_{2\beta }^2 \lambda_5+2 \lambda_3+\lambda
	_4+4 \lambda_5\right)+4 \lambda_6 s_{2\beta }^2 v^2+2 \lambda_{11} v_R^2} 
%
\varpropto \mathcal{O}\left(\frac{v^2}{v_R^2}\right) \ll1. \nn
\end{align}

 In the limit given in Eq. \eqref{eq:Higgscondition}, the masses and mixing parameters of all Higgs bosons are:
 \begin{align}
 \label{eq:Hsimple}
 m_A^2&= -\frac{c_{\beta } s_{\beta } v^2 (\lambda_4+4 \lambda_5)+\lambda_{11} v_R^2}{c_{\beta } s_{\beta }},
\;  m_{h^+}^2= -\frac{\lambda_{11} \left( v_R^2 +  c_{2\beta }^2   v^2\right)}{c_{\beta } s_{\beta }}, 
\crn  m_{h}^2&=   M^2_{11} + M^2_{12}t_{\delta},  \;  m_{h_2}^2=  M^2_{22} -M^2_{12}t_{\delta}, 
 \end{align} 
and $h_1\equiv h$ with
\begin{align}
\label{eq:rmix}
\begin{pmatrix}
r_1\\
r_2
\end{pmatrix}=&  \begin{pmatrix}
s_{\alpha}& c_{\alpha}\\
c_{\alpha}&-s_{\alpha}
\end{pmatrix} \begin{pmatrix}
h\\
h_2
\end{pmatrix},\; \alpha =\beta +\delta,\;  |\delta| \ll1. 
\end{align}
It is worth noting that the trace of the squared-mass matrix of the neutral CP-even Higgs bosons is invariant under unitary transformations, namely,
\begin{align}
m^2_h+m^2_{h_2}=M^2_{11}+M^2_{22}= v^2 \left( \lambda_4 +4 \lambda_5 +8 c_{\beta }  s_{\beta }\lambda_6 +2 \lambda_3\right) -\frac{\lambda_{11} v_R^2}{c_{\beta } s_{\beta }}. 
\end{align}
   In addition, since $M^2_{12}t_{\delta}\varpropto v^2\times \mathcal{O}(v^2/v_R^2)\ll M_{11}^2 \varpropto \mathcal{O}(v^2)$, we have  $m_h^2\simeq M_{11}^2$, which is consistent with identifying $h$ as the SM-like Higgs boson. The three remaining neutral CP-even Higgs bosons are heavy and are irrelevant to our analysis; we therefore omit them here. It is also worth noting that two of these states have masses proportional to $u^2+u'^2$, implying that at least one of the corresponding vacuum expectation values must be sufficiently large.
   
 \section{\label{app:LFVB} Analytic formulas for decay rates LFV$Z$ and LFV$h$ in the LRiss}
 
 The analytic formulas for the one-loop contributions considered here are expressed in terms of Passarino–Veltman (PV) functions, using the results of Ref. \cite{Hue:2024rij}, which are consistent with previous works \cite{Jurciukonis:2021izn, Hong:2023rhg} and with the LoopTools package \cite{Hahn:1998yk}. In particular, for each Feynman diagram containing one lepton propagator in the loop, the product of two LFV couplings is denoted by $g^{XY}_{iBB'} \equiv  g^{X*}_{aiB} g^{Y}_{biB'}$, where $X,Y=L,R$ and $B,B'=W,W',h^\pm$ denote the charged gauge and Higgs bosons. The corresponding arguments of the one-loop three-point PV functions are  $C_{x}=C_x(m_a^2,q^2,m_{e_b}^2; m_{n_i}^2,m_B^2,m_{B'}^2)$ with $x=0,i,00, ij$ ($i,j=1,2$), and $q^2=m_h^2,m_Z^2$ for LFV$h$ and LFV$Z$ decays, respectively.  The one-loop two point PV functions are  $B^{(1)}_{0,1}=B_{0,1}(m_a^2;m_{n_i}^2,m_B^2)$, $B^{(2)}_{0,1}=B_{0,1}(m_{e_b}^2;m_{n_i}^2,m_{B'}^2)$, and $B^{(12)}_{0,1}=B_{0,1}(q^2;m_B^2,m_{B'}^2)$.   For Feynman diagrams containing two lepton propagators in the loop, we use the LFV coupling product $g^{XY}_{Bij} \equiv  g^{X*}_{aiB} g^{Y}_{bjB}$.  The corresponding PV-functions  are  $C_{x}=C_x(m_a^2,q^2,m_{e_b}^2; m_B^2, m_{n_i}^2,m_{n_j}^2)$,  $B^{(1)}_{0,1}=B_{0,1}(m_a^2;m_B^2, m_{n_i}^2)$, $B^{(2)}_{0,1}=B_{0,1}(m_{e_b}^2;m_{B}^2, m_{n_j}^2)$, and $B^{(12)}_{0,1}=B_{0,1}(q^2;m_{n_i}^2,m_{n_j}^2)$.  In the following, for the specific formulas of LFV$h$ or LFV$Z$ decays, where $q^2$ is fixed,  we suppress the first three arguments $(m_a^2,q^2,m_{e_b}^2)$ for the two respective notations defined above. We also use the notation for combinations of PV functions with the same arguments,
 $X_0\equiv C_0 +C_1 +C_2$, $X_1\equiv C_{11} +C_{12} +C_1$, $X_2\equiv C_{22} +C_{12} +C_2$, $X_{(i\pm j\pm\dots)} \equiv X_i \pm X_j \pm \dots$with $i,j,\dots=0,1,2$, and  $C_{(i \pm j)}=C_{i} \pm C_j$ with $i,j,\dots=0,1,2,11,12,22$. To verify the finiteness  of the total LFV amplitude, we  list here the nonzero divergent parts of the relevant PV functions:
  \begin{align}
  \label{eq:divPV}
 \mathrm{div}[C_{00}]=& 
 \frac{C_{UV} }{4},\; \mathrm{div}[B^{(1)}_0]=\mathrm{div}[B^{(2)}_0]=\mathrm{div}[B^{(12)}_0]=C_{UV},
 \crn  \mathrm{div}[B^{(1)}_1]=&\mathrm{div}[B^{(2)}_1]=\mathrm{div}[B^{(12)}_1] =- \frac{C_{UV}}{2},
  \end{align}
where $C_{UV}=(1/\varepsilon) -\gamma_{E}+\ln (4\pi \mu^2) +\mathcal{O}(\varepsilon)$ where $\varepsilon =(4-d)/2\to 0$ when $d\to 4$.  
  
The formula of the partial decay width for LFV$Z$ decays is \cite{Jurciukonis:2021izn}
\begin{align} 
\label{eq_GAZeba}
\Gamma (Z\to e^+_b e^-_a)= 	\frac{\sqrt{\lambda}}{16\pi m_Z^3}\times \left(\frac{e}{16\pi^2}\right)^2 \left( \frac{\lambda N_0}{12 m^2_Z} +N_1 +\frac{N_2}{3 m^2_Z}\right),
\end{align}
where $q^2=m_Z^2$, $\lambda= m^4_Z +m^4_{b} +m^4_{a} -2(m^2_Zm^2_{a} +m^2_Zm^2_{b} +m^2_{a}m^2_{b})$ and 
\begin{align}
\label{eq_Ni}
N_0= & (m^2_Z -m_{a}^2 -m_{b}^2)\left(|\tilde{y}_L|^2 +|\tilde{y}_R|^2\right)  -4 m_{a} m_{b} \mathrm{Re}\left[ \tilde{y}_L  \tilde{y}^*_R\right]
\crn&
- 4m_{b} \mathrm{Re}\left[ \tilde{x}^*_R \tilde{y}_L   + \tilde{x}^*_L \tilde{y}_R  \right] -  4m_{a}\mathrm{Re}\left[ \tilde{x}^*_L \tilde{y}_L   + \tilde{x}^*_R  \tilde{y}_R  \right] , 
\crn N_1 = & 4 m_{a}m_{b} \mathrm{Re}\left[\tilde{x}_L\tilde{x}_R^* \right],
\crn  N_2 = &  \left[ 2 m^4_Z - m_Z^2\left( m_{a}^2 + m_{b}^2\right) - \left( m_{a}^2 - m_{b}^2\right)^2  \right] \left( |\tilde{x}_L|^2 +|\tilde{x}_R|^2\right),
\end{align}
where 
\begin{align}
\label{eq_abZeba}
\tilde{x}_{L(R)}&= \sum_{m=1}^{10}\tilde{x}^{(m)}_{L(R)},   
\;   \tilde{y}_{L(R)} =\sum_{m=1}^{6}\tilde{y}^{(m)}_{L(R)}.
\end{align}
Here  $\tilde{x}^{(m)}_{L(R)}$ denote  the one-loop contributions from  diagam $(m)$ in Fig. \ref{fig:ZebaU}. We omit the LFV index $(ab)$ for simplicity, for example $\tilde{x}^{ X}_{L(R)}\equiv \tilde{x}^{(ab) X}_{L(R)}$ for all $X= iVV',\dots$. The explicitly analytic formulas of all $\tilde{x}^{ (m)}_{L(R)}$ and $\tilde{y}^{ (m)}_{L(R)}$ in terms of the PV-functions are derived based on previous results \cite{Hue:2024rij}. Namely, diagram (1) in Fig. \ref{fig:ZebaU} gives the following one-loop form factors:   
\begin{align}
\label{eq:xyLR1}
&\tilde{x}^{iVV'}_L =  g_{ZVV'} \left\{ g^{LL}\left[\dots \right] +g^{RR} m_am_{e_b}\left[\dots\right] - g^{RL} m_a m_{n_i} \left[\dots\right]	 - g^{LR}m_{e_b} m_{n_i} \left[ \dots\right]	
\right\},  
\crn & \tilde{y}^{i VV'}_L  = g_{ZVV'} \left\{ g^{LL} m_a\left[\dots \right] + g^{RR}m_{e_b} \left[\dots \right]	  - g^{RL}  m_{n_i} \left[\dots\right]	 - g^{LR} \left[ \dots\right]
\right\},
\crn &\tilde{x}_R^{iVV'} \left( \tilde{y}_R^{iVV'}\right)=   \tilde{x}_L^{iVV'}\left( \tilde{y}_L^{iVV'}\right) \left[ g^{LL} \leftrightarrow  g^{RR},  g^{RL} \leftrightarrow  g^{LR} \right].
\end{align} 
Here,   $g_{ZVV'}$ is given in Eq. \eqref{eq:gZVV} and $g^{XY}\equiv g^{XY}_{iVV'}= g^{X*}_{aiV}g^{Y}_{biV'}$ with $X,Y=L,R$. The notation $[\dots]$  represents lengthy formulas that can be readily obtained from Ref. \cite{Hue:2024rij}. The formulas of $g^{XY}$ are   derived from couplings given in  Eq. \eqref{eq:gX} as follows: 
\begin{align}
\label{eq:gXYiVVp}
g^{LL}_{iWW}=& \frac{g^2 c^2_{\theta}}{2} U^{\nu}_{ai} U^{\nu*}_{bi},\; g^{RR}_{iWW}= \frac{g^2s^2_{\theta} t^2_W }{2 s^2_{\zeta}} U^{\nu*}_{(a+3)i} U^{\nu}_{(b+3)i},\;  
\crn g^{LR}_{iWW}=&  -g^{LR}_{iW'W'}= \frac{g^2 s_{\theta} c_{\theta} t_W}{2 s_{\zeta}} U^{\nu}_{ai} U^{\nu}_{(b+3)i},\;  g^{RL}_{iWW}=-g^{RL}_{iW'W'}= \frac{g^2 s_{\theta} c_{\theta} t_W}{2 s_{\zeta}} U^{\nu*}_{(a+3)i} U^{\nu*}_{b i},
\crn g^{LL}_{iW'W'}=& \frac{g^2 s^2_{\theta}}{2} U^{\nu}_{ai} U^{\nu*}_{bi},\; g^{RR}_{iW'W'}= \frac{g^2 c^2_{\theta} t^2_W }{2 s^2_{\zeta}} U^{\nu*}_{(a+3)i} U^{\nu}_{(b+3)i},\;  
\crn g^{LL}_{iWW'}=& g^{LL}_{iW' W}=-\frac{g^2 s_{\theta} c_{\theta} }{2 } U^{\nu}_{ai} U^{\nu*}_{bi},\;  g^{RR}_{iWW'}= g^{RR}_{iW' W}= \frac{g^2 s_{\theta} c_{\theta} t_W^2}{2 s^2_{\zeta}} U^{\nu*}_{(a+3)i} U^{\nu}_{(b+3) i},
\crn g^{LR}_{iWW'}=& \frac{g^2 c^2_{\theta} t_W}{2 s_{\zeta}} U^{\nu}_{ai} U^{\nu}_{(b+3)i},\; g^{RL}_{iWW'}= -\frac{g^2 s^2_{\theta} t_W }{2 s^2_{\zeta}} U^{\nu*}_{(a+3)i} U^{\nu*}_{bi},
\crn g^{LR}_{iW'W}=& -\frac{g^2 s^2_{\theta} t_W}{2 s_{\zeta}} U^{\nu}_{ai} U^{\nu}_{(b+3)i},\; g^{RL}_{iW' W}= \frac{g^2 c^2_{\theta} t_W }{2 s^2_{\zeta}} U^{\nu*}_{(a+3)i} U^{\nu*}_{bi}.  
\end{align}
Since $\sum_{i=1}^9 g^{LL}= \sum_{i=1}^9 g^{RR} \varpropto \delta_{ab}, \delta_{(a+3)(b+3)}=0$ with $a\neq b$, the PV functions appearing in the original formulas in Ref. \cite{Hue:2024rij}, namely  $B^{(12)}_0$, $A_0(m_V)$, $A_0(m_{V'})$, and divergent part of $C_{00}$,  vanish in the model under consideration. In addition, $\sum_{i=1}^9 g^{LR}m_{n_i}\varpropto \mathcal{M}^{\nu}_{a(b+3)}=(m_D)^T_{ab} =0, \;  \sum_{i=1}^9 g^{RL}m_{n_i}=\mathcal{M}^{\nu}_{(a+3)b}=(m_D)^{\dagger}_{ab}=0$. Therefore these terms can be omitted, and we set $d=4$ in all expressions in Eq. \eqref{eq:xyLR1}.

 The general form of $U^{\nu}$ given in Eq. \eqref{eq:Usimeq1} leads to the following important relations for $i>3$:
 \begin{align}
 \label{eq:Unu2}
 U^{\nu}_{ai} U^{\nu*}_{bi}, U^{\nu*}_{(a+3)i} U^{\nu}_{(b+3)i}\propto&  V^{R}_{ac}V^{R*}_{bc} ;\;  U^{\nu}_{ai} U^{\nu}_{(b+3)i}, U^{\nu*}_{(a+3)i} U^{\nu}_{bi}\propto  V^{R*}_{ac}V^{R}_{bc},
 \end{align}
 where  $c=i-3$($c=i-6$) for $i<7$ ($i\geq 7$). 
Therefore, $g^{XY}_{i VV'}=0$ for all $b\neq a$. The only nonzero contributions arise from $g^{LL}_{iVV}\propto  U^{\nu}_{ai} U^{\nu*}_{bi} \simeq \left(U_{\mathrm{PMNS}}\right)_{ai} \left(U_{\mathrm{PMNS}}\right)^*_{bi}$ with $i\leq3$. Therefore, the nonzero one-loop contributions to $c_{(ab)R}$ and cLFV decay rates are:
\begin{align}
\label{eq:cLL}
c^{LL}_{(ab)R}=&\frac{g^2c_{\theta}e m_{e_b}}{32\pi^2 m_V^2}\times \sum_{c=1}^3\left(U_{\mathrm{PMNS}}\right)_{ac}\left(U_{\mathrm{PMNS}}\right)^*_{bc} \tilde{f}_V\left( \frac{m^2_{n_c}}{m_V^2}\right),
\crn \mathrm{Br}^{LL}(e_b\to e_a \gamma)=&\frac{3c_{\theta}^2m_W^4}{2 \pi^2m_V^4} \left| \sum_{c=1}^3\left(U_{\mathrm{PMNS}}\right)_{ac}\left(U_{\mathrm{PMNS}}\right)^*_{bc} \tilde{f}_V\left( \frac{m^2_{n_c}}{m_V^2}\right)\right|^2\mathrm{Br}(e_b\to e_a \overline{\nu_a}\nu_b).
\end{align}
Therefore we can ignore $m_W^4/m_V^4\ll 1$ and set $c_{\theta}\simeq1$, then obtain the following constant values given in Eq. \eqref{eq:cLFVrates}. 

The couplings given in Eq. \eqref{eq:gXYiVVp} also lead to the one-loop form factors obtained from the sum of diagrams (7) and (8) in Fig. \ref{fig:ZebaU}:  
\begin{align}
\tilde{x}^{(7+8)}_{L(R)}= & \sum_{V=W,W'} \sum_{i=1}^{9} \tilde{x}^{iV}_{L(R)}, 
\crn \tilde{x}_L^{iV} =&  \dfrac{ t_L}{(m_a^2 -m^2_b) m_V^2}
%
\left\lbrace  g^{LL} \left[ \dots \right]  +g^{RR} m_am_{e_b} \left[ \dots \right]  +3\left( m_a g^{RL} +m_{e_b} g^{LR}\right)m_{n_i} m_V^2  \left(\dots \right) 
\right\rbrace,
\crn \tilde{x}_R^{iV} =& \tilde{x}_L^{iV} \left[ t_L\to t_R, g^{LL} \leftrightarrow  g^{RR},  g^{RL} \leftrightarrow  g^{LR} \right],
\end{align}
where  $g^{XY} \equiv  g^{X*}_{iV}g^{Y}_{iV}$ given in Eq. \eqref{eq:gXYiVVp}  and 
\begin{align}
\label{eq:tLR}
t_L= \frac{c_{\xi}(2s_W^2 -1) -s_Wt_{\zeta}s_{\xi}}{2s_Wc_W},\; t_R=\frac{s_Wc_{\xi}}{c_W}  -\frac{s_{\xi}c_{2\zeta}}{2c_Ws_{\zeta} c_{\zeta}}.
\end{align}
We note that both $t_L$ and $t_R$ reduce to the corresponding SM expressions \cite{Jurciukonis:2021izn} in the limit $\xi\to 0$. We also omit terms that vanish as a consequence of  $\sum_i g^{XX}=0$ for $X=L,R$..

One-loop form factors relating to  diagram (2) in Fig. \ref{fig:ZebaU} are:
\begin{align}
&\tilde{x}^{(2)}_{L(R)}=  \sum_{V=W,W'}\sum_{i,j =1}^9 \tilde{x}^{Vij}_{L(R)}, \; \tilde{y}^{(2)}_{L(R)}=  \sum_{V=W,W'}\sum_{i,j =1}^9 \tilde{y}^{Vij}_{L(R)}. 
\crn &\tilde{x}^{Vij}_L = \frac{g^{LL}}{m_V^2} \left\lbrace g^L_{Zij} \left[\dots\right] +g^L_{Zji} m_{n_i} m_{n_j}[\dots]\right\rbrace - \frac{g^{RR}g^L_{Zji} m_am_{e_b}}{m_V^2} \left[\dots \right],  \label{eq:xl2u} 
\\ &\tilde{y}^{Vij}_L  = \frac{2g^{LL} m_a}{m_V^2}  \left[\dots \right]
+  \frac{2g^{RR} m_{e_b}}{m_V^2}  \left[\dots\right],   \label{eq:yl2u} 
\\& \tilde{x}^{Vij}_R \left(\tilde{y}^{Vij}_R \right)  = \tilde{x}^{Vij}_L \left(\tilde{y}^{Vij}_L \right)\left[g^{LL} \leftrightarrow g^{RR},  g^L_{Zij} \leftrightarrow  -g^L_{Zji}\right], \label{eq:xyr2u}
\end{align} 
where $g^L_{Zij}$ is given in Eq. \eqref{eq:gLZij}, arguments for PV-funtions are $( m_V^2,m_{n_i}^2, m_{n_j}^2)$, and  $g^{XY}\equiv g^{XY}_{Vij}= g^{X*}_{aiV} g^{Y}_{bjV}$ is listed as follows:
\begin{align}
\label{eq:gXYVij}
g^{LL}_{Wij}=& \frac{g^2 c^2_{\theta}}{2} U^{\nu}_{ai} U^{\nu*}_{bj},\; g^{RR}_{Wij}= \frac{g^2 t^2_W s^2_{\theta}}{2 s^2_{\zeta}} U^{\nu*}_{(a+3)i} U^{\nu}_{(b+3)j},\;  
\crn g^{LR}_{Wij}=& \frac{g^2 s_{\theta} c_{\theta} t_W}{2 s_{\zeta}} U^{\nu}_{ai} U^{\nu*}_{(b+3)j},\;  g^{RL}_{Wij}= \frac{g^2 s_{\theta} c_{\theta} t_W}{2 s_{\zeta}} U^{\nu*}_{(a+3)i} U^{\nu}_{b j},
\crn g^{LL}_{W'ij}=& \frac{g^2 s^2_{\theta}}{2} U^{\nu}_{ai} U^{\nu*}_{bj},\; g^{RR}_{W'ij}= \frac{g^2 t^2_W c^2_{\theta}}{2 s^2_{\zeta}} U^{\nu*}_{(a+3)i} U^{\nu}_{(b+3)j},\;  
\crn g^{LR}_{W'ij}=& -\frac{g^2 s_{\theta} c_{\theta} t_W}{2 s_{\zeta}} U^{\nu}_{ai} U^{\nu}_{(b+3)j},\;  g^{RL}_{W'ij}= -\frac{g^2 s_{\theta} c_{\theta} t_W}{2 s_{\zeta}} U^{\nu*}_{(a+3)i} U^{\nu*}_{b j}.
\end{align}
Because of  the  following properties,
\begin{align}
\label{eq:sumgVij}
&\sum_{i,j=1}^9 g^L_{Zij}g^{LL} \varpropto \delta_{ab},\delta_{(a+3)b}=0, 
\crn &\sum_{i,j=1}^9 g^L_{Zji}g^{LL} m_{n_i}m_{n_j} \varpropto \sum_{j=1}^9 U^{\nu*}_{aj}m_{n_j} U^{\nu*}_{bj},\; \sum_{j=1}^9 U^{\nu*}_{(a+3)j}m_{n_j} U^{\nu*}_{bj}\to \mathcal{M}^{\nu*}_{ab},\mathcal{M}^{\nu*}_{(a+3)b} \left( =(m_D)^{\dagger}_{ab}\right)=0,
\crn &\sum_{i,j=1}^9 g^L_{Zji}g^{RR} \varpropto \delta_{a(b+3)},\delta_{(a+3)(b+3)}=0, 
\end{align}
 many divergent parts appearing in the original formulas in Ref.  \cite{Hue:2024rij}  vanish, leading to the simplified expression in Eq. \eqref{eq:xl2u},  fixing $d=4$. The intermediate steps leading to these results were discussed in Ref. \cite{Thao:2017qtn, Nguyen:2018rlb}. 

One-loop form factors from diagram (3) in Fig. \ref{fig:ZebaU} are:   
\begin{align*}
 \tilde{x}^{(3)}_{L(R)} =&  \sum_{V=W,W'} \sum_{i=1}^9 \tilde{x}^{iVh^+}_{L(R)},
 \quad \tilde{y}^{(3)}_{L(R)} = \sum_{V=W,W'} \sum_{i=1}^9 \tilde{y}^{iVh^+}_{L(R)},
\crn \tilde{x}^{iVh^+}_L =& \frac{g^*_{h^-V^+Z}}{m_V^2} \left[ \frac{}{}g^{LL} m_{n_i} (\dots ) + g^{RL} m_a (\dots ) -g^{LR} (\dots )\right],  
\\ \tilde{y}^{iVh^+}_L  =& \frac{g^*_{h^-V^+Z}}{m_V^2} \left[ -m_{n_i} \left(g^{LL}(\dots ) +g^{RR}(\dots )\right) 
 + g^{RL} (\dots )+ g^{LR} (\dots ) \right],   
\\ \tilde{x}^{iVh^+}_R \left(\tilde{y}^{iVh^+}_R\right)=& \tilde{x}^{iVh^+}_L\left(\tilde{y}^{iVh^+}_L\right)\left[\frac{}{} g^{LL} \leftrightarrow  g^{RR},  g^{RL} \leftrightarrow  g^{LR} \right], 
%
\end{align*} 
where $g_{h^-V^+ Z}$ given in Eq. \eqref{eq:gZBB} for particular gauge bosons $V=W,W'$, and $g^{XY} \equiv g^{XY}_{iVh^+}= g^{X*}_{aiV}g^{Y}_{bih^+}$, which is listed as follows.  
\begin{align}
\label{eq:giVhLR}
g^{LL}_{iWh^+}=&  -  \frac{gc_{\theta} c_{\kappa}}{vc_{2\beta}} \times    \left( \hat{\mathcal{M}}_{\ell} - m_D s_{2\beta}\right)_{bb} U^{\nu}_{ai} U^{\nu}_{(b+3)i},  
\crn  g^{RR}_{iWh^+}=& - \frac{g}{\sqrt{2}}\frac{t_Ws_{\theta}}{s_{\zeta}} U^{\nu*}_{(a+3)i} \left\{ \frac{\sqrt{2} c_{\kappa}}{vc_{2\beta}} \times  U^{\nu*}_{bi}\left(\hat{\mathcal{M}}_{\ell}s_{2\beta} - m_D\right)_{bb}
+  s_{\kappa}  U^{\nu*}_{(b+6)i} \left(Y^*_R\right)_{bb} \right\}, 
\crn   g^{LR}_{iWh^+}=& -  \frac{g c_{\theta}}{\sqrt{2}}U^{\nu}_{ai} \left\{ \frac{\sqrt{2} c_{\kappa}}{vc_{2\beta}} \times  U^{\nu*}_{bi}\left(\hat{\mathcal{M}}_{\ell}s_{2\beta} - m_D\right)_{bb}
+  s_{\kappa}  U^{\nu*}_{(b+6)i} \left(Y^*_R\right)_{bb} \right\},
\crn  g^{RL}_{iWh^+}=&  -  \frac{gt_W s_{\theta} c_{\kappa}}{vc_{2\beta} s_{\zeta}} \times    \left( \hat{\mathcal{M}}_{\ell} - m_D s_{2\beta}\right)_{bb} U^{\nu*}_{(a+3)i} U^{\nu}_{(b+3)i},
\crn g^{LL}_{iW'h^+}=&  \frac{g s_{\theta} c_{\kappa}}{vc_{2\beta}} \times    \left( \hat{\mathcal{M}}_{\ell} - m_D s_{2\beta}\right)_{bb} U^{\nu}_{ai} U^{\nu}_{(b+3)i},  
\crn  g^{RR}_{iW'h^+}=& - \frac{g}{\sqrt{2}}\frac{t_Wc_{\theta}}{s_{\zeta}} U^{\nu*}_{(a+3)i} \left\{ \frac{\sqrt{2} c_{\kappa}}{vc_{2\beta}} \times  U^{\nu*}_{bi}\left(\hat{\mathcal{M}}_{\ell}s_{2\beta} - m_D\right)_{bb}
+  s_{\kappa}  U^{\nu*}_{(b+6)i} \left(Y^*_R\right)_{bb} \right\},    
\crn   g^{LR}_{iW'h^+}= &  \frac{g s_{\theta}}{\sqrt{2}}U^{\nu}_{ai} \left\{ \frac{\sqrt{2} c_{\kappa}}{vc_{2\beta}} \times  U^{\nu*}_{bi}\left(\hat{\mathcal{M}}_{\ell}s_{2\beta} - m_D\right)_{bb}
+  s_{\kappa}  U^{\nu*}_{(b+6)i} \left(Y^*_R\right)_{bb} \right\}, 
\crn  g^{RL}_{iW'h^+}= &  -  \frac{gt_W c_{\theta} c_{\kappa}}{vc_{2\beta}s_{\zeta}} \times    \left( \hat{\mathcal{M}}_{\ell} - m_D s_{2\beta}\right)_{bb} U^{\nu*}_{(a+3)i} U^{\nu}_{(b+3)i},
\end{align}
One-loop form factors from diagram (4)    in Fig. \ref{fig:ZebaU}  are: 
\begin{align}
\tilde{x}^{(4)}_{L(R)} = &  \sum_{V=W,W'} \sum_{i=1}^9 \tilde{x}^{ih^+V}_{L(R)},
 \quad \tilde{y}^{(4)}_{L(R) } = \sum_{V=W,W'} \sum_{i=1}^9 \tilde{y}^{ih^+V}_{L(R)},
\crn \tilde{x}^{ih^+V}_L =& \frac{g_{h^-V^+Z}}{m_V^2} \left[ g^{LL} (\dots ) -g^{LR} (\dots )\right],  \label{eq:x4L} 
\\ \tilde{y}^{ih^+V}_L  =& \frac{g_{h^-V^+ Z}}{m_V^2} \left[ -m_{n_i} \left(g^{RR}(\dots ) +g^{LL} (\dots )\right)   + g^{LR}(\dots )+ g^{RL} (\dots ) \right],   \label{eq:y4L} 
%
\\ \tilde{x}^{ih^+V}_R \left( \tilde{y}^{ih^+V}_R\right)=& \tilde{x}^{ih^+V}_L \left( \tilde{y}^{ih^+V}_L\right) \left[ g^{LL} \leftrightarrow  g^{RR},  g^{RL} \leftrightarrow  g^{LR} \right], \label{eq:y4R}
\end{align} 
where $g_{h^-V^+ Z}$ given in Eq. \eqref{eq:gZBB} for particular gauge bosons $V=W,W'$, and  $g^{XY} \equiv g^{XY}_{ih^+V}= g^{X*}_{aih^+}g^{Y}_{biV}$ expresses precisely as follows:   
\begin{align}
\label{eq:gihVLR}
g^{LL}_{ih^+W}=& -  \frac{g c_{\theta} c_{\kappa}}{vc_{2\beta}}    \left( \hat{\mathcal{M}}_{\ell} - m_D^* s_{2\beta}\right)_{aa}  U^{\nu*}_{(a+3)i} U^{\nu*}_{bi},  
\crn   g^{RR}_{ih^+W}=&  - \frac{g  t_Ws_{\theta}}{\sqrt{2} s_{\zeta} } \left[ \frac{\sqrt{2} c_{\kappa}}{vc_{2\beta}} \times  U^{\nu}_{ai} \left(\hat{\mathcal{M}}_{\ell}s_{2\beta} - m_D^* \right)_{aa}
+  s_{\kappa}  U^{\nu}_{(a+6)i} \left(Y_R\right)_{aa}\right]   U^{\nu}_{(b+3)i} , 
\crn   g^{LR}_{ih^+W}=& -  \frac{g  t_W c_{\kappa} s_{\theta}}{vc_{2\beta} s_{\zeta} }   \left( \hat{\mathcal{M}}_{\ell} - m_D^* s_{2\beta}\right)_{aa}  \times U^{\nu*}_{(a+3)i}  U^{\nu}_{(b+3)i},
\crn  g^{RL}_{ih^+W}= & - \frac{g  c_{\theta}}{ \sqrt{2}} \left[ \frac{\sqrt{2} c_{\kappa}}{vc_{2\beta}}\left(\hat{\mathcal{M}}_{\ell}s_{2\beta} - m_D^* \right)_{aa}  U^{\nu}_{ai}
+  s_{\kappa}  U^{\nu}_{(a+6)i} \left(Y_R\right)_{aa}\right]  U^{\nu*}_{bi},
\crn g^{LL}_{ih^+W'}=&  \frac{g c_{\theta} s_{\kappa}}{vc_{2\beta}}    \left( \hat{\mathcal{M}}_{\ell} - m_D^* s_{2\beta}\right)_{aa}  U^{\nu*}_{(a+3)i} U^{\nu*}_{bi},  
\crn g^{RR}_{ih^+W'}= &  - \frac{g  t_W c_{\theta}}{\sqrt{2}s_{\zeta} } \left[ \frac{\sqrt{2} c_{\kappa}}{vc_{2\beta}} \times  U^{\nu}_{ai} \left(\hat{\mathcal{M}}_{\ell}s_{2\beta} - m_D^* \right)_{aa}
+  s_{\kappa}  U^{\nu}_{(a+6)i} \left(Y_R\right)_{aa}\right]   U^{\nu}_{(b+3)i},
\crn   g^{LR}_{ih^+W'}= &  -  \frac{g  t_W c_{\kappa} c_{\theta}}{vc_{2\beta} s_{\zeta} }   \left( \hat{\mathcal{M}}_{\ell} - m_D^* s_{2\beta}\right)_{aa}  \times U^{\nu*}_{(a+3)i}  U^{\nu}_{(b+3)i},
\crn  g^{RL}_{ih^+W'}=&  \frac{g  s_{\theta}}{ \sqrt{2}} \left[ \frac{\sqrt{2} c_{\kappa}}{vc_{2\beta}}\left(\hat{\mathcal{M}}_{\ell}s_{2\beta} - m_D^* \right)_{aa}  U^{\nu}_{ai}
+  s_{\kappa}  U^{\nu}_{(a+6)i} \left(Y_R\right)_{aa}\right]  U^{\nu*}_{bi},
\end{align}

Form factors corresponding to diagram (5) are:
\begin{align}
\label{eq_ab4LR}	
 \tilde{x}^{(5)}_{L} &= - 2g_{Zh^+h^-} \times \sum_{i=1}^9 g^{LL}_{ih^+h^-} C_{00},
\crn  \tilde{x}^{(5)}_{R} &= -2g_{Zh^+h^-} \times \sum_{i=1}^9 g^{RR}_{ih^+h^-} C_{00},
\crn  \tilde{y}^{(5)}_{L} &= -2g_{Zh^+h^-} \sum_{i=1}^9  \left[   m_a  g^{LL}_{ih^+h^-} X_1  +  m_{e_b}  g^{RR}_{ih^+h^-} X_2  -m_{n_i} g^{RL}_{ih^+h^-}  X_0  \right] ,
\crn   \tilde{y}^{(5)}_{L} &= -2g_{Zh^+h^-} \sum_{i=1}^9  \left[  m_a  g^{RR}_{ih^+h^-}  X_1  +  m_{e_b}  g^{LL}_{ih^+h^-} X_2   -m_{n_i} g^{LR}_{ih^+h^-} X_0  \right],
\end{align}
where $g_{Zh^+h^-}$ is given in Eq. \eqref{eq:gZBB}, arguments for PV-funtions are $(m_{n_i}^2, m_{h^+}^2, m_{h^-}^2)$, and $g^{XY}_{ih^+h^-}=g^{X*}_{aih^+}g^{Y}_{bih^+}$ 
with the LFV couplings of $h^+$ are given in Eq. \eqref{eq:gX}:
\begin{align}
\label{eq:gXYihh}
g^{LL}_{ih^+h^-}\propto& U^{\nu*}_{(a+3)i} U^{\nu}_{(b+3)i}\propto \delta_{ab},
\\ g^{RR}_{ih^+h^-}=&\left[ \frac{ c_{\kappa}}{c_{\beta}} \left(\frac{\sqrt{2}m_{e_a}s_{\beta}}{v} - \tilde{Y}^{a}\right)U^{\nu}_{ai}
+  \frac{\sqrt{2}s_{\kappa}}{v_R}  \sum_{c=1}^{3}U^{\nu}_{(c+6)i} \left(M_R\right)_{ca}\right]
\crn &\times \left[ \frac{ c_{\kappa}}{c_{\beta}} \left(\frac{\sqrt{2}m_{e_b}s_{\beta}}{v} - \tilde{Y}^{b}\right)U^{\nu*}_{bi}
+  \frac{\sqrt{2}s_{\kappa}}{v_R}  \sum_{c=1}^{3}U^{\nu*}_{(c+6)i} \left(M_R^*\right)_{cb}\right],
\crn g^{LR}_{ih^+h^-}\propto&U^{\nu}_{(a+3)i}  \left[ \frac{ c^2_{\kappa}}{c_{\beta}} \left(\frac{\sqrt{2}m_{e_b}s_{\beta}}{v} - \tilde{Y}^{b}\right)U^{\nu*}_{bi}
+  \frac{\sqrt{2}s_{2\kappa}}{2v_R}  \sum_{c=1}^{3}U^{\nu*}_{(c+6)i} \left(M_R^*\right)_{cb}\right] \propto \delta_{ab},
\crn g^{RL}_{ih^+h^-}\propto&\left[ \frac{ c^2_{\kappa}}{c_{\beta}} \left(\frac{\sqrt{2}m_{e_b}s_{\beta}}{v} - \tilde{Y}^{a*}\right)U^{\nu}_{ai}
+  \frac{\sqrt{2}s_{2\kappa}}{2v_R}  \sum_{c=1}^{3}U^{\nu}_{(c+6)i} \left(M_R\right)_{ca}\right]U^{\nu}_{(b+3)i} \propto \delta_{ab}.\nn
\end{align}
It can be seen that $g^{RR}_{ih^+h^-}=0$ for $i>3$. For $i\leq 3$, we have: $\sum_{c=1}^{3}U^{\nu}_{(c+6)i} \left(M_R\right)_{ca}=-m^a_D \left(U^{\nu}_3\right)_{ai}$. Ignoring suppressed terms proportional to $s_{\kappa}m_D^a/v_R$ and $R^0R^{0\dagger}$, we get:
 \begin{align}
\label{eq:gRRihhab}
g^{RR}_{ih^+h^-} \simeq &\frac{ 2c_{\kappa}^2}{c^2_{\beta}} \tilde{Y}'^a \tilde{Y}'^b (U_{\mathrm{PMNS}})_{ai} (U^*_{\mathrm{PMNS}})_{bi},\; \tilde{Y}'^a\equiv \frac{m_{e_a}s_{\beta}}{v} - \frac{\tilde{Y}^{a}}{\sqrt{2}}.
\end{align} 
Denoting  $\tilde{F}_{S,ab}=\sum_{c=1}^3 (U_{\mathrm{PMNS}})_{ac} (U^*_{\mathrm{PMNS}})_{bc} \tilde{f}_S(\frac{M^2_{c}}{m_{h^+}^2})$, this gives:
\begin{align}
\label{eq:cRR}
c^{RR}_{(ab)R}  \simeq &\frac{ ec_{\kappa}^2 m_{e_a}}{c^2_{\beta} 4\pi^2 m_{h^+}^2} \tilde{Y}'^a \tilde{Y}'^b  \tilde{F}_{S,ab},
\crn 
\Delta a^{RR}_{e_a}\simeq &\frac{ c_{\kappa}^2 m^2_{e_a}(\tilde{Y}'^a)^2}{c^2_{\beta} \pi^2 m_{h^+}^2}  \mathrm{Re}\left(\tilde{F}_{S,aa}\right) ,
\crn \mathrm{Br}^{RR}(e_b\to e_a \gamma)=&\frac{3e^2\pi^2}{G_F^2 m^2_{e_a}m^2_{e_b}}\times  
\frac{\left|  \tilde{F}_{S,ba}\right|^2 \mathrm{Br}(e_b\to e_a \overline{\nu_a}\nu_b)\times \left| \Delta a^{RR}_{e_a} \Delta a^{RR}_{e_b}\right| }{\left|\mathrm{Re}\left(\tilde{F}_{S,aa}\right) \mathrm{Re}\left(\tilde{F}_{S,bb}\right)\right|}.
\end{align}
Because $m_{n_i}\ll m_{h^+}$ and the unitary of $U_{\mathrm{PMNS}}$, we have $\tilde{F}_{S,11}\simeq 0.042$, while $|\tilde{F}_{S,12}|<\mathcal{O}(10^{-27})$, $|\tilde{F}_{S,13}|<\mathcal{O}(10^{-27})$, and $|\tilde{F}_{S,23}|<\mathcal{O}(10^{-26})$ for $m_{h^+}>100$ GeV. Therefore, $ \mathrm{Br}^{RR}(\mu\to e \gamma)<\mathcal{O}(10^{-32})\left| \Delta a^{RR}_{e_a} \Delta a^{RR}_{e_b}\right|<\mathcal{O}(10^{-54})$.

Forms factors corresponding to diagram (6) are 
\begin{align}
\label{eq_ab6LR}	
 \tilde{x}^{(6)}_{L(R)} = &\sum_{i,j=1}^9 \tilde{x}^{h^+ij}_{L(R)},
\quad \tilde{y}^{(6)}_{L(R)} = \sum_{i,j=1}^9 \tilde{y}^{h^+ij}_{L(R)}, 
\crn \tilde{x}^{h^+ij}_{L}=& - \left\{  g^{L}_{Zij}\left[ g^{LL}  m_{n_i} m_{n_j}C_0 + g^{RL}  m_{a} m_{n_j} C_{(0+1)} 
%
+  g^{LR}  m_{b} m_{n_i} C_{(0+2)} +  g^{RR} m_{a} m_{e_b}X_0 \frac{}{}\right]   
\right.\crn  &\left. \hspace{0.6 cm}+g^{L}_{Zji} \left[ g^{LL} \left(2C_{00} +m_a^2 X_1 +m_{e_b}^2X_2 -m_Z^2 C_{12}\right) \frac{}{}
%
+m_am_{n_i}g^{RL}C_1 + m_{e_b}m_{n_j}g^{LR}C_2 \right]\right\}
\crn  \tilde{x}^{h^+ij}_{R}=& - \left\{  -g^{L}_{Zij} \left[ g^{RR} \left( 2 C_{00} +m_a^2 X_1 +m_{e_b}^2X_2 -m_Z^2 C_{12}\right) \frac{}{}
%
 +g^{LR} m_am_{n_i}C_1 +g^{RL}m_{e_b}m_{n_j}C_2 \right]
\right.\crn  &\left. \hspace{0.5 cm} 
- g^{L}_{Zji}\left[\frac{}{}  g^{RR}  m_{n_i} m_{n_j}C_0 + g^{LR}  m_{a} m_{n_j} C_{(0 +1)} 
%
+  g^{RL}  m_{b} m_{n_i} C_{(0+2)} +  g^{LL}  m_{a} m_{e_b}X_0\right] 	\right\},
\crn \tilde{y}^{h^+ij}_{L}=& -2 \left[\frac{}{} g^{L}_{Zij}  \left( g^{RL} m_{n_j}C_2 +  g^{RR}  m_{b} X_2\right) 
%
-g^{L}_{Zji}  \left(  g^{RL}  m_{n_i}C_1 +  g^{LL}  m_{a} X_1\right) 
\right]	,
\crn\tilde{y}^{h^+ij}_{R} =&- 2 \left[\frac{}{} g^{L}_{Zij}   \left(  g^{LR} m_{n_i}C_1 +  g^{R R}  m_{a} X_1\right)
%
-g^{L}_{Zji} \left( g^{LR} m_{n_j}C_2 +  g^{LL}  m_{b} X_2\right) 
\right], 
\end{align}
where the arguments for PV-functions are $(m_{h^+}^2,m^2_{i},m^2_{j})$ and $g^{XY} \equiv g^{X*}_{aih^+}g^{Y}_{bjh^+}$ with $X,Y=L,R$, and particular expressions are given in  Eq. \eqref{eq:gX}. Because $\sum_{i,j=1}^9 g^L_{Z_{ji}}g^{LL}= \sum_{i,j=1}^9 g^L_{Z_{ij}}g^{RR}=0$ the divergent parts in Eq. \eqref{eq_ab6LR} vanish.

Sum of two diagrams (9) and (10) gives the following form factors 
\begin{align}
\label{eq_a910LR}
 \tilde{x}^{(9+10)}_{L}  = & - \frac{ t_{L}}{m_a^2 -m_{e_b}^2}  \sum_{i=1}^9 \left[  m_{n_i} \left( m_a g^{RL}  + m_{e_b} g^{LR}   \right) (\dots )  -m_am_{e_b} g^{RR}(\dots )  - g^{LL}(\dots )
\right],
\crn \tilde{x}^{(9+10)}_{R} =& \tilde{x}^{(9+10)}_{L}  \left[ t_L\to t_R,   g^{LL} \leftrightarrow  g^{RR},  g^{RL} \leftrightarrow  g^{LR} \right],
\end{align}
where $g^{XY}\equiv g^{X*}_{aih^+} g^{Y}_{bih^+}$  derived from Eq. \eqref{eq:gX}, $t_{L(R)}$ is given in Eq. \eqref{eq:tLR},  and  $B^{(k)}_{0,1}=B_{0,1}(p_k^2;m_{n_i}^2,m_{h^+}^2)$.   Note that the last four diagrams in Fig. \ref{fig:ZebaU} do not contribute to $\tilde{y}_{L(R)}$.

The respective partial decay width in the limit  $m_{h}\gg m_{a,b}$  is   \cite{Arganda:2004bz}
\begin{equation}
\Gamma (h \rightarrow e_ae_b)\equiv\Gamma (h\rightarrow e_a^{-} e_b^{+})+\Gamma (h \rightarrow e_a^{+} e_b^{-})
\simeq   \fr{ m_{h}}{8\pi}\left(|\Delta_L|^2+|\Delta_R|^2\right), \label{eq_LFVwidth}
\end{equation}
where $\Delta_{L(R)} = \sum_{m=1}^{10}\Delta^{(m)}_{L(R)}$ is loop-contributions 
and the index $(ab)$ was omitted for simplicity, namely $\Delta^{(m)}_{L(R)}\equiv \Delta^{(m)ab }_{L(R)}$ for all $m=1,2,\dots, 10$. The corresponding branching ratio is  Br$(h\rightarrow e_ae_b)= \Gamma (h\rightarrow e_ae_b)/\Gamma^{\mathrm{total}}_{h}$ where $\Gamma^{\mathrm{total}}_{h}\simeq 4.1\times 10^{-3}$ GeV \cite{LHCHiggsCrossSectionWorkingGroup:2016ypw} with  $ q^2 \equiv( p_1+p_2)^2=m^2_{h}$.   
One-loop contributions to the LFV$h$ amplitude from diagram (1) in Fig. \ref{fig:hebaU} are:
\begin{align}
\Delta^{(1)}_{L(R)} =& \sum_{V=W,W'} \sum_{i=1}^9\Delta^{(iVV')}_{L(R)} , 
\crn \Delta^{iVV'}_L=&-g_{hVV'} \left( \frac{ g^{LL} m_a}{16\pi^2}\left[\dots\right]
+ \frac{  g^{RR} m_{e_b}}{16\pi^2}\left[\dots\right] +\frac{ g^{RL} m_{n_i}}{32\pi^2 m_V^2 m_{V'}^2} \left[\dots\right]
+ \frac{  g^{LR} m_{e_a}m_{e_b}m_{n_i}}{32 \pi^2 m_V^2 m_{V'}^2}  \left[\dots\right]\right)
, 
\crn  \Delta^{iVV'}_R=&  \Delta^{iVV'}_L\left[ g^{LL} \leftrightarrow  g^{RR},  g^{RL} \leftrightarrow  g^{LR} \right],\label{eq:DeFVVR}
\end{align}
where  $g_{hVV'}$ is given in Eq. \eqref{eq:hZBB}, $g^{XY}=g^{XY}_{iVV'}$ is given in Eq. \eqref{eq:gXYiVVp}. We omit many terms using the similar comments mentioned in discussion for diagram (1) of the LFV$Z$ decay.

The one-loop form factors corresponding to the diagram (2) in Fig. \ref{fig:hebaU} are:
\begin{align}
\label{eq:DeVFFL}
\Delta^{(2)}_{L(R)} &=\sum_{V=W,W'} \sum_{i,j=1}^9  \Delta^{Vij}_{L(R)}, 
\crn \Delta^{Vij}_L=&  \left\{ g^{LL} m_{e_a} \left[  g^{L}_{hij}  m_{n_i} \left(\dots\right)  +g^{L*}_{hij}  m_{n_j} \left( \dots \right)\right] +g^{RR} m_{e_b} \left[  g^{L}_{hij}  m_{n_j} \left( \dots \right)  + g^{L*}_{hij}  m_{n_i} \left( \dots \right)\right]
\right. \crn&\left. \hspace{.1cm} -g^{RL} \left[ g^L_{hij}\left(\dots \right) + g^{L*}_{hij} m_{n_i}m_{n_j} \left(\dots\right)
\right]  - g^{LR} g^{L*}_{hij} m_{e_a} m_{e_b} \left[ \dots \right] 
\right\} \frac{1}{16\pi^2 m_V^2},
\crn  \Delta^{Vij}_R=&  \Delta^{Vij}_L\left[ g^L_{hij} \leftrightarrow g^{L*}_{hij}, g^{LL}  \leftrightarrow g^{RR},\; g^{LR}  \leftrightarrow g^{RL}\right],
\end{align}
where  $g^L_{hij}$ and  $g^{XY}=g^{XY}_{Vij} = g^{X*}_{aiV}g^{Y}_{bjV}$ are  given in Eqs. \eqref{eq:lahij} and \eqref{eq:gXYVij}, respectively. 

 Two diagrams (7) and (8) gives the following one-loop form factors:
\begin{align}
\label{eq:DeLR78}
 \Delta^{(7) }_L = &  -\sum_{i=1}^9\frac{gm_{e_b}r_{hbb} \left\{ 
 	\left( g^{LL}  m_{e_b} +   g^{RR} m_{e_a}\right)m_{e_a}  \left[\dots \right]  + \left( g^{RL} m_{e_b} +g^{LR} m_{e_a} \right) m_{n_i} \left[  \dots\right]
 	\right\}}{32\pi^2 m_W m_V^2 (m_{e_a}^2 -m_{e_b}^2)}  ,
\crn \Delta^{(8) }_L =&\sum_{i=1}^9 \frac{gm_{e_a}r_{haa}  \left\{ 
	\left( g^{LL}  m_{e_b} +  g^{RR} m_{e_a} \right)  m_{e_b}\left[\dots\right]   + \left( g^{RL} m_{e_a} +g^{LR} m_{e_b} \right) m_{n_i} \left[ \dots\right]
	\right\}}{32\pi^2 m_W m_V^2 (m_{e_a}^2 -m_{e_b}^2)}, 
\crn \Delta^{(7+8) }_R= & \left( \Delta^{(7)}_L + \Delta^{(8)}_L\right) \left[ g^{LL} \leftrightarrow g^{RR},  g^{RL} \leftrightarrow  g^{LR} \right],
\end{align}
where $g^{XY}=g^{X*}_{aiV} g^{Y}_{biV}$ is given in Eq. \eqref{eq:gXYiVVp}   and 
\begin{align}
\label{eq:rhaa} r_{haa} = \left[c_{\delta} -t_{2\beta}s_{\delta}+ \frac{(m_D^T)_{aa}s_{\delta}}{m_{e_a}c_{2\beta}} \right] .
\end{align}
Note that the limit  $\delta=0$ gives $r_{haa}=1$, consistent with the SM result. We  choose here the simple case that $\delta=0$, then $r_{haa}=r_{hee}$ for all $a=1,2,3$. 

The  form factors corresponding to  diagram $(3)$ in Fig. \ref{fig:hebaU}  are:
\begin{align}
\label{eq:DelFVSL}
\Delta^{(3)}_{L(R)} =& \sum_{V=W,W'} \sum_{i=1}^9 \Delta^{ iVh^+}_{L(R)},
\crn \Delta_L^{iVh^+}= &  \frac{g_{Vh^+h}}{16\pi^2 m_V^2}\left\{m_{n_i} \left[ g^{LL} m_{e_a}   \left(\dots\right) -g^{RR} m_{e_b} \left(\dots\right)  \right]   + g^{RL} \left[\dots\right]  - g^{LR} m_{e_a}m_{e_b} \left[ \dots \right]   \right\},
\crn 	\Delta_R^{iVh^+}= &	\Delta_L^{iVh^+}\left[ g^{LL} \leftrightarrow  g^{RR},  g^{RL} \leftrightarrow  g^{LR} \right],
\end{align}
where $g^{XY}=g^{X*}_{aiV}g^{Y}_{bih^+}$ in Eq. \eqref{eq:gihVLR} and the arguments of PV-functions are  $(m_{n_i}^2,m_V^2,m_{h^+}^2)$.

Final results for form factors corresponding to diagram (4) in Fig. \ref{fig:hebaU}: 
\begin{align}
\label{eq:Delta4LR}
\Delta^{(4)}_{L(R)}= & \sum_{V=W,W'} \sum_{i=1}^9 \Delta^{ih^+V}_{L(R)},
\crn 
\Delta^{ih^+V}_L= & \frac{g^*_{Vh^+h}}{ 16\pi^2 m_V^2}   \left\{  m_{n_i}  \left[ -g^{LL} m_{e_a}  \left(\dots\right) +g^{RR}m_{e_b} \left(\dots\right) \right]   +   g^{RL} \left[\dots \right]  -g^{LR} m_{e_a}m_{e_b}  \left(\dots\right)   \right\} ,
\crn 	\Delta^{ih^+V}_R= &	\Delta^{ih^+V}_L\left[g^{LL} \leftrightarrow  g^{RR},  g^{RL} \leftrightarrow  g^{LR} \right],
\end{align}
where $g^{XY}=g^{X*}_{aih^+} g^{Y}_{biV}$ given in Eq. \eqref{eq:gihVLR}.

Final results for form factors corresponding to diagram (5) in Fig. \ref{fig:hebaU}: 
\begin{align}
\label{eq:De5LR}
 \Delta^{(5)}_{L} =& \sum_{i=1}^9 \frac{\lambda_{hh^+h^-}}{ 16\pi^2}  \left[ g^{RL}m_{n_i}  C_0  -(g^{LL} m_{e_a}C_1 +g^{RR} m_{e_b}C_2)\right] ,
\crn 	 \Delta^{(5)}_{R}= & \Delta^{(5)}_{L}\left[ g^{LL} \leftrightarrow  g^{RR},  g^{RL} \leftrightarrow  g^{LR} \right],
\end{align}
where  $g^{XY}=g^{X*}_{aih^+}g^{Y}_{bih^+}$ derived from  Eq. \eqref{eq:gX} and the PV arguments are $(m_{n_i}^2,m_{h^+}^2,m_{h^+}^2)$.

Final results for form factors corresponding to diagram (6) in Fig. \ref{fig:hebaU}: 
\begin{align}
\label{eq:De6LR}
 \Delta^{(6)}_{L} &=  \sum_{i,j=1}^9  \left\{  g^L_{hij} \left[ m_{n_i}  \left( g^{RL}   m_{n_j} C_0  + g^{RR}  m_{e_b}C_{(0+2)} \right)  
%
+  m_{e_a}\left( g^{LL} m_{n_j}  C_{(0+1)} +g^{LR}   m_{e_b} X_0\right)\right]
\frac{}{} \right. \crn& \left. \frac{}{} \hspace{1.2 cm} + g^{L*}_{hij} \left[  g^{RL} \left( \dots  \right)  + g^{LL} m_{e_a} m_{n_i} C_1 +g^{RR} m_{e_b} m_{n_j} C_2 \right]\right\}  \frac{1}{ 16\pi^2},
\crn 	\Delta^{(6)}_{R}= &	\Delta^{(6)}_{L} \left[ g^{LL} \leftrightarrow  g^{RR},  g^{RL} \leftrightarrow  g^{LR} \right],
\end{align}
where $g^{XY}=g^{X*}_{aih^+}g^{Y}_{bjh^+}$ derived from  Eq. \eqref{eq:gX}  and the PV  arguments  $(m_{h^+}^2,m_{n_i}^2,m_{n_j}^2)$.

Form factors corresponding to sum of two diagrams (9) and (10):
\begin{align}
\label{eq:De910LR}
& \Delta^{(9+10)}_{L(R)} = \sum_{i=1}^9 \left( \Delta^{ (9)ih^+}_{L(R)} + \Delta^{ (10)ih^+}_{L(R)}\right),
\crn & \Delta^{(9) ih^+}_L=       \frac{g m_{e_b} r_{hbb} }{32\pi^2 m_W(m_{e_a}^2 -m_{e_b}^2)}   \left[  m_{n_i} \left( g^{LR}m_{e_a} + g^{RL}m_{e_b}\right)B^{(1)}_0  
 -m_{e_a} (g^{LL}  m_{e_b} + g^{RR} m_{e_a})  B^{(1)}_1   \frac{}{}\right] ,
\crn& \Delta^{(10) ih^+}_L=       \frac{-g m_{e_a} r_{haa} }{32\pi^2 m_W(m_{e_a}^2 -m_{e_b}^2)}   \left[ m_{n_i} \left( g^{LR}m_{e_b} + g^{RL}m_{e_a}\right)B^{(2)}_0    - m_{e_b}(g^{LL}  m_{e_b} + g^{RR} m_{e_a}) B^{(2)}_1   \frac{}{}\right] ,
\crn& 	\Delta^{(9)ih^+}_R +\Delta^{(10)ih^+}_R= 	\left( \Delta^{(9)ih^+}_L + \Delta^{(10)ih^+}_L\right)\left[ g^{LL} \leftrightarrow  g^{RR},  g^{RL} \leftrightarrow  g^{LR} \right],
\end{align}
where  $g^{XY}=g^{X*}_{aih^+}g^{Y}_{bih^+}$  given in  Eq. \eqref{eq:gX},    $B^{(k)}_{0,1}=B_{0,1}(p_k^2; m_{n_i}^2,m_{h^+}^2)$ with $k=1,2$, and $r_{haa}$ is given in Eq. \eqref{eq:rhaa}.

\end{document}